%% file: Main.tex
\documentclass[sigconf]{acmart}

\renewcommand\footnotetextcopyrightpermission[1]{}
\usepackage{booktabs} 

\usepackage{soul}

\usepackage{mathptmx}
\usepackage{subcaption}
\usepackage{graphicx}
\usepackage{epsfig}
\usepackage{epstopdf}
\usepackage{multirow}
\usepackage{hyperref}
\hypersetup{
    colorlinks = true,
    citecolor = blue,
    linkcolor = red,
    urlcolor = cyan,
}

\usepackage{enumitem}
\usepackage{amsmath}
\usepackage{fixltx2e}
\usepackage{fancyhdr}

\usepackage{tikz}
\newcommand*\circled[1]{\tikz[baseline=(char.base)]{
            \node[fill=black,shape=circle,draw,inner sep=0.5pt] (char) {#1};}}

\setcopyright{none}
\setcopyright{acmcopyright}
\setcopyright{acmlicensed}
\setcopyright{rightsretained}
\setcopyright{usgov}
\setcopyright{usgovmixed}
\setcopyright{cagov}
\setcopyright{cagovmixed}

\acmDOI{10.475/123_4}
\acmISBN{123-4567-24-567/08/06}
\acmConference[SIGMETRICS'17]{ACM SIGMETRICS}{June 2017}{Champaign-Urbana, Illinois USA} 
\acmYear{2017}
\copyrightyear{2017}
\acmPrice{15.00}

\begin{document}

\title{Exploiting Data Longevity for Enhancing the Lifetime of Flash-based Storage Class Memory}

\author{Wonil Choi}
\affiliation{
  \institution{Pennsylvania State University}
}
\email{wuc138@cse.psu.edu}

\author{Mohammad Arjomand}
\affiliation{
  \institution{Pennsylvania State University}
}
\email{mxa51@psu.edu}

\author{Myoungsoo Jung}
\affiliation{
  \institution{Yonsei University}
}
\email{m.jung@yonsei.ac.kr}

\author{Mahmut Kandemir}
\affiliation{
  \institution{Pennsylvania State University}
}
\email{kandemir@cse.psu.edu}



\begin{abstract}
Storage-class memory (SCM) combines the benefits of a solid-state memory, such as high-performance and robustness, with the archival capabilities and low cost of conventional hard-disk magnetic storage. Among candidate solid-state nonvolatile memory technologies that could potentially be used to construct SCM, flash memory is a well-established technology and have been widely used in commercially available SCM incarnations.
Flash-based SCM enables much better tradeoffs between performance, space and power than disk-based systems.
However, write endurance is a significant challenge for a flash-based SCM (each act of writing a bit may slightly damage a cell, so one flash cell can be written 10\textsuperscript{4}--10\textsuperscript{5} times, depending on the flash technology, before it becomes unusable). 
This is a well-documented problem and has received a lot of attention by manufactures that are using some combination of write reduction and wear-leveling techniques for achieving longer lifetime. 
In an effort to improve flash lifetime, first, by quantifying data longevity in an SCM, we show that a majority of the data stored in a solid-state SCM do not require long retention times provided by flash memory (i.e., up to 10 years in modern devices); second, by exploiting retention time relaxation, we propose a novel mechanism, called Dense-SLC (D-SLC), which enables us perform multiple writes into a cell during each erase cycle for lifetime extension; and finally, we discuss the required changes in the flash management software (FTL) in order to use this characteristic for extending the lifetime of the solid-state part of an SCM. Using an extensive simulation-based analysis of a flash-based SCM, we demonstrate that D-SLC is able to significantly improve device lifetime (between 5.1$\times$ and 8.6$\times$) with no performance overhead and also very small changes at the FTL software.
\end{abstract}


\maketitle

\input{Sec-1-Intro-Mhmd}

\input{Sec-2-bkg-Mhmd}

\input{Sec-3-tech-Mhmd}

\input{eval-setup-Mhmd}

\input{eval-base}

\input{eval-sense-sum}

\input{related-works}

\input{conclusion}

\bibliographystyle{ACM-Reference-Format}
\bibliography{sigproc} 

\newpage


\end{document}

%% file: Sec-1-Intro-Mhmd.tex
\section{Introduction}
\label{intro}
During the last decade, the CPUs have become power constrained, and scaling of the logic devices no longer results in substantial performance improvement of computers. 
Therefore, it is imperative to consider developing additional ways for performance improvement.
For instance, one might target the memory wall problem and consider how to achieve higher overall performance by changing the memory-storage hierarchy.
Looking at the conventional memory-storage hierarchy, we observe that there is large performance-cost gap between DRAM (located near processor) \cite{dram-1,dram-2,dram-3,dram-4} and HDD \cite{hdd-1,hdd-2}, and this gap has become larger with the recent technology advances. 
Bridging this gap has the potential to boost system performance in all kinds of computing systems. This is possible with a high-performance, high-density and low-cost non-volatile memory (NVM) technology whose access time (as well as its cost-per-bit) falls between DRAM and HDD, and is called Storage Class Memory (SCM) \cite{scm-1,scm-2}. 
Despite the recent advances in NVM technologies (such as Phase Change Memory~\cite{pcm} and Magnetic RAM~\cite{mram}), it is quite unlikely to exploit them in SCM in any near future (because of their high fabrication costs). 
Instead, this paper assumes a NAND flash-based SCM which has been widely used in various kinds of commercial systems ranging from laptops and desktops to enterprise computers.

Flash memory stores binary data in the form of a charge, i.e., the amount of electrons it holds.
There are two types of popular flash memories: Single-Level Cell (SLC) and Multi-Level Cell (MLC). 
An SLC flash cell has two voltage states used for storing one-bit information, while an MLC flash cell has more than two states and stores 2 or more bit data at each time.
SLC is fast and has a long lifetime, but MLC trades off these metrics for higher density.
In order to have the benefits of both the technologies in the same system, a flash-based SCM typically has a hierarchal internal structure: there is a SLC Solid State Drive (SSD)~\cite{ssd-general-1,ssd-general-2,ssd-general-3,ssd-general-4} with tens of gigabytes capacity at the upper level, and an MLC SSD with terabyte capacity at the lower level.
Write endurance is a significant challenge for the SLC SSD in this setup. The reason is that the SLC SSD services a great portion of the incoming traffic which poses high write pressure on it (flash memory suffers from low cell endurance, i.e., each cell can tolerate 10\textsuperscript{4}--10\textsuperscript{5} program/erase cycles).

In this paper, we target the lifetime problem of SLC SSD in an SCM and discuss the opportunity for improving it by relaxing the \emph{retention time} of the flash, i.e., the period of time that a flash cell can correctly hold the stored data.
The flash devices traditionally have long retention times and are expected to retain data for one or more years. Although this long-term non-volatility is a must for many flash memory applications, there are also some cases where the stored data does \emph{not} require it. 
For example, within a memory-storage hierarchy, we expect the SLC SSD to handle the I/O requests with short-term longevity, while the I/O requests with long-term longevity are normally handled by the underlying MLC SSD or HDD.
Our characterization study in this work confirms this behavior -- we measure the longevity of written data into an SLC SSD for a wide range of enterprise (and OLTP) workloads taken from the MSR Cambridge I/O suite~\cite{cambridge}. 
We observe that a majority of written data into the SCM for all evaluated workloads exhibits very short longevity. In fact, more than 90\% of written data in these workloads have a longevity of up to 10 hours (it is less than 3 minutes for some applications and less than 10 hours for some others). 

The retention time relaxation for flash memory is previously studied by some works~\cite{relax-1,relax-2}. 
They have shown that, by controlling the write process in flash device, it is possible to reduce its retention time (more details on the theory behind this property are given in Section~\ref{drift}). 
The prior work mostly use this property for performance improvement of the flash by reducing its write execution time. In this paper, however, we use the retention time relaxation of flash to enhance its lifetime.
The main idea is that, by relaxing the retention time of an SLC device, we can have more than two states in a cell. At each given time,
similar to the baseline SLC, we use every two states to write one bit information. In this way, a device stores multiple bits (one bit at each time) before it needs an erase, increasing the number writes to cell during one erase cycle, or simply increasing the PWE\footnote{We use the term ``page writes per erase cycle'' (PWE) as the maximum number of \emph{logical pages} stored in \emph{one physical page} during \emph{one P/E cycle}.} of the device beyond the conventional SLC flash (i.e., one).
Increasing PWE of a device directly translates into lifetime improvement.


Our proposed flash memory design is called Dense-SLC (D-SLC) and its implementation needs two minor changes at FTL. First, the block-level allocation algorithm in FTL should be modified to enable having blocks with different retention times and use them for storing data values of different longevities. Our proposed block-level allocation algorithm does not require any specific mechanism for retention time estimation. Instead it uses a simple and yet effective page migration scheme that imposes negligible lifetime and bandwidth overhead.
Second, the garbage collection algorithm, which is needed for the flash management, is modified to ease the system-level implementation of writing multiple bits in one erase cycle. 
These modifications are simple to implement in an FTL and need two-bit metadata information per one block.
Using a detailed implementation of D-SLC flash memory in DiskSim simulator~\cite{framework1,framework2}, we evaluate its lifetime and performance efficiency for a large workload set. Our experimental evaluation confirms that a typical implementation of D-SLC is able to improve SLC SSD's lifetime by up to 8.6$\times$ (6.8$\times$, on average) with no degradation in the overall system performance.


%% file: Sec-2-bkg-Mhmd.tex
\begin{figure*}
  \centering
  \includegraphics[scale=.33]{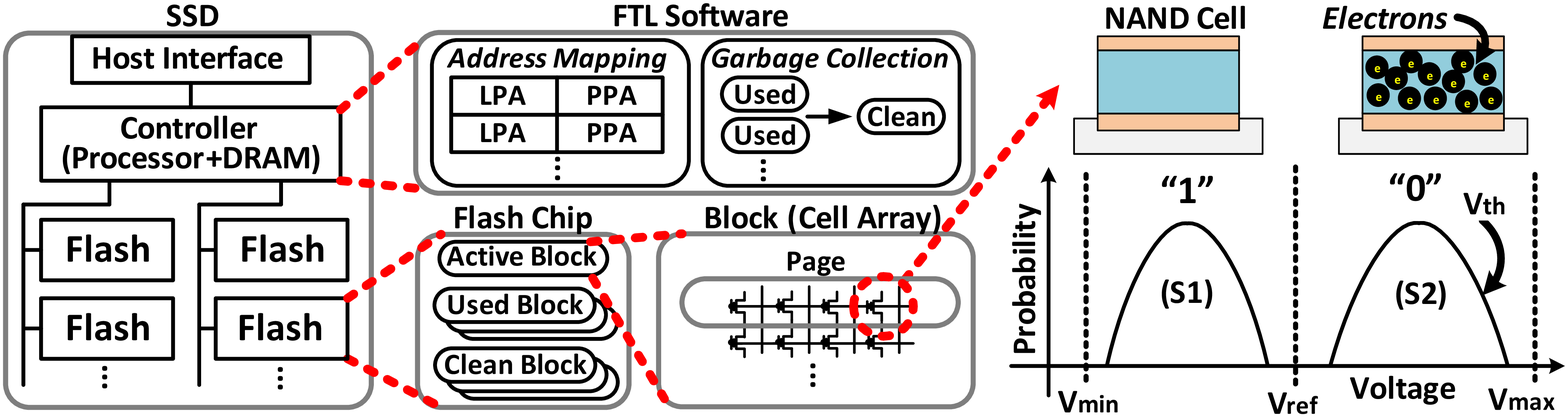}
  \caption{The internal architecture of an SSD with SLC NAND flash chips.}
  \label{figure:ssd-Internals}
\end{figure*}

\section{Preliminaries}
\label{bkg}
\subsection{SSD and Flash Memory}
Figure~\ref{figure:ssd-Internals} illustrates the internal architecture of an SSD which is composed of three components: 1) \emph{Host Interface} communicates with the host system, queues the incoming requests, and schedules them for services; 2) The \emph{SSD controller} is responsible for processing I/O requests and managing SSD resources by executing \emph{Flash Translation Layer} (FTL) software; 3) A set of NAND flash memory chips as the storage medium, which are connected to the controller via multiple buses.

\textbf{NAND flash chip:} A flash memory has thousands of \emph{blocks} and each block has hundreds of \emph{pages}. Each page is a row of NAND flash cells. Binary values of a cell is represented by its charge holding property. Flash memory has three main operations: \emph{read}, \emph{program (write)}, and \emph{erase}. 
Page is the unit of a read or a write operation, and reprogramming a cell needs to be preceded by an erase. Erase is performed at unit of a block.
Due to the need for erase-before-write operation and high latency of an erase, flash memory usually employs an out-of-place update policy, i.e., when updating a data, the page containing the old data is marked as \emph{invalid}, and the new data is written to an arbitrary clean page. The new page is marked as \emph{valid}.

\textbf{FTL:} 
The FTL implements some important functionalities for flash memory management. We go over two main FTL's functionalities in below.

\begin{itemize}[leftmargin=*]
\item \textbf{Address mapping:} On receiving an I/O request, FTL segments it into multiple pages and maps each page onto flash chips separately. Address mapping for a write request is a two-step process. First, a chip-level allocation strategy~\cite{page-alloc} determines which chip each page should be mapped to. Then, the page is mapped to one of the blocks and a page inside it (block allocation). For each chip, FTL always keeps one clean block as the active block for new page writes. Within an active block, clean pages are used to program data in a sequential order. Once the clean pages run out in the active block, a new active block is assigned.
Finally, the mapping information of each page (i.e., chip number, block number in the chip, and page index in the block) is stored in a \emph{mapping table} which is kept by FTL.
On receiving a read request, the FTL looks up the mapping table for finding its physical page location.
\item \textbf{Garbage collection (GC):} When the number of free pages falls below a specific threshold, the FTL triggers a GC procedure to reclaim the invalid pages and make some pages clean. When a GC is invoked, the target blocks are selected, their valid pages are moved (written) to other places, and finally the blocks are erased. Due to the page migrations and erase operation, a GC generally takes a long time and consumes significant SSD bandwidth \cite{gc-1,gc-2,gc-3}.
\end{itemize}

\subsection{SLC-based SSD} 
Flash memory conventionally stores one-bit information in each cell (Single-Level Cell or SLC). However, during the last few years, manufactures leverage the ability to store multiple bits in a single cell -- cells in recent products can store 3 bits (called Triple-Level Cell or TLC) before which 2-bit cell (Multiple-Level Cell or MLC) was the norm. The multi-bit capability of a cell is provided by enabling multiple voltage states in it -- MLC has four different voltage states, whereas TLC has eight different voltage states (sometimes called voltage levels). 
Despite of their low cost per bit, the TLC/MLC flash memories have higher access latencies and lower endurance than the SLCs\cite{flash-basic-1}. 

In order to have the benefits of both technologies in the same system, current SCM designs usually rely on a two-level and hybrid flash-based hierarchy. At upper level, there is a small and fast SLC flash-based SSD (with tens of gigabyte capacity), and a dense MLC flash-based SSD (with few terabyte capacity) is used at lower level. 
In such an architecture, the SLC-based SSD is responsible for servicing most of the I/O traffic and hence its lifetime is very crucial (because of writes). In this work, we focus on enhancing lifetime of the SLC SSD in SCM. However, the studied characterization and our proposed optimization mechanism is general and can be applied to MLC/TLC SSDs. This is left as the future work.

\begin{figure*}
  \centering
  \includegraphics[scale=.33]{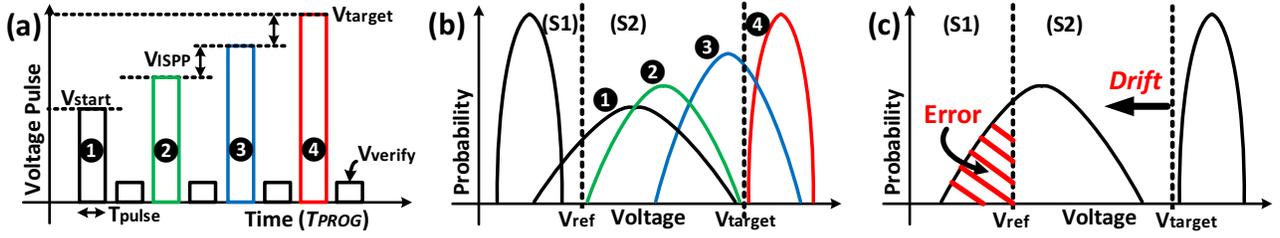}
  \caption{(a) The ISPP write process for SLC programming; (b) A demonstration of changes in threshold voltage during ISPP; (c) The case for data error because of threshold voltage drift.}
  \label{figure:ispp-drift}
\end{figure*}

\subsection{SLC Flash Memory}
\label{sec:slc-flash-memory}
Data in a flash cell is stored in the form of a threshold voltage ($V_{th}$), i.e., the amount of electrons captured in the cell represents different states. The threshold voltage is formed within a fixed-sized voltage window, bounded by a minimum voltage ($V_{min}$) and a maximum voltage ($V_{max}$). For instance, in SLC the entire voltage window is divided into two non-overlapping ranges (two voltage states \verb"S1" and  \verb"S2"), separated by one reference voltage ($V_{ref}$), as shown in Figure~\ref{figure:ssd-Internals}.

\textbf{Write operation:} When the written data is ``1'', no action is needed as the cell is initially in the no-charging state or erase state (i.e., State \verb"S1" in Figure~\ref{figure:ssd-Internals}). 
On writing ``0'', the flash memory employs a specific scheme called Incremental Step Pulse Programming (ISPP)~\cite{ispp-1}. 
As shown in Figures~\ref{figure:ispp-drift}a and \ref{figure:ispp-drift}b, the ISPP applies a sequence of voltage pulses with a fixed duration ($T_{pulse}$) and staircase-up amplitude ($V_{ISPP}$) to the cell, in order to form the desired threshold voltage ($V_{target}$). After triggering each pulse, the cell state is verified to check if the programmed threshold voltage reaches $V_{target}$. This process is repeated until the desired voltage is reached. 
The program time ($T_{PROG}$) is a proportional to the number of ISPP loops, that is inversely proportional to $V_{ISPP}$, and can be expressed as follows~\cite{ispp-2}:
$$T_{PROG} \propto \frac{V_{taregt} - V_{start}}{V_{ISPP}}$$
Under a fixed $V_{ISPP}$, the higher the target voltage ($V_{target}$) is, the longer the program time is.

\textbf{Read operation:}
Reading a SLC flash is realized by applying a reference voltage ($V_{ref}$) and inferring the threshold voltage ($V_{th}$). 
If the threshold voltage is larger than the reference voltage ($V_{th} {\le} V_{ref}$), the cell state is \verb"S1" and its value is ``1''; otherwise, the cell state is \verb"S2" and its value is ``0''.
As the flash read time is a proportional to the number of voltage sensing/comparisons, reading from SLC is very fast since it needs only one sensing/comparison. 

\textbf{Errors in SLC flash:}
Right after a cell is programmed as ``0'', the threshold voltage is around the target voltage ($V_{target}$). However, as time goes by, due to the charge loss, the threshold voltage in the cell drifts and it will finally overlap with the neighboring voltage state. As a result, the cell data is interpreted as ``1'' when it is read. We call this data corruption \emph{retention error}~\cite{retention-error-1,flash-failure-1,program-error-1}, which is illustrated in Figure~\ref{figure:ispp-drift}c. In this error model, the lower tail of the state \verb"S2" overlaps the part of the state \verb"S1" after a specific elapse time, called retention time.
In order to avoid fast data corruption and provide years of retention time in current flash products, the reference voltage is conservatively calibrated to be far from the erase state. This is shown in Figure~\ref{figure:ispp-drift}c.

As a flash block experiences more and more erases (or P/E cycles), the voltage drift (charge loss) accelerates. To enable data integrity for long time, vendors specify a guaranteed retention times (e.g., 10 years) and endurance limit (e.g., 100K P/E cycles) for their flash products.

%% file: Sec-3-tech-Mhmd.tex
\section{Retention Relaxation for Lifetime Enhancement}
\label{proposal}
SLC flash products normally guarantee one long-term retention duration throughout the whole flash lifetime. Such a long-term reliability requirement is critical, when a flash-based SSD is used as a main I/O storage and a replacement for hard disk drives. However, when employing SSD in the intermediate layers of a storage system (e.g., as the SCM which is the focus of this work), such a long retention time guarantee can be an overkill. Hence, if the retention time guarantee could be \emph{relaxed} under a specific condition, one could have an opportunity to improve other system requirements such as performance and endurance without least concern about the data loss.

\begin{figure*}[t]
\centering
	\begin{subfigure}{.24\linewidth}
		\centering
		\includegraphics[width=0.99\linewidth]{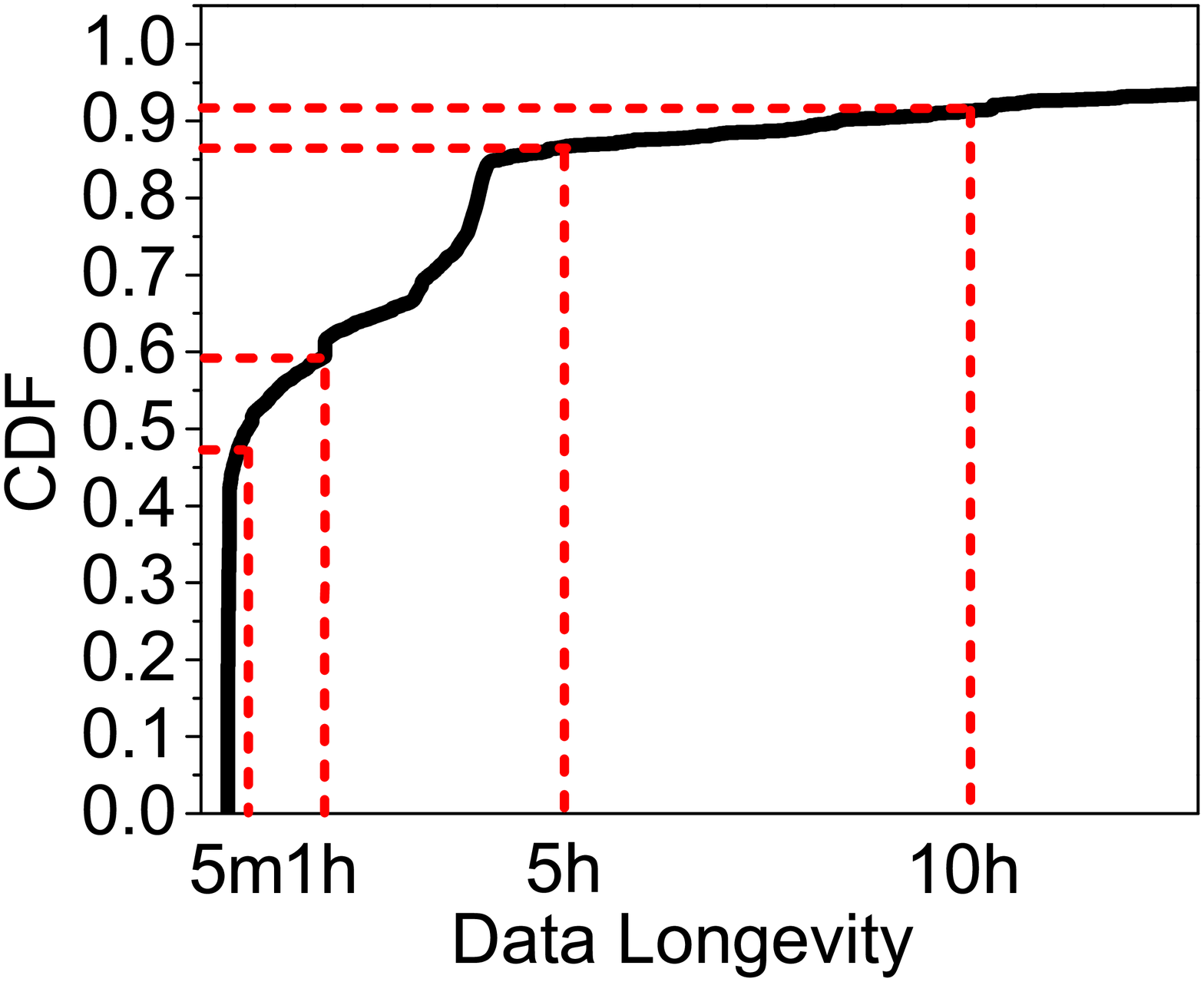}
		\caption{hm\textunderscore0}\label{fig:hm_0}
	\end{subfigure}
	\begin{subfigure}{.24\linewidth}
		\centering
		\includegraphics[width=0.99\linewidth]{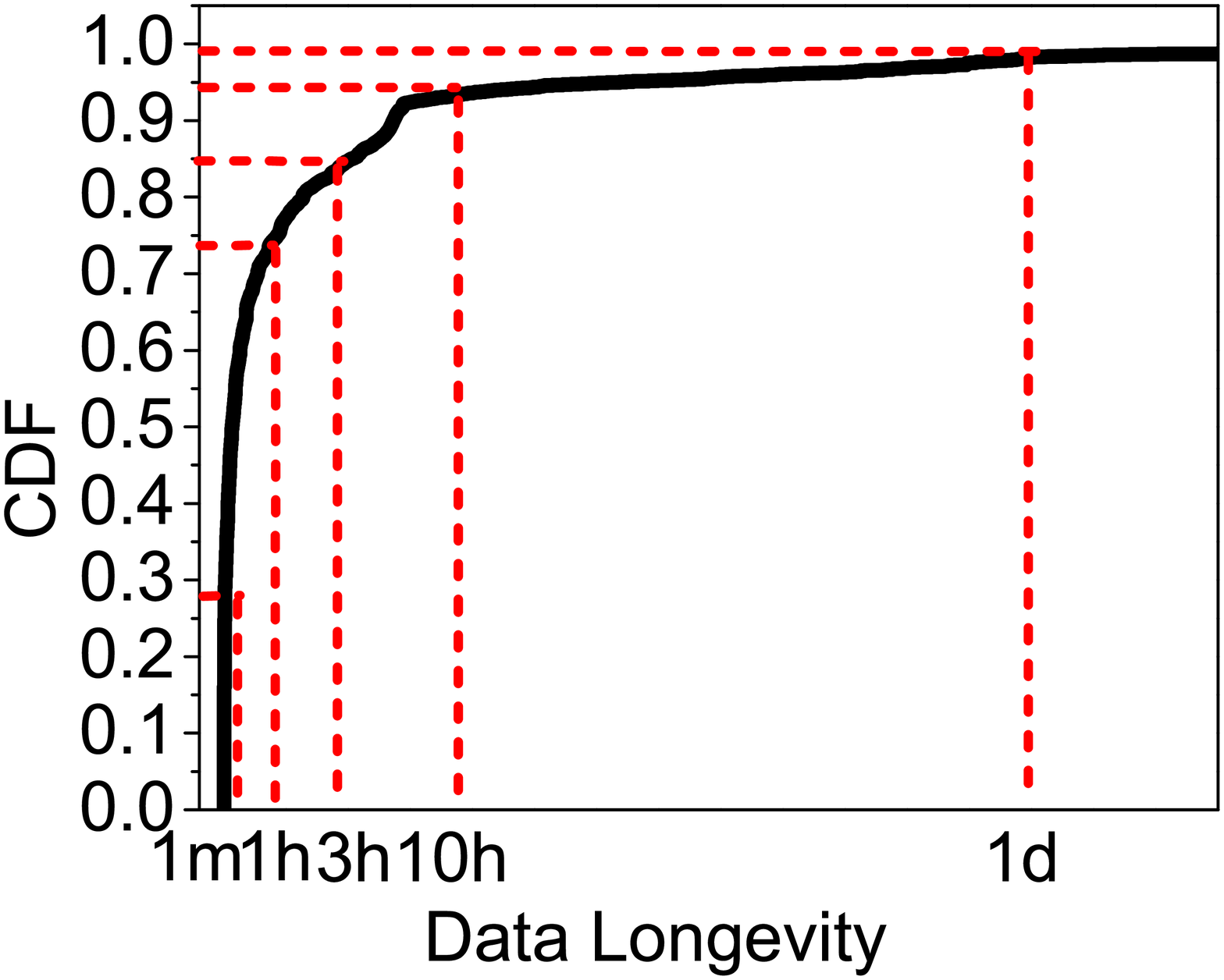}
		\caption{prn\textunderscore0}\label{fig:prn_0}
	\end{subfigure}
	\begin{subfigure}{.24\linewidth}
		\centering
		\includegraphics[width=0.99\linewidth]{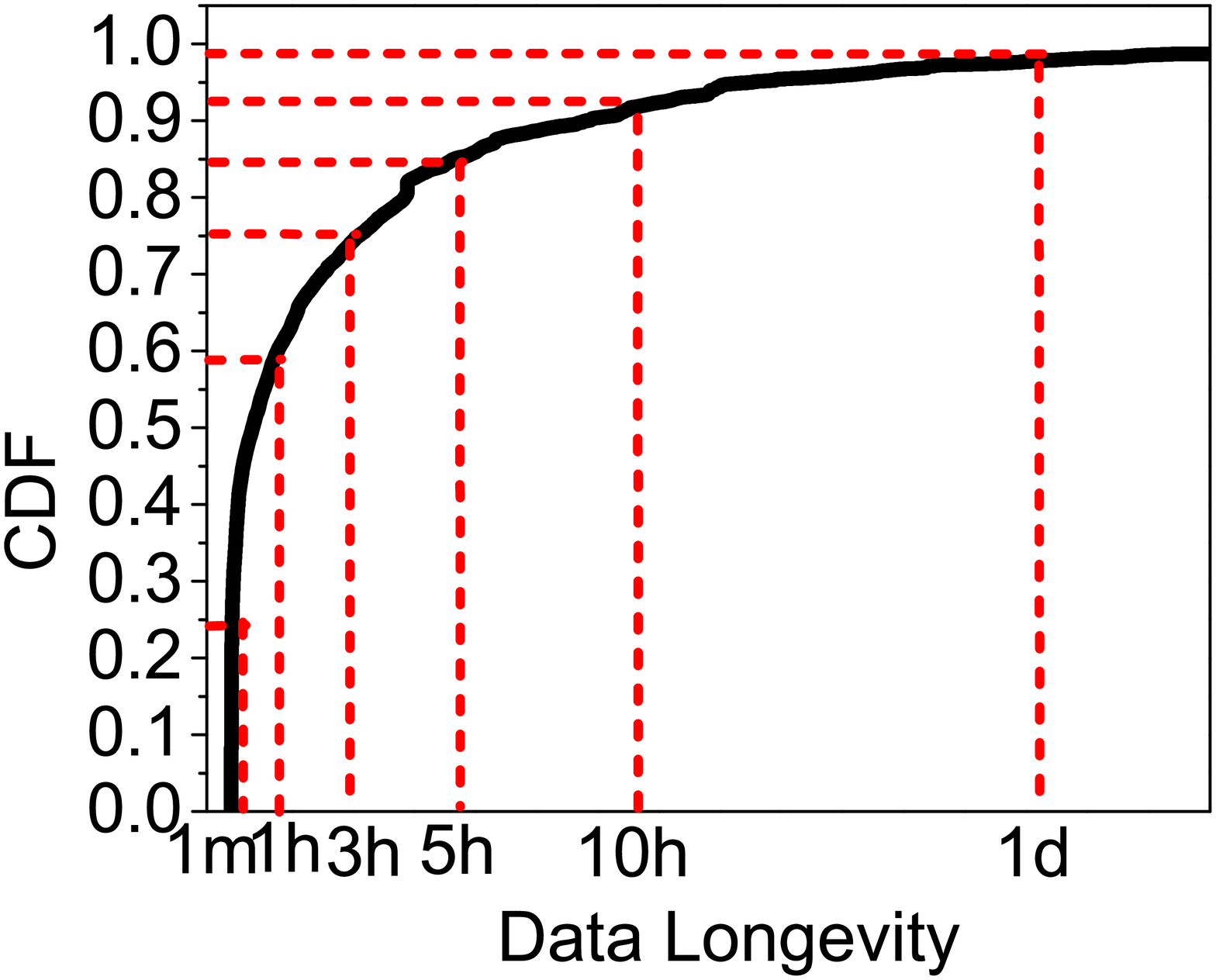}
		\caption{prn\textunderscore1}\label{fig:prn_1}
	\end{subfigure}
	\begin{subfigure}{.24\linewidth}
		\centering
		\includegraphics[width=0.99\linewidth]{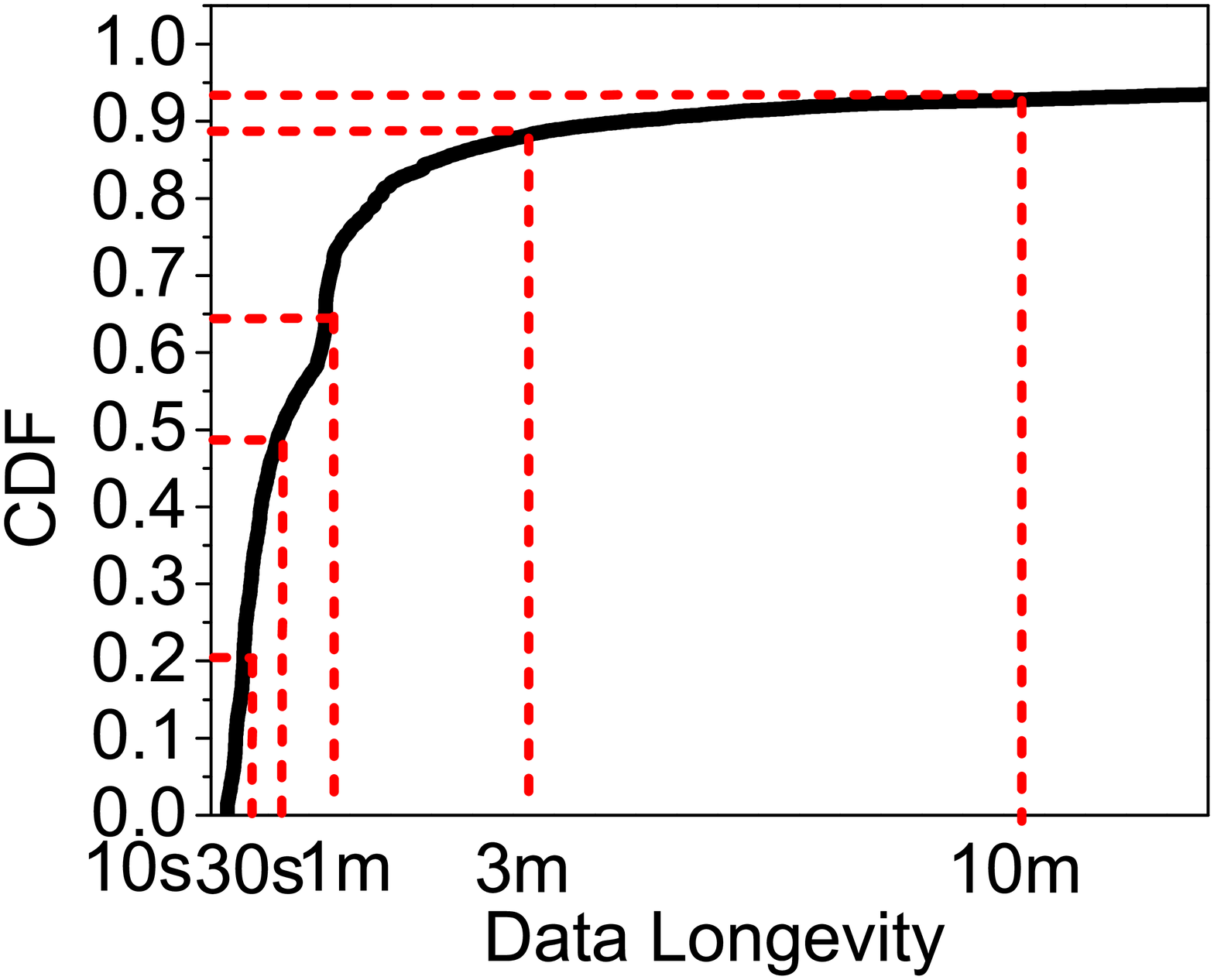}
		\caption{proj\textunderscore0}\label{fig:proj_0}
	\end{subfigure}
	\begin{subfigure}{.24\linewidth}
		\centering
		\includegraphics[width=0.99\linewidth]{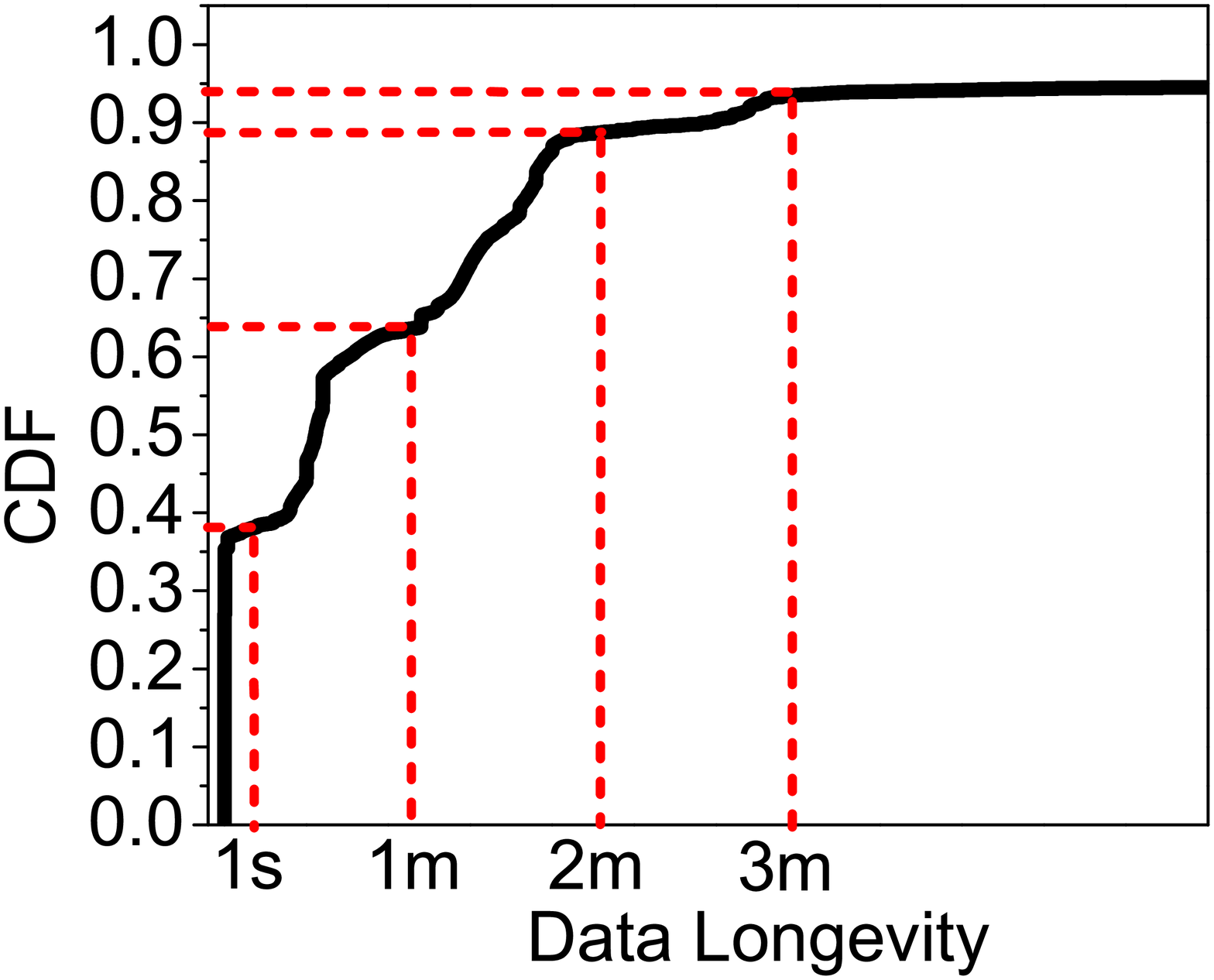}
		\caption{prxy\textunderscore0}\label{fig:prxy_0}
	\end{subfigure}
	\begin{subfigure}{.24\linewidth}
		\centering
		\includegraphics[width=0.99\linewidth]{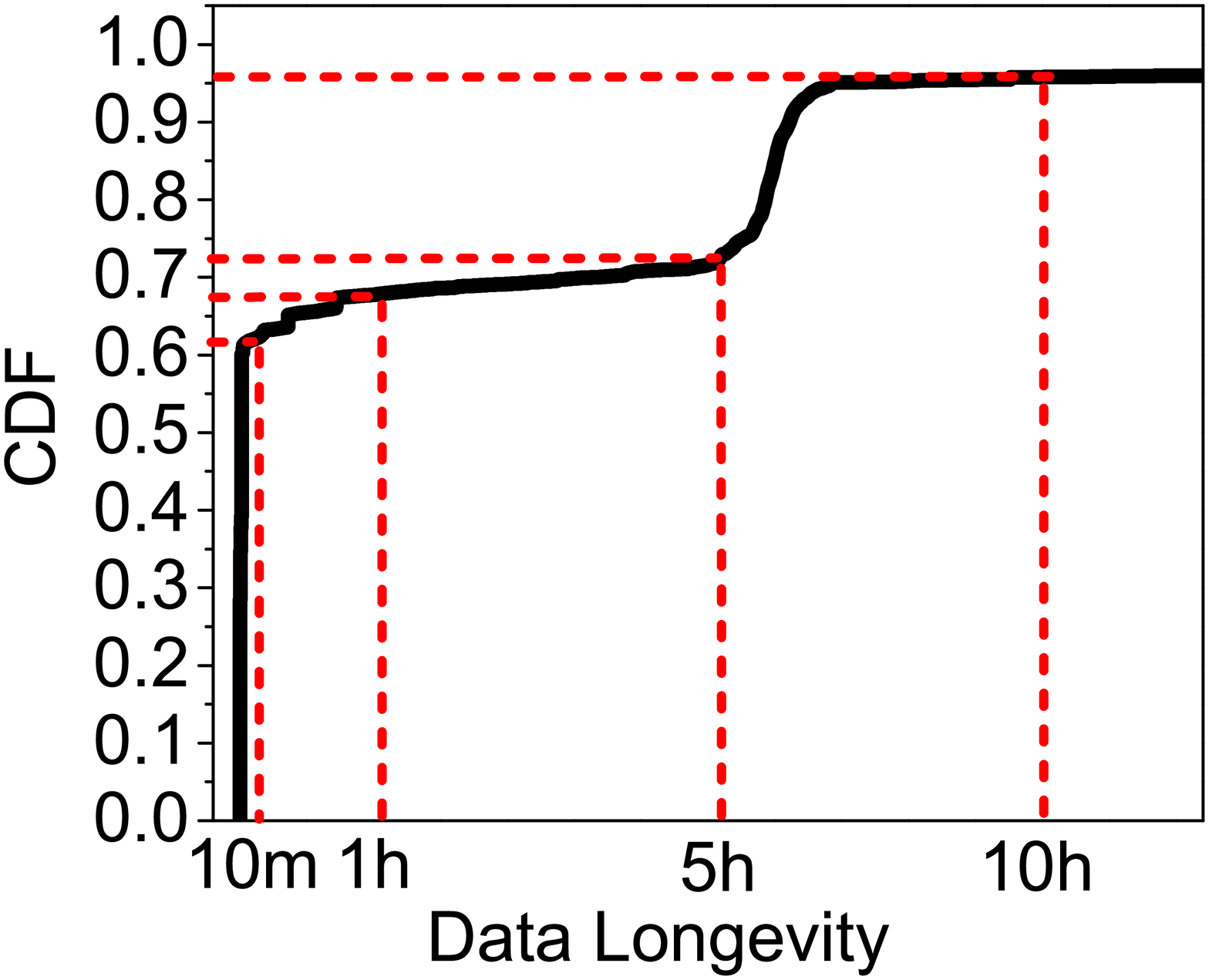}
		\caption{mds\textunderscore0}\label{fig:mds_0}
	\end{subfigure}
	\begin{subfigure}{.24\linewidth}
		\centering
		\includegraphics[width=0.99\linewidth]{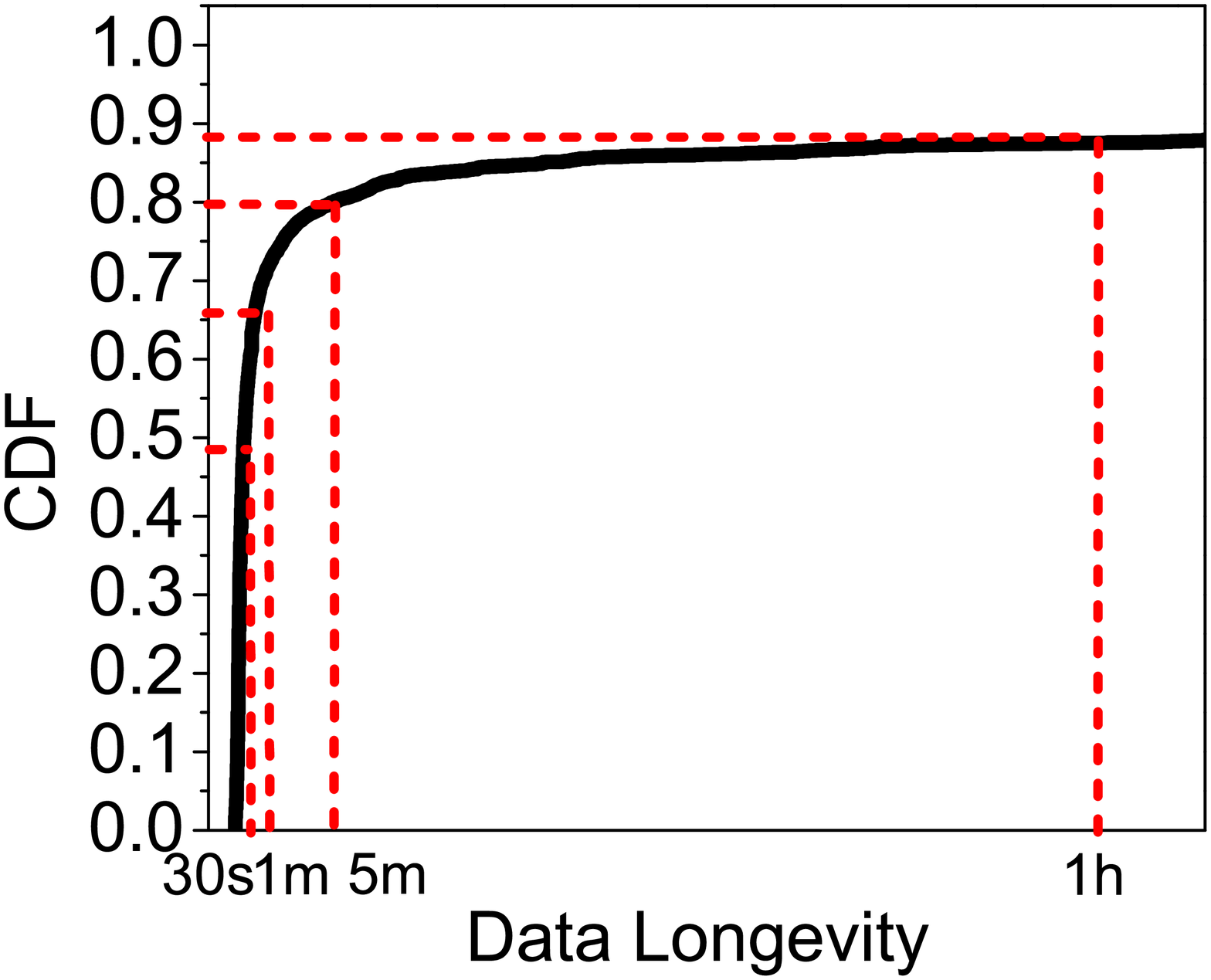}
		\caption{src1\textunderscore2}\label{fig:src1_2}
	\end{subfigure}
	\begin{subfigure}{.24\linewidth}
		\centering
		\includegraphics[width=0.99\linewidth]{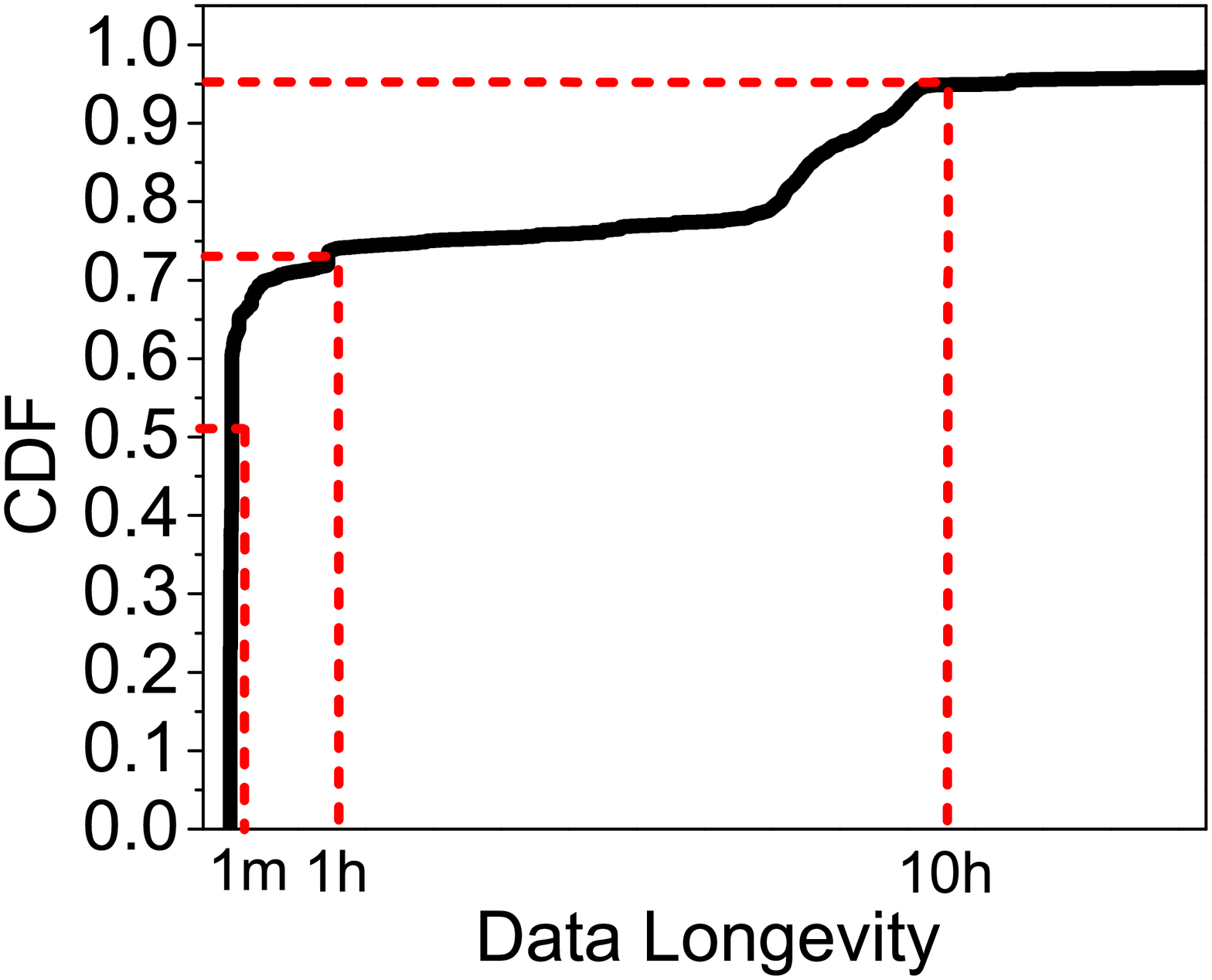}
		\caption{src2\textunderscore0}\label{fig:src2_0}
	\end{subfigure}
	\begin{subfigure}{.24\linewidth}
		\centering
		\includegraphics[width=0.99\linewidth]{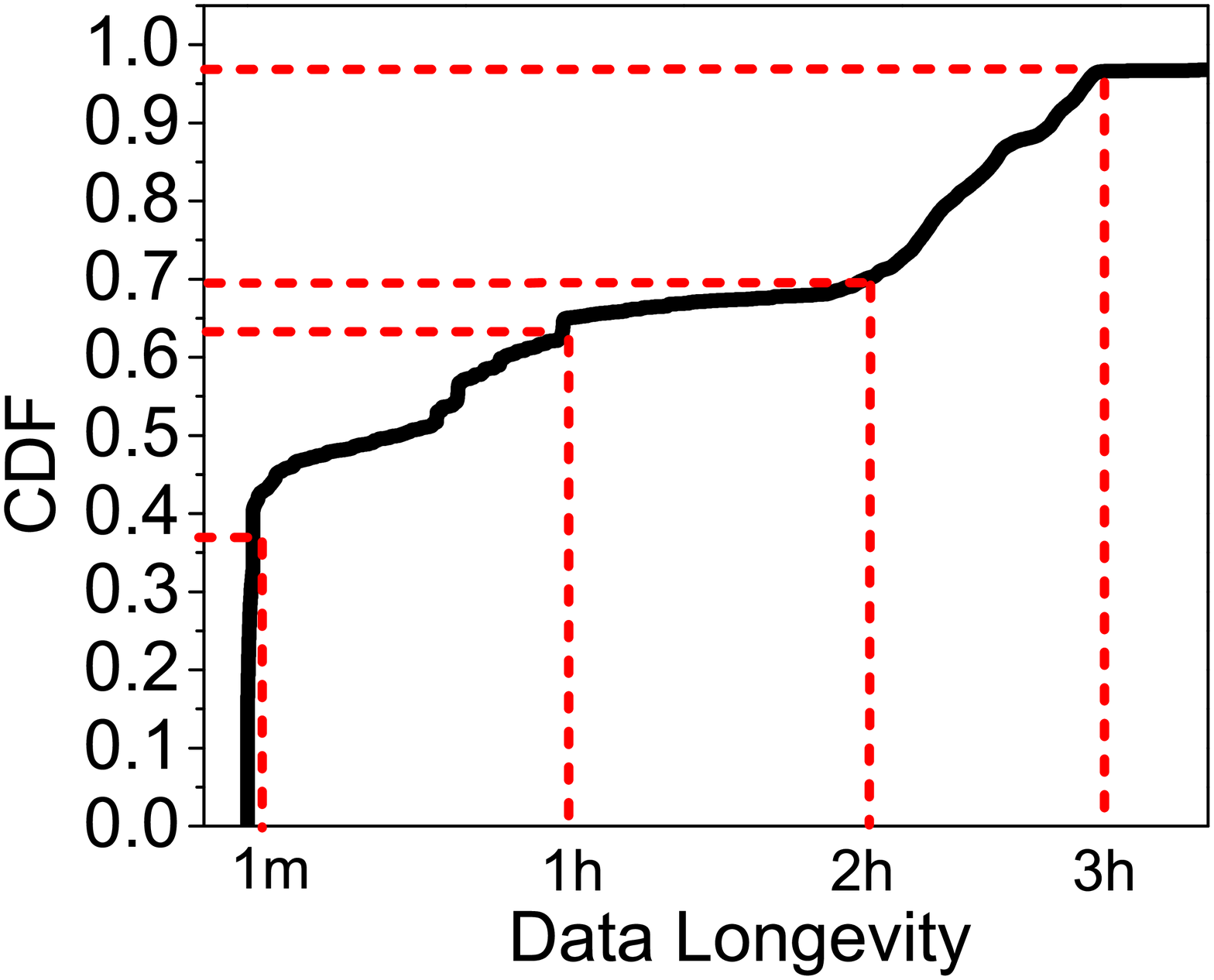}
		\caption{stg\textunderscore0}\label{fig:stg_0}
	\end{subfigure}
	\begin{subfigure}{.24\linewidth}
		\centering
		\includegraphics[width=0.99\linewidth]{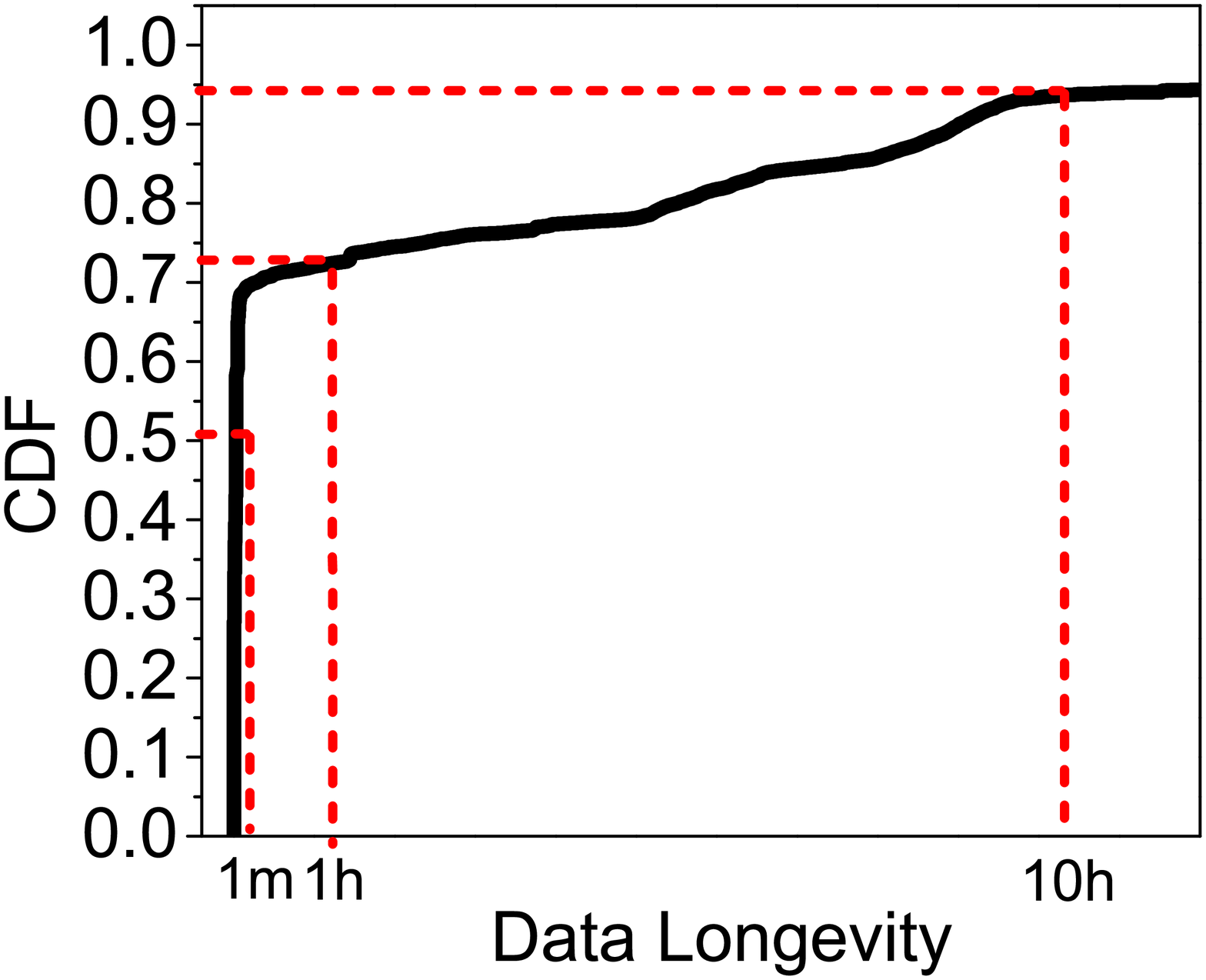}
		\caption{usr\textunderscore0}\label{fig:usr_0}
	\end{subfigure}
	\begin{subfigure}{.24\linewidth}
		\centering
		\includegraphics[width=0.99\linewidth]{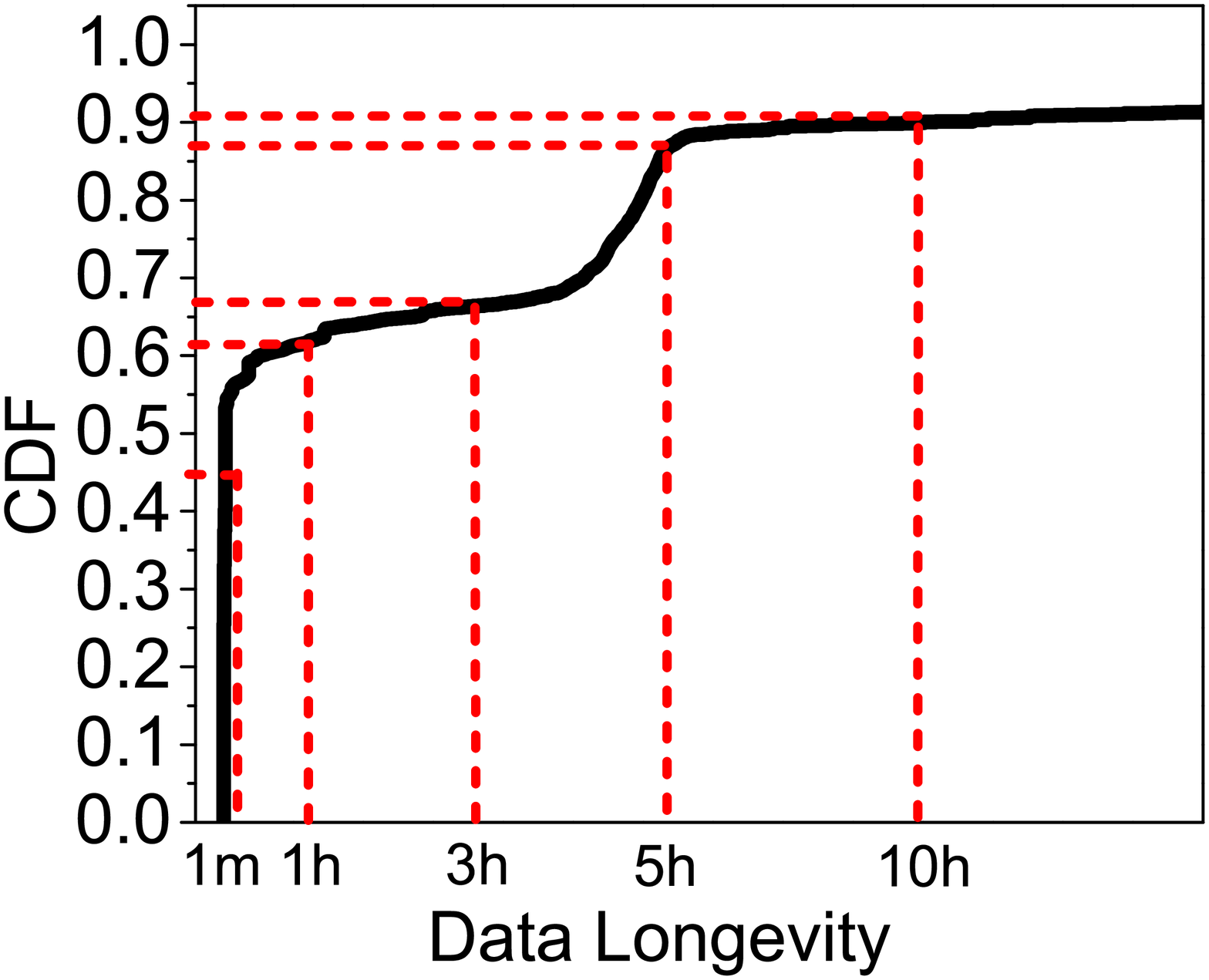}
		\caption{web\textunderscore0}\label{fig:web_0}
	\end{subfigure}
	\begin{subfigure}{.24\linewidth}
		\centering
		\includegraphics[width=0.99\linewidth]{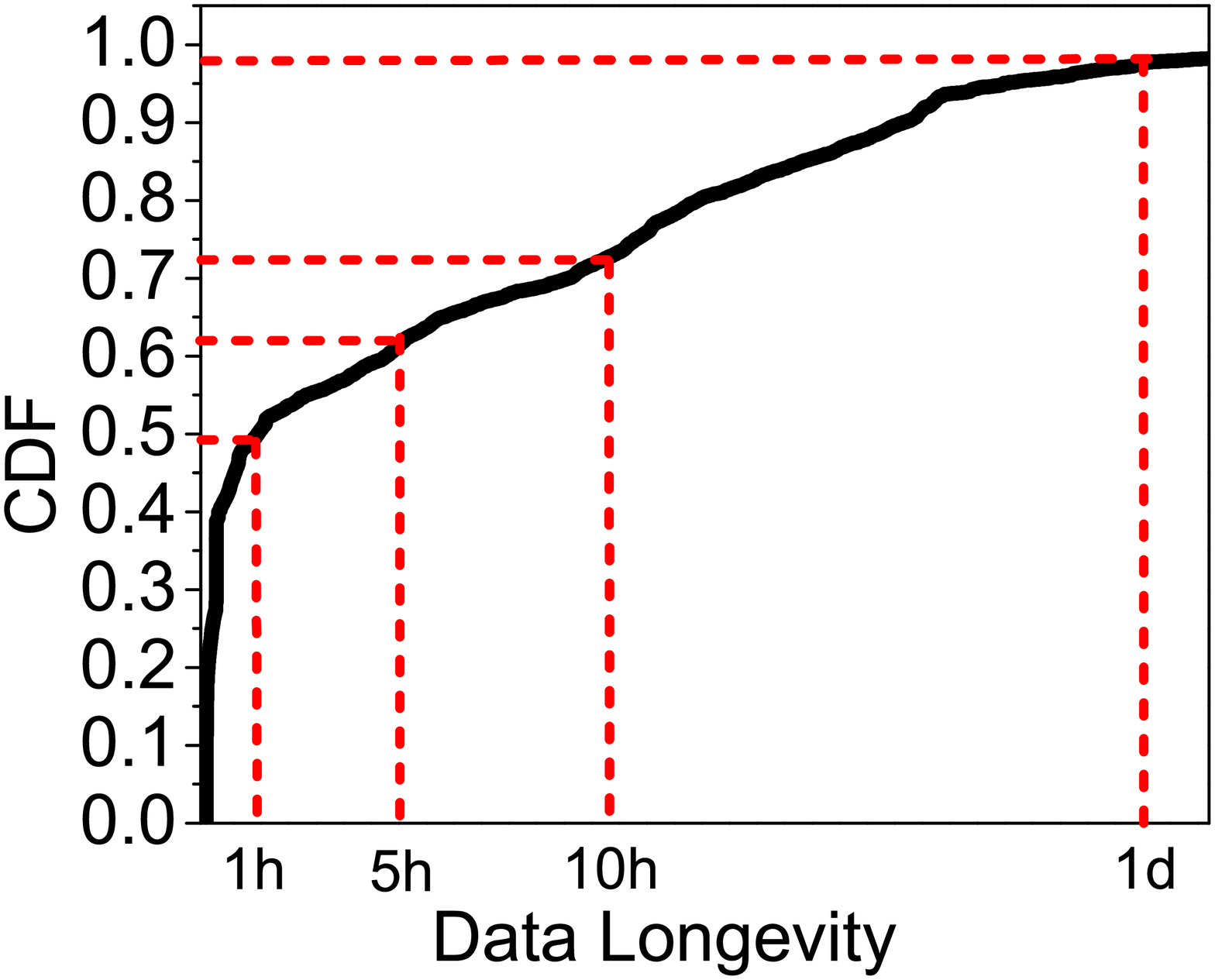}
		\caption{web\textunderscore1}\label{fig:web_1}
	\end{subfigure}
	\begin{subfigure}{.24\linewidth}
		\centering
		\includegraphics[width=0.99\linewidth]{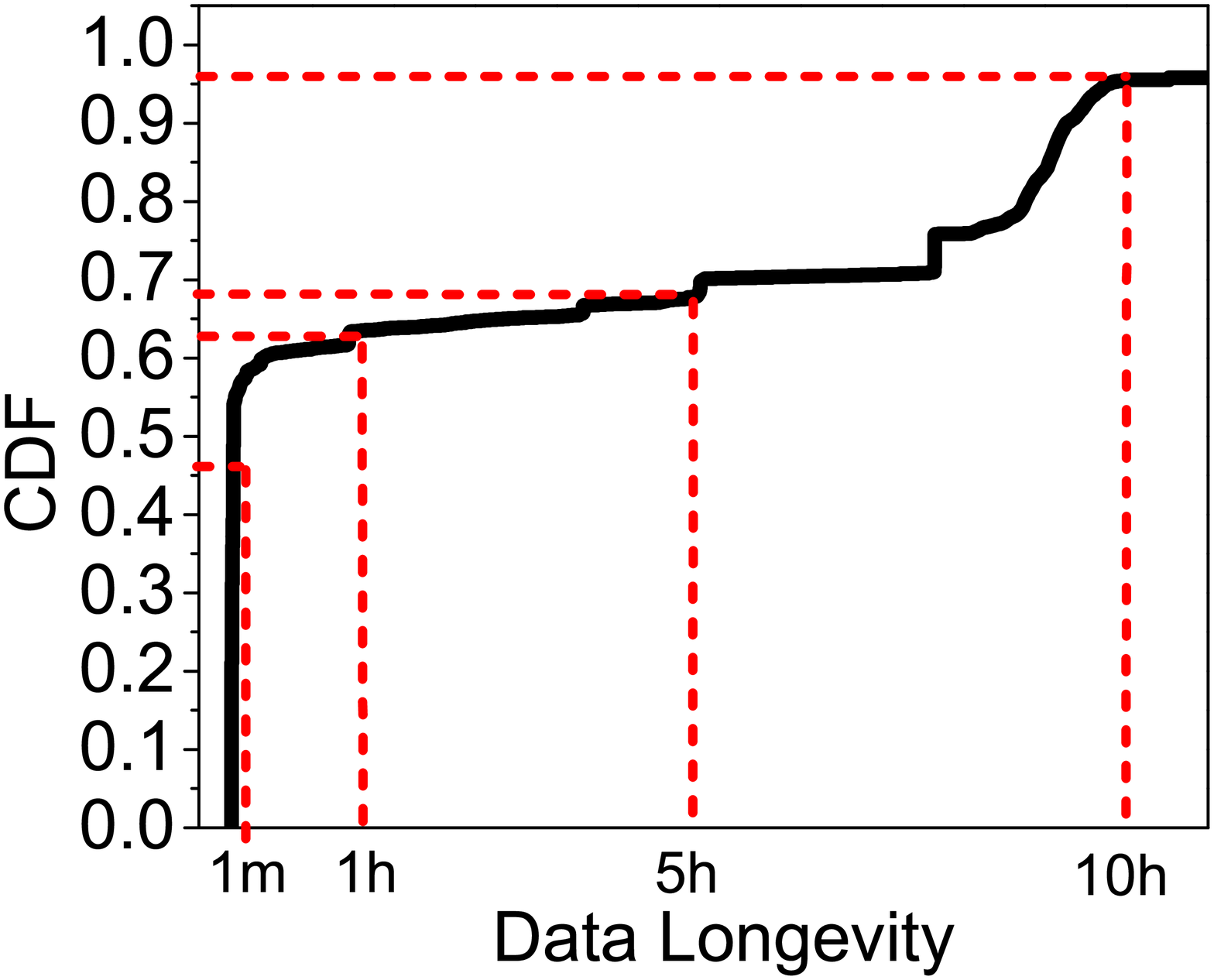}
		\caption{wdev\textunderscore0}\label{fig:wdev_0}
	\end{subfigure}
	\begin{subfigure}{.24\linewidth}
		\centering
		\includegraphics[width=0.99\linewidth]{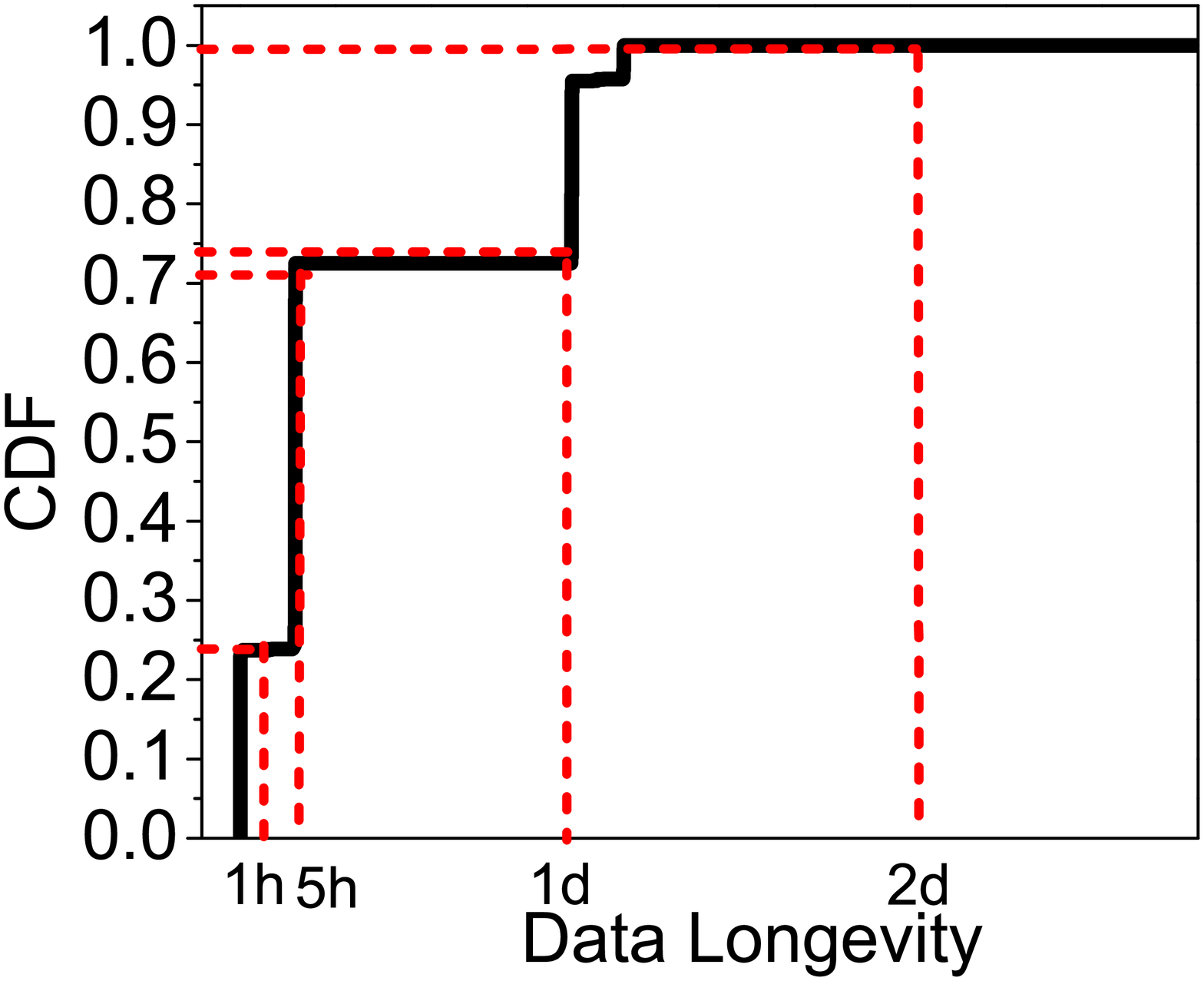}
		\caption{wdev\textunderscore2}\label{fig:wdev_2}
	\end{subfigure}
	\begin{subfigure}{.24\linewidth}
		\centering
		\includegraphics[width=0.99\linewidth]{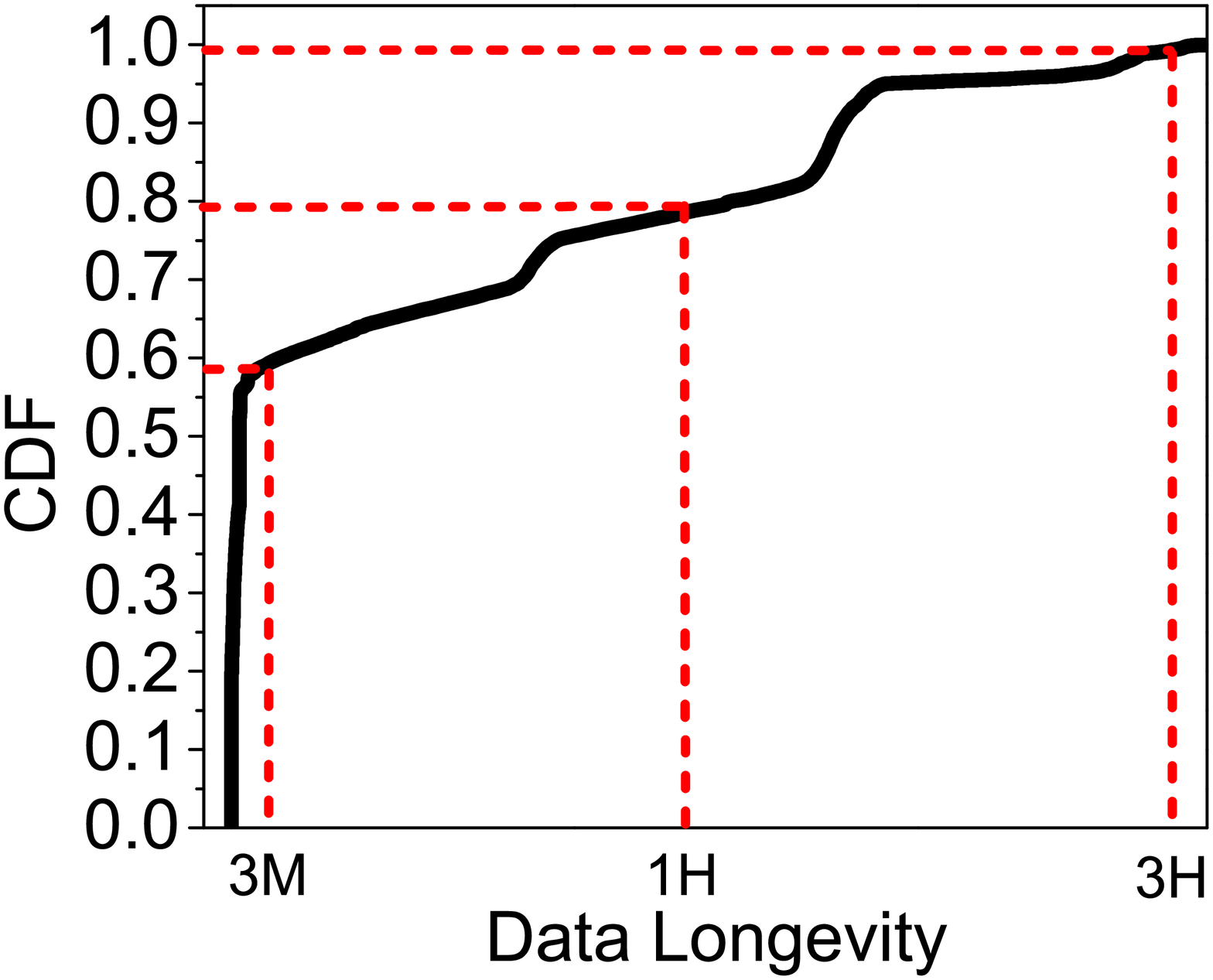}
		\caption{rsrch\textunderscore0}\label{fig:rsrch_0}
	\end{subfigure}
\caption{The CDF of data longevity for I/O blocks in 15 write-intensive workloads taken from the MSR Cambridge suite.}
\label{fig:cdf-longevity}
\end{figure*}

Relaxing the guaranteed retention time has been explored in various kinds of non-volatile memories~\cite{relax-nvm-1,relax-nvm-2}. Some prior works exploited retention relaxation for improving the write performance of flash memories~\cite{relax-1,relax-2}. 
The principle behind most of these works is to form the threshold voltage less accurately, and by doing so, they would reduce the number of loops in the ISPP process -- that would reduce (improve) the device program time. In this work, however, we leverage retention time relaxation for enhancing the lifetime of SLC flash memories in an SCM. 
To the best of our knowledge, this paper is among the first works that exploits retention time relaxation for lifetime enhancement in SLC SSDs. We believe that our findings give insights to SSD developers for developing highly-reliable flash storage.

In the next three subsection, we introduce our mechanism by answering the following questions:
\begin{enumerate}[leftmargin=*]
\item \emph{What is the distribution of data longevity values in a flash-based SCM? Do all the data written into an SSD need the long retention time guarantee of flash memory?} (Section~\ref{DL})
\item \emph{Is it practically possible to relax the guaranteed retention time of a flash memory? What is the theory behind it?} (Section~\ref{drift})
\item \emph{How can we exploit the retention relaxation for improving the lifetime of flash memory? What kind of architectural and software support is required to implement such a relaxation?} (Section~\ref{tech})
\end{enumerate}

\subsection{Distribution of Data Longevity in I/O Workloads}
\label{DL}

In a well-managed SCM-based memory hierarchy, we expect that data blocks with short retention times get stored in the solid-state part, while the other data blocks (i.e., those with long retention times) will normally be kept in the HDD (at the lowest level of storage hierarchy). To examine the distribution of data longevity (i.e., the time between two consecutive update of the data) in a typical flash-based SSD, we configured a 64GB SSD (consisting of eight 8GB SLC flash chip). Details of the evaluated configuration and its parameters are given later in Section~\ref{sec:config}. On this SSD, we ran 15 workloads from the MSR Cambridge suite~\cite{cambridge}. The selected workloads are \emph{write-intensive}, as our interest is to investigate the data longevity and improve the storage lifetime (read traffics do not have an impact on SSD's lifetime). Table~\ref{tab:workload-cha} characterizes our workloads.


Figure~\ref{fig:cdf-longevity} shows the cumulative distribution function (CDF) of data longevity of I/O blocks stored in the SSD. For the I/O blocks written once in a workload, we set their data longevity to the maximum (e.g., 10 years) and assume that we are not allowed to relax the retention time for them. One can observe from this figure that, for all the examined workloads, a large portion of the written data blocks have a short longevity in the range of few minutes, few hours or few days. Specifically, the 95th-percentile of I/O blocks written in \verb"prxy_0" have a longevity of 3 minutes; it is 10 minutes for \verb"proj_0"; 1 hour for \verb"src1_2"; 10 hours for \verb"hm_0", \verb"mds_0", \verb"src2_0", \verb"usr_0", \verb"web_0" and \verb"wdev_0"; 1 day for \verb"prn_0", \verb"prn_1" and \verb"web_1"; and 10 days for \verb"wdev_2".

To sum up, a majority of data blocks (95th percentile) in all our examined workloads are frequently-updated; hence they do not need such a long retention time guarantee (up to 10 years) provided by the commercial SLC flash memories. In contrast, a small fraction of the write data need a retention time larger than 10 days (the percentage varies between 1\% to 10\% across our workloads). Using these characteristics of our workloads, the next section demonstrates how one can trade off the short retention times for a prolonged storage lifetime.

We note that reading from a flash cell multiple times may affect its voltage level and reduce its retention time. However, the probability of data disturbance due to the intensive reads is quite low (e.g., less than 0.01\% in \cite{read-disturb}). Moreover, we observed that our workloads do not exhibit such excess reads on data. Therefore, read disturbance is not a big issue in a retention-relaxed flash cell and the focus of our work is retention times related to data longevity. 


\subsection{Retention Time Relaxation for NAND Flash}
\label{drift}
To relax the retention time in flash memory, we first need to investigate how long the threshold voltage drifts due to the charge loss. Some prior work have shown that the voltage drift of a flash device is affected by multiple parameters including the \emph{initial threshold voltage}, the \emph{current device wear-out level} (in P/E cycles), and the \emph{fabrication technology}. Pan et al.~\cite{relax-2} proposed a detailed model of the voltage drift distance ($D_{drift}$) for NAND flash memory. We simplify this model by considering the critical factors as below and use it for our model throughout this work:

\begin{equation} \label{eq:drift-model}
D_{drift} = K_{scale} \times N_{PE} \times \ln \big(1 + T_{RT} \big),
\end{equation}
where $N_{PE}$ and $T_{RT}$ are the number of P/E cycles the cell (block) experienced and the retention time in hour, respectively. $K_{scale}$ is a device-specific constant.

\begin{figure}
  \centering
  \includegraphics[scale=.265]{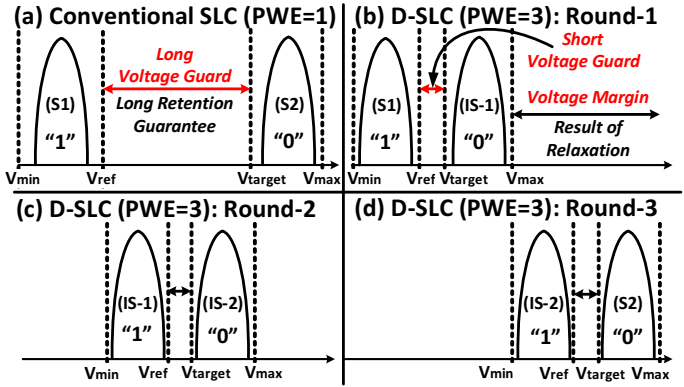}
  \caption{(a) conventional SLC with PWE of ``1''; (b), (c) and (d) our D-SLC with a PWE of ``3''.}
  \label{figure:d-slc}
\end{figure}

Being aware of the voltage drift behavior, we can reduce the ``voltage guard'' between the two states by shifting the threshold voltage of the program state  (\verb"S2" in Figure~\ref{figure:ssd-Internals}) to the left. By doing so, we can decrease (relax) the retention time. 
An example of this process is given in Figure~\ref{figure:d-slc}. Figures~\ref{figure:d-slc}a shows the baseline SLC NAND flash, in which there is a large voltage guard between two states and a long retention time (e.g., 10 years) is guaranteed. 
Figure \ref{figure:d-slc}b shows the case where, by shifting the program state (\verb"S2") to the left, we could achieve smaller voltage guard between the two states, which results in shorter retention time compared to the baseline. In this figure, the new program state is named as \verb"IS-1" (intermediate state).

Achieving a program state with lower threshold voltage (like \verb"IS-1" in Figure \ref{figure:d-slc}b) is easy in flash memory -- we need to tune the ISPP's parameter (i.e., duration and/or amplitude of each pulse) such that we can program the cell to the new threshold voltage level.
ISPP controller is a programmable circuit inside a flash chip; that is, it can give appropriate commands to the flash chip to program pulse duration and/or amplitude of the ISPP pulses. 
Note that we want to keep the program latency of a retention-time-relaxed flash memory (e.g., Figure \ref{figure:d-slc}b) same as in the conventional SLC memory (Figure \ref{figure:d-slc}a). Thus, we should keep overall duration of the ISPP process (i.e., sum of duration of all pulses) identical in both designs. 
To achieve this goal, we assume that the number of pulses and duration of a pulse are fixed and same as the baseline, and they work with $V_{ISPP}$.
In the baseline, the voltage step (difference between amplitude of two consecutive pulses) is set to large values, i.e., the cell receives large charges at each step, which helps us reach \verb"S2" in Figure \ref{figure:d-slc}a very quickly. If we use this large voltage step for programming a retention-time-relaxed cell, it is very likely that we jump over the intermediate threshold voltage.
Intuitively, to have fine-grain threshold voltage jumps in our design while the number of pulses is fixed, we need to keep the voltage steps smaller than the baseline.

\subsection{Trading-off Retention Time for Higher Lifetime of SLC Flash}
\label{tech}
The discussion in Section~\ref{DL} reveals that, on the one hand, the data longevity of the written data blocks into SSD (as SCM) is mostly limited to few minutes, few hours or few days in the transactional and enterprise applications, i.e., much shorter than 10 years provided by the current flash products. On the other hand, we showed that it is possible to relax the retention time guarantee of flash memory by calibrating the voltage guard and the ISPP parameters (Section~\ref{drift}). In this section, we propose a novel mechanism and show how retention time relaxation can be exploited for achieving longer lifetime in flash memories. We start by defining a new metric for lifetime analysis that helps us describe our proposal more clearly.



\subsubsection{Page Write per Erase cycle (PWE) Metric}
\label{sec:pwe-define}
We define the term ``page writes per erase cycle'' (PWE) as the maximum number of \emph{logical pages} stored in \emph{one physical page} during \emph{one P/E (erase) cycle}. 
The conventional SLC flash memory stores one bit data in each cell during each erase cycle, and hence, its PWE is one.

If one can write more than one bit during an erase cycle, and hence increase the PWE, the device stores a larger amount of data during its whole lifetime, or in other words, the device lifetime gets improved (i.e., 50K P/Es for an SLC flash memory in our setting).
The increase in the number of writes in an erase cycle does not accelerate the cell wear-out, due to two reasons~\cite{wear-out-1,wear-out-2}. First, the total amount of electrons that go in and out of a cell in an erase cycle determines the cell wear-out. Second, the amount of electrons that pass through a cell in an erase cycle is limited, no matter how many writes are applied. 
Note also that increasing PWE of an SLC device does not necessarily mean that it stores more than one bit information at each moment -- the device is still SLC (single bit storage at each given time); rather, it means that the device does not need to be erased before reprogramming it.
Our main objective is to enhance the SLC flash lifetime by increasing its PWE to values higher than one.


\subsubsection{Overview of the Proposed Mechanism}
Figure~\ref{figure:d-slc} shows a high-level view of our proposed design versus the conventional SLC flash memory.
The conventional SLC flash cell (shown in Figure~\ref{figure:d-slc}a) has two states: \verb"S1" or the erase state (value ``1'), and \verb"S2" (value ``0''). There is a large \emph{voltage gap} between these two states, which results in a very long retention time (10 years in this example). This cell stores one bit at each time and reprogramming it requires first erasing it. Thus, during each erase cycle, it stores one bit -- its PWE is one.

Figure~\ref{figure:d-slc}b shows the initial state of our proposed SLC flash design.
Similar to the conventional design, it has two states: \verb"S1" (value ``1'') and \verb"IS-1" (value ``0''). However, the voltage gap between these two states is \emph{small}, and hence the device retention time is \emph{relaxed} to smaller durations (say few minutes, few hours or few days).
In contrast to the conventional SLC, in our proposed design, we do \emph{not} need to erase the cell before reprogramming. Instead, when the current values gets invalid, the cell can store the new value by using higher voltage values. For example, as shown in Figure~\ref{figure:d-slc}c, the new binary states are \verb"IS-1" and \verb"IS-2", representing the new binary values ``1'' and ``0'', respectively. 
As before, the cell stores one bit data at each time (similar to the baseline SLC) and also, the voltage distribution of the new binary states (\verb"IS-1" and \verb"IS-2") is calibrated for short retention time.
Repeating this procedure, the device stores one more bit in the cell by programming it into states \verb"IS-2" and \verb"S2" for binary values ``1'' and ``0'', respectively.
As this example demonstrates, by calibrating the voltage states in an SLC device and having two intermediate states \verb"IS-1" and \verb"IS-2", one can store three bits (one bit at each time) in one cell before erasing it. This increases the PWE of the SLC flash from \emph{one} in conventional design to \emph{three} in this example, which directly translates to a longer device lifetime.

We want to emphasize two points. 
First, a few prior works~\cite{wom-empirical-1,related-empirical-2} have experimentally demonstrated that we can gradually increase the threshold voltage of a cell by repeating the process of electron injection.
Second, achieving higher lifetime is not free in this approach. In fact, one would need to adjust/program the ISPP's parameters to take advantage of the intermediate states -- that would increase the controller's complexity (even though it is not that much). As discussed in Section~\ref{drift}, to keep the write performance of our design similar to that of the baseline SLC, we do not touch the number of pulses and duration of each pulse in ISPP with respect to the baseline. Thus, the only overhead of such a design is related to adjusting the voltage steps between the ISPP pulses.
The other cost of this design is the need for some changes at FTL's page/block's status management as well as garbage collection. We describe changes at FTL in Section~\ref{sec:ftl-design}.


In short, the proposed mechanism, named \textbf{Dense-SLC} or \textbf{D-SLC}, archives a longer lifetime compared to the conventional design by exploiting the relaxed retention time.
The only potential problem with D-SLC is that it may increase the number of page migrations inside the SSD. Indeed, if the written value at each round has a longevity longer than the device retention time (now relaxed to few minutes for example), we need to move it to another location to avoid data loss.
In the following, we describe the required changes at the FTL and SSD controller that help to get most of the potential benefits of D-SLC while avoiding the potential overheads related to unwanted page migrations.

\subsubsection{Detailed Design of D-SLC}
The D-SLC flash design is highly scalable, i.e., by controlling the ISPP parameters and calibrating distribution of the voltage states, it is possible to increase the number of voltage states in D-SLC and hence enhance its PWE.
However, this is not always beneficial since, by increasing the number of states, either a more accurate write mechanism (or finer-grain ISPP) is required or the inter-state voltage gap is reduced. The former increases the controller's complexity (or write latency if we do not want to keep the D-SLC's performance similar to the baseline SLC). And, the latter results in an exponential decrease in device retention time which in turn increases the number unwanted page migrations.

In order to provide sufficient retention time for the majority of I/O blocks while keeping the PWE level of D-SLC high, we make use of the data longevity characterization presented in Section~\ref{DL} and the drift model in Section~\ref{drift} for the threshold voltage calibration in D-SLC. 
We categorize the I/O blocks of each workload into four groups based on their longevity (or retention time): longevity of a block is either \emph{less than 1 hour}, \emph{between 1 hour and 10 hours}, \emph{between 10 hours and 3 days}, or \emph{more than 3 days}. For the I/O blocks which are only written once in a workload during the examined duration, we assume the maximum longevity and they belong to the last group (i.e., that with longevity larger than 3 days).
Table~\ref{tab:longevity} reports the ratio of the I/O blocks belonging to the four retention time categories.


We determine the voltage threshold distribution in an SLC flash by using the drift model in Section~\ref{drift} with two optimization goals. 
First, we want to increase the PWE of the SLC flash for each data longevity category in Table~\ref{tab:longevity} during the entire lifetime of the device. 
Second, we want to keep the performance of our SLC design close to that of the conventional SLC. By assuming a fixed duration for each pulse in ISPP, we determine $V_{ISPP}$ to keep the number of ISPP loops close to that of the baseline SLC.
Following these optimization goals, Table~\ref{tab:mode-assign-3-base} reports the number of voltage states used for storing I/O blocks of our four longevity categories during the entire lifetime of the device. Similar to the baseline, the block's endurance limit is 50K P/Es.

\begin{table}[t]
\centering
\caption{Classification of I/O blocks in the studied workloads based on their longevity.}
\label{tab:longevity}
\small
\begin{tabular}{|l||c|c|c|c|}
\hline
& \multicolumn{4}{|c|}{\textbf{Percentage of I/O Blocks (\%)}} \\ \cline{2-5}
 \textbf{Name}& $\le$1Hour & 1Hour$\sim$10Hours & 10Hours$\sim$3Days & $\ge$3Days \\
\hline\hline
	hm\textunderscore0 &	59.8	&	33.7	&	6.4	&	0.1	\\ \hline
        prn\textunderscore0 &      73.3	&	21.9	&	4.8    &	0	\\ \hline
	prn\textunderscore1 &	59.3	&	33.3	&	7.4	&	0	\\ \hline
	proj\textunderscore0 &      96.7	&      2.7	&	0.5	&	0.1	\\ \hline
	prxy\textunderscore0 &      96.1	&	3.1	&	0.7	&	0.1	\\ \hline
	mds\textunderscore0 &  66.4	&	29.6	&	3.6	&	0.4	\\ \hline
	src1\textunderscore2 & 87.9	&	7.9	&	4.1	&	0.1	\\ \hline
	src2\textunderscore0 & 72.5	&	23.3	&	4.0	&	0.2	\\ \hline
	stg\textunderscore0 & 62.8	&	35.1	&	2.0	&	0.1	\\ \hline    
	usr\textunderscore0 & 72.9	&	21.9	&	4.8	&	0.4	\\ \hline
	web\textunderscore0 & 62.7	&	28.7	&	8.4	&	0.2	\\ \hline
	web\textunderscore1 & 48.3	&	24.0	&	27.7	&	0	\\ \hline		
	wdev\textunderscore0 & 62.3	&	33.7	&	3.4	&	0.6	\\ \hline
	wdev\textunderscore2 & 23.7	&	48.8	&	27.5	&	0	\\ \hline
	rsrch\textunderscore0 & 79.7	&	20.3	&	0	&	0	\\ \hline
\end{tabular}
\end{table}

\begin{table}[!t]
\caption{Block state mode assignment (2, 4, and 8-state mode) for different I/O retention times as a function of the block age (i.e., the number of erases a block experiences).}
\label{tab:mode-assign-3-base}
\small
\begin{tabular}{|p{0.7in}||c|c|c|c|c|  }
\hline 
\textbf{Retention}	& \multicolumn{5}{|c|}{\textbf{Block Age (Number of Erases to the Block)} } \\ \cline{2-6}
\textbf{Time}	&	0$\sim$10K 	&	10$\sim$20K	&	20$\sim$30K	&	30$\sim$40K	&	40$\sim$50K	 \\ 
\hline\hline 
$\le$1Hour			&	8	& 	8	&	8	&	8	&	8	\\
\hline 
1$\sim$10Hours		&	8	&   	8	&	8	&	4	&	4	\\
\hline 
10Hours$\sim$3Days	& 	4	&	4	&	4	&	2	&	2	\\
\hline 
$\ge$3Days  			& 	2	&	2	&	2	&	2	&	2	\\
\hline
\end{tabular}
\end{table}

In this study, we limit our calculation to three modes for each cell: it is either in the 2-state mode (i.e., exactly same as the conventional SLC), 4-state mode (i.e., shown by the example in Figure~\ref{tab:mode-assign-3-base}), or 8-state mode (i.e., it has 6 tightly-arranged intermediate states). The 8-state mode has the shortest retention time and very suitable for storing data values with short longevity (like those with ``less than an hour longevity''). The 2-state mode has the longest retention time and suitable for data values with long longevity (like those with ``greater than 3 days longevity''). The 4-state mode has a moderate retention time and is mostly used for values with ``10 hours to 3 days longevity''. 
One can observe form this table that, as the device wears out, the drift rate increases and we need to decrease the device state to lower levels to avoid (unwanted) migrations. As an example, this behavior happens for I/O blocks with a longevity of ``1--10 hours'' that are targeted to 8-state mode in early cycles of the device lifetime, but later are targeted to 4-state mode for P/E cycles larger than 30K.

We use the three modes described above for our FTL design and main evaluation results. We later analyze the sensitivity of the D-SLC's efficiency to different parameters including the number of voltage states.

\subsubsection{FTL Design for D-SLC Support.}
\label{sec:ftl-design}
FTL has three main responsibilities (see Section~\ref{bkg}): address mapping, garbage collection, and wear-leveling. 
To support D-SLC in an SSD, two changes are required at the FTL -- the block allocation algorithm needs to be modified to enable multiple blocks/pages with different modes, and the garbage collection algorithm needs to be redesigned to enable reprogramming a page without erasing that.
The new FTL is called DSLC-FTL. We describe the modifications to DSLC-FTL for D-SLC with three modes (2-state, 4-state, and 8-state modes). However, our methodology is general and can be applied to D-SLC with a different mode configuration.

\begin{figure}
  \centering
  \includegraphics[scale=.23]{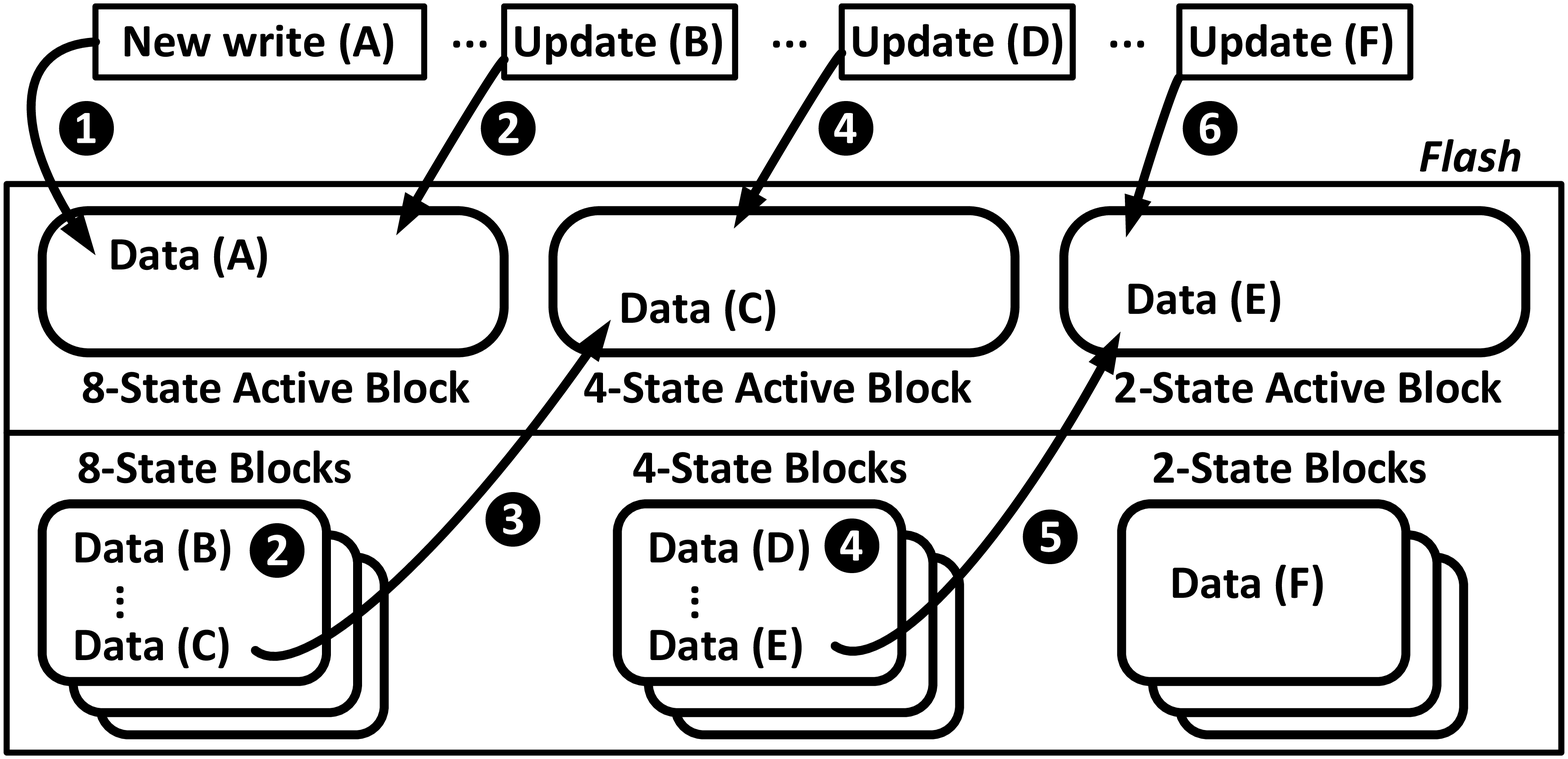}
  \caption{Block allocation examples; DSLC-FTL maintains three different active blocks, and write (update) data is stored in one of these. For mis-assigned data, the data scrubbing mechanism moves them to a safer block.}
  \label{figure:blk-alloc}
\end{figure}

\textbf{Block Allocation in DSLC-FTL: }
Due to the limitations of the write and erase operations, all cells in a single page and all pages in a single block have to be in the same mode in D-SLC. Thus, as opposed to the conventional SLCs that have two block types (clean or used) at each time, D-SLC has four block types in a flash chip -- each block is either clean (or empty), a 2-state mode, a 4-state mode, or a 8-state mode block.
Also, at each time, D-SLC has three active blocks and active write points corresponding to the three state modes it has.
Figure~\ref{figure:blk-alloc} shows the block allocation algorithm used in DSLC-FTL. 
On arrival of a new I/O block, the FTL assumes that it will have a short longevity and maps it to the 8-state mode active block (\circled{\color{white}1}). The heuristic behind this assumption is that, as shown in Table~\ref{tab:longevity}, a majority of the written data have ``less than one hour'' longevity which, irrespective of device wear-out level, is always mapped to 8-state mode based on mode-assignment in Table~\ref{tab:mode-assign-3-base}.
If this I/O block gets updated in less than an hour, i,e., the retention time of a 8-state mode block, the new update is also allocated in the (current) 8-state mode active block (\circled{\color{white}2}); so we do not change the block mode, as its history admits its short data longevity.
Otherwise, on expiration of the block's retention time, we read its all valid pages and migrate them to the 4-state mode active block (\circled{\color{white}3}); so the controller downgrades mode of these pages/blocks because of retention time violation. We call this mechanism \emph{data scrubbing} and implemented it in our DSLC-FTL.
We follow the same procedure for the I/O blocks mapped to 4-state mode: if their updates come before 4-state mode expiration, we keep rewriting them in the (current) 4-state mode active block (\circled{\color{white}4}); otherwise, on expiration of the block's retention time, we move its all valid pages to the 2-state mode active block by invoking data scrubbing (\circled{\color{white}5}).
The I/O data in 2-state mode block always remains in this mode (\circled{\color{white}6}).

This simple heuristic is easy to implement -- it needs two minor changes at the FTL metadata. 
\begin{enumerate}[leftmargin=*]
\item FTL needs keep the retention time information at block granularity (instead of page granularity). Indeed, when the first page is allocated to a block, FTL records the clock tic for that block, and periodically monitors it for expiration. 
\item FTL also needs 2-bit information per each block to indicate its status: ``00'' for the clean mode, ``01'' for the 2-state mode, ``10'' for the 4-state mode, and ``11'' for the 8-state mode.
\end{enumerate}

Finally, we note that (i) this configuration gives the maximum flexibility to DSLC-FTL and allows to write any incoming page into either of the blocks, depending on its data longevity, and (ii) we can employ a retention time predictor (like the one proposed in~\cite{relax-3}) to avoid the data scrubbing cost. However, we found that such mechanism brings a negligible lifetime gain and the data scrubbing in our scheme imposes a very small overhead, as will be discussed in Section~\ref{sec:eval-scrub}. Accordingly, the current version of D-SLC exploits data scrubbing, instead of a retention-time predictor.


\begin{figure}
  \centering
  \includegraphics[scale=.23]{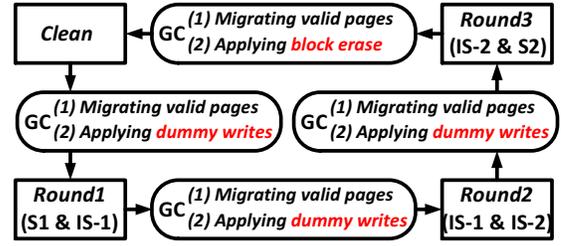}
  \caption{The state diagram of a 4-state mode block. The transition between any two state modes is performed by a GC invocation.}
  \label{figure:gc-state}
\end{figure}

\textbf{Garbage Collection in DSLC-FTL: }
We now describe the garbage collection procedure employed for a 4-state mode block, as an example, in D-SLC. This mechanism can be generalized to other modes as well.
The diagram in Figure~\ref{figure:gc-state} depicts the life-cycle of a 4-state mode block. At any given time, a block can be in one of the four states:
\begin{enumerate}[leftmargin=*]
    \item \textbf{Clean:} A block is initially clean or empty. All pages are erased.
    \item \textbf{Round1:} Starting with a clean block, at this state, we write data into the pages of the block in an in-order fashion (i.e., page \emph{i+1} has to be written after page \emph{i}). In this state, we use two first states, i.e., the states \verb"S1" and \verb"IS-1", for writing one bit data in each cell.
    \item \textbf{Round2:} When all the pages in a block are used up in Round1, the block state is changed to Round2 and we store one new page in the target page frame. Again, due to the constraint imposed by the in-order page writes in a block, the next three following actions have to be sequentially applied: (1) All valid pages, programmed in Round1, have to be relocated to elsewhere; (2) We apply dummy write pulses to all page frames to change their voltage states to \verb"IS-1" (pseudo-erase state for Round2) -- so, all the pages (cells) in the block will have the state \verb"IS-1"; and (3) We start the second round by using two intermediate states (\verb"IS-1" and \verb"IS-2"). The writes are again performed in an in-order fashion.
    \item \textbf{Round3:} When all the pages in a block are used up in Round2, the block state is changed to Round3 by following a procedure very similar to what described for Round1 to Round2 transition. The only difference is that, during Round3, FTL uses two last states (\verb"IS-2" and \verb"S2") for writing one new data.
\end{enumerate}

Here are some salient points to keep in mind about this block diagram:
\begin{itemize}[leftmargin=*]
\item Dummy write is the process during which all cells in all pages of a block are initialized to the state ``1'' in Round2. In fact, when the controller decides to change the status of a block from Round1 to Round2, it needs to make sure that all the cells in the block have the state \verb"IS-1" (i.e., like erase state for Round2). Implementing dummy write is easy -- at the end of Round1, if the content of a cell is ``1'' (\verb"S1"), the controller writes into the cell to make its state \verb"IS-1"; otherwise, i.e. the cell's content is ``0'' (\verb"IS-1"), no action is required. \\
The same procedure (dummy write) applies at the end of Round2, in order to make sure that all cells have the state \verb"IS-2" (i.e., like erase state for Round3).
\item Changing the block status from Round1 to Round2 and Round2 to Round3 is carried out by garbage collection (GC). This is because we need to move all valid pages to elsewhere, prior to applying our dummy writes. However, in these cases, we do \emph{not} erase the block. 
\item When all the pages of a block in Round3 are used up, we invoke a normal GC in order to erase the block (making it ready for Round1 programming).
\item D-SLC can work with any GC algorithm available for flash memories; however, the GC select algorithm has to be changed. When FTL invokes GC, it chooses one of the already-used blocks, regardless of its current state (i.e., the selected block can be either in Round1, Round2 or Round3). After moving the valid pages of the victim block, FTL applies an erase pulse (if the current state is Round3) or a dummy write (if the current state is Round1 or Round2). So, we do \emph{not} distinguish among the blocks in Round1, Round2 and Round3 during the victim block selection. 
\end{itemize}

%% file: eval-setup-Mhmd.tex
\section{Evaluation Methodology}

\subsection{Evaluation Framework}
We used DiskSim simulator~\cite{framework2} with the SSD extensions by Microsoft~\cite{framework1} to model a SLC-based SSD as an SCM.
This simulator is highly-parametrized and modularized which enables us configure various parameters including the number of flash chips, the flash internal components (i.e., the number of blocks, the number of pages in a block, and page size), and different timing values (i.e., page read and write latencies, block erase time, and data transfer time in/out of the flash chip).

On top of the DiskSim+SSD simulator\footnote{Our analysis and reported results in this paper are based on simulation. As a part of future work, we plan to have a more realistic implementation of D-SLC and D-SLC-FTL by using OpenNVM~\cite{impl-real}.}, we added one function (\emph{data scrubbing}) and modified two existing functions (\emph{block allocation} and \emph{garbage collection}) for D-SLC and its FTL implementation.
\begin{itemize}[leftmargin=*]
\item The \emph{data scrubbing} function implements the data scrubbing mechanism (i.e., when the retention time of a block is expired, valid pages in it, if any, are moved to a new block).
\item The \emph{block allocation} algorithm is modified to (i) support and maintain multiple active blocks for each flash chip in D-SLC, and (ii) implement the block allocation algorithm in Section~\ref{sec:ftl-design}.
\item The \emph{garbage collection} algorithm is also modified to support our multiple-round GC policy in Section~\ref{sec:ftl-design}.
\end{itemize}


\subsection{Configuration of the Baseline System}
\label{sec:config}
Table~\ref{tab:base-config} gives the details of the baseline SSD configuration. It is a 64GB SSD with eight 8GB SLC flash chips. The flash memory parameters are taken from a modern Micron device~\cite{baseline-slc} -- each chip has 8K blocks, each block has 128 pages and each page is 8KB. The read, write and erase latencies are 35 microseconds, 230 microseconds, and 1.5 milliseconds, respectively. The block endurance is 50K P/E cycles. 
We also assume that its FTL uses GREEDY algorithm~\cite{gc-greedy} for victim block selection during garbage collection, and the chip-level allocation strategy is static~\cite{page-alloc}.

\begin{table}[!t]
\caption{Configuration of the SLC NAND flash SSD in the baseline SCM.}
\label{tab:base-config}
\small
\begin{tabular}{|p{0.6in}||p{2.45in}|}
\hline 
\textbf{SSD} & 64GB capacity, eight 8GB SLC NAND flash chips~\cite{baseline-slc} \\ 
\hline
\textbf{Flash chip} & 8192 blocks per chip, 128 pages per block, 8KB pages, block endurance of 50K P/E cycles  \\
\hline 
\textbf{Timing parameters} & 35us for page read, 350us for page write, 1.5ms for block erase, 200MB/sec data transfer rate of the chip. \\
\hline 
\textbf{FTL} & GREEDY garbage collection, and round-robin policy for block allocation. \\
\hline
\end{tabular}
\vspace{-5pt}
\end{table}

\begin{table}[!t]
\centering
\caption{Important characteristics of our workloads.}
\label{tab:workload-cha}
\small
\begin{tabular}{|l|l|c|c|c|}
\hline
	\textbf{Name} & \textbf{Description} & \textbf{Write} & \textbf{Write} & \textbf{Read}  \\ 
 & &  \textbf{Ratio (\%)} &  \textbf{Size (KB)} &  \textbf{Size (KB)}     \\
\hline\hline
	hm\textunderscore0 & Hardware monitoring		&   64.5   &	  7.4	&   8.3	\\ \hline
        prn\textunderscore0 & Print server			&   89.2   &	  22.8	&   9.7	\\ \hline
	prn\textunderscore1 & Print server			&   24.7   &	  22.5	&   11.7	\\ \hline
	proj\textunderscore0 & Project directories		&   87.5   &	  17.8	&   40.9	\\ \hline
	prxy\textunderscore0 & Firewall/web proxy		&   96.9   &	  8.3	&   4.6	\\ \hline
	mds\textunderscore0 & Media server			&   88.1   &	  23.7	&   7.2       \\ \hline
	src1\textunderscore2 & Source control 		&   74.6   &	  19.1	&   32.5	\\ \hline
	src2\textunderscore0 & Source control			&   88.7   &	  8.1 	&   7.1	\\ \hline
	stg\textunderscore0 & Web staging			&   84.8   &	  24.9	&   9.2       \\ \hline    
	usr\textunderscore0 & User home directories	&   59.6   &	  40.9	&   10.3	\\ \hline
	web\textunderscore0 & Web/SQL server		&   70.1   &	  29.9	&   8.6      	\\ \hline
	web\textunderscore1 & Web/SQL server		&   45.9   &	  45.9	&   9.2	\\ \hline		
	wdev\textunderscore0 & Test web server		&   79.9   &	  12.6	&   8.2	\\ \hline
	wdev\textunderscore2 & Test web server		&   99.9   &	  6.1	&   8.1	\\ \hline
	rsrch\textunderscore0 & Research projects		&   90.7   &	  10.9	&   8.7	\\ \hline
\end{tabular}
\end{table}


\begin{figure*}[t]
\centering
\includegraphics[width=0.98\linewidth, bb= 0 0 3312 720]{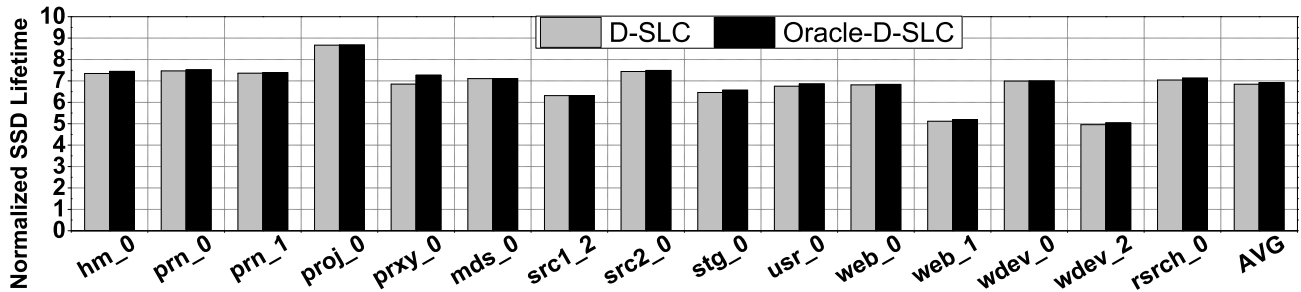}
\caption{Storage lifetimes of D-SLC and Oracle-D-SLC, normalized to that of the baseline SLC.}
\label{fig:lifetime-base}
\end{figure*}

\subsection{Workloads}
We use the I/O traces provided in the MSR Cambridge suite~\cite{cambridge}. 
These I/O traces are collected from different transactional and enterprise applications (or different disk volumes in a system running one single application) running multiple consecutive days, which allows us capture the longevity of I/O blocks for long time durations. 
Among the 36 traces in this benchmark suite, we used 15 traces for our evaluations. Our workloads are listed in Table~\ref{tab:workload-cha} (different indices refer to different volumes of the same application). The 21 excluded traces are either read-intensive (write ratio less than 20\%) where lifetime of the baseline SSD is not a concern (the endurance enhancement is the main goal of our technique), or many blocks in them are accessed once during the trace collection time (that is one week in these traces). 
Table~\ref{tab:workload-cha} gives the important characteristics of the studied workloads in terms of the \emph{write ratio}, \emph{average write request size}, and \emph{average read request size} (note that Table~\ref{tab:longevity} reports the retention time categorization of data blocks in these workloads).

\subsection{Evaluated Systems}
We evaluated and compared the results of three systems:
\begin{enumerate}[leftmargin=*]
	\item \textbf{Baseline:} This uses the conventional 2-state mode for all blocks. 
	\item \textbf{D-SLC:} This uses the proposed D-SLC technique with three state modes for the blocks (i.e., 2-state mode, 4-state mode, or 8-state mode). It also uses DSLC-FTL for flash management, which includes necessary functions to implement the scrubbing mechanism, block allocation and garbage collection in D-SLC. As explained before, we assume that D-SLC's read and write latencies in all the block modes are comparable to those in the baseline SLC (and hence performance optimization is not an objective in this analysis).
	\item \textbf{Oracle-D-SLC:} This system is mostly similar to the D-SLC and uses all DSLC-FTL functionalities except its block allocation algorithm. For block allocation, this system assumes that retention time information for each incoming I/O block is known ahead and hence it is allocated to the most suitable block mode. Doing this, Oracle-D-SLC removes the need for scrubbing and, as a result, gives the upper limit of the lifetime and performance improvement by D-SLC.
\end{enumerate}

During our analysis, the results of the evaluated systems are normalized to the \emph{baseline} system for comparison.

\subsection{Metrics}
We use the following metrics for our evaluation:
\begin{enumerate}[leftmargin=*]
	\item \textbf{Lifetime:} It refers to the lifespan of the SLC SSD system and is measured as the total data volume (in KBs) written to it up to the point that its all chips/blocks reach their endurance limit. Under a fixed endurance limit, the more data written to an SSD, the longer lifetime it has.
	\item \textbf{PWE:} Section~\ref{sec:pwe-define} defines our PWE metric. Note that the PWE of the conventional SLC is ``1'' during its entire lifetime. However, the proposed D-SLC results in various PWE values for each block during its lifetime (it can be ``1'', ``3'' or ``7'' for the 2-state, 4-state or 8-state modes, respectively).
	\item \textbf{GC rate and GC cost:} The GC rate refers to the average number of GC invocations in a time unit, and GC cost represents the average execution time of a GC. The higher GC rate and cost  results in lower available bandwidth for normal I/O operations.
	\item \textbf{Scrubbing rate and scrubbing cost:} The scrubbing rate indicates how often our data scrubbing mechanism is triggered (i.e., the ratio of blocks on which the data scrubbing is actually triggered, as a fraction of the total number of blocks used). The scrubbing cost is the average number of page migrations required for each scrubbing initiation.
	\item \textbf{Throughput:} It is measured as the amount of data (in KBs) read from or written to the SSD in a time unit.
\end{enumerate}


%% file: eval-base.tex
\section{Evaluation Results}
\label{sec:eval-base}

\subsection{Lifetime Enhancement}
\label{sec:eval-life}
Figure \ref{fig:lifetime-base} shows that the lifetimes of D-SLC and Oracle-D-SLC, \emph{normalized} to the baseline SLC. Compared to the baseline SLC, D-SLC and Oracle-D-SLC increase the lifetime by 6.8x and 6.9x, respectively. Exploiting short retention times in workloads and introducing multiple state modes (i.e., additional 4 and 8 state modes) are quite effective in prolonging the storage lifespan. Specifically, the significant increase in PWEs allows more and more data to be written in the same P/E cycles, which is analyzed in Section \ref{sec:eval-pwe}. Meanwhile, D-SLC achieves a lifetime improvement that is very close (only 1.1\% less) to that brought by Oracle-D-SLC. This implies that D-SLC does not need frequent data scrubbing invocations. (Section \ref{sec:eval-scrub} provides an analysis on the data scrubbing overheads). In the following sections, we analyze the behavior of the feasible system (D-SLC) only.


\begin{figure*}[t]
\centering
	\begin{subfigure}{.24\linewidth}
		\centering
		\includegraphics[width=0.99\linewidth]{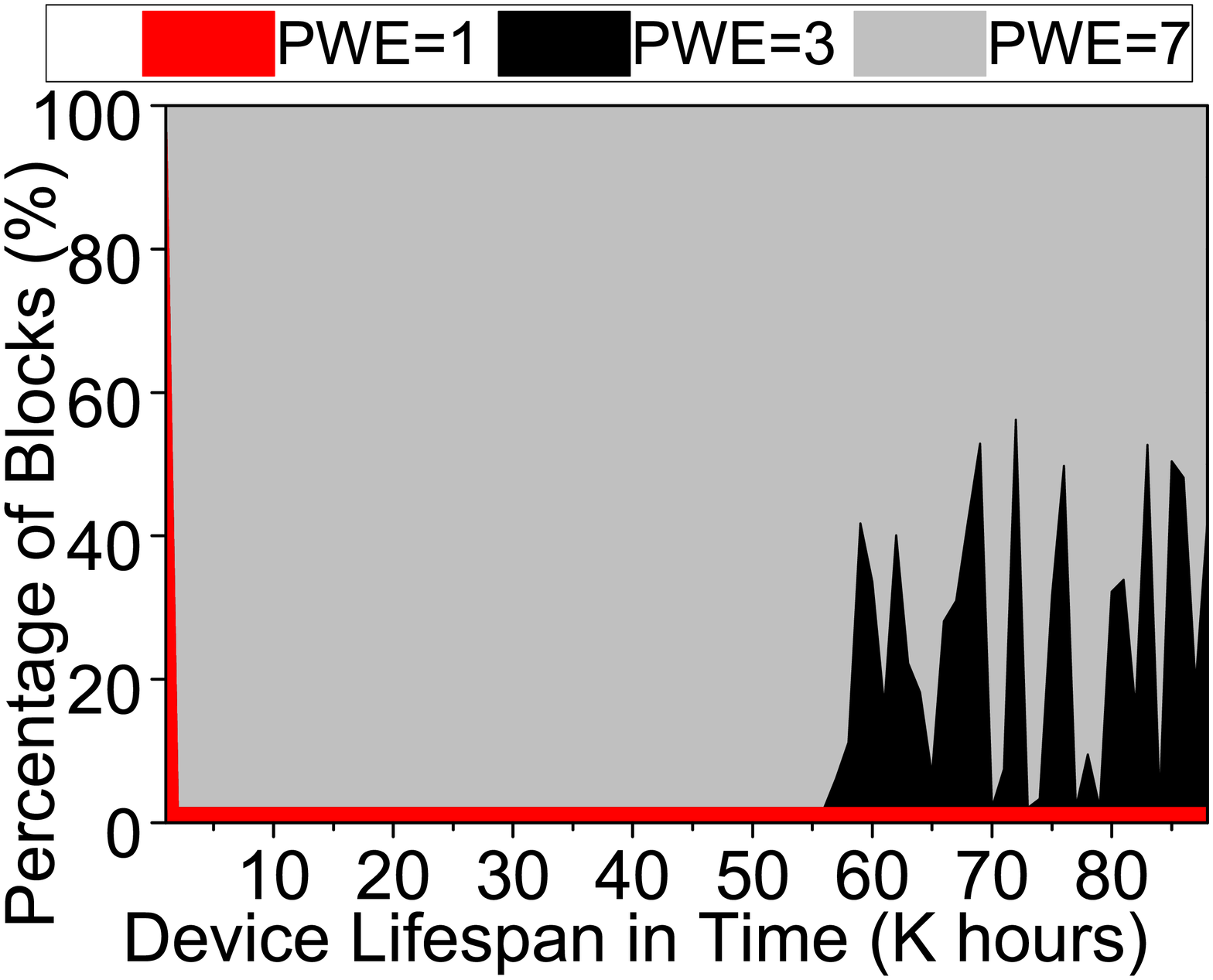}
		\caption{hm\textunderscore0}\label{fig:pwe-base-hm_0}
	\end{subfigure}
	\begin{subfigure}{.24\linewidth}
		\centering
		\includegraphics[width=0.99\linewidth]{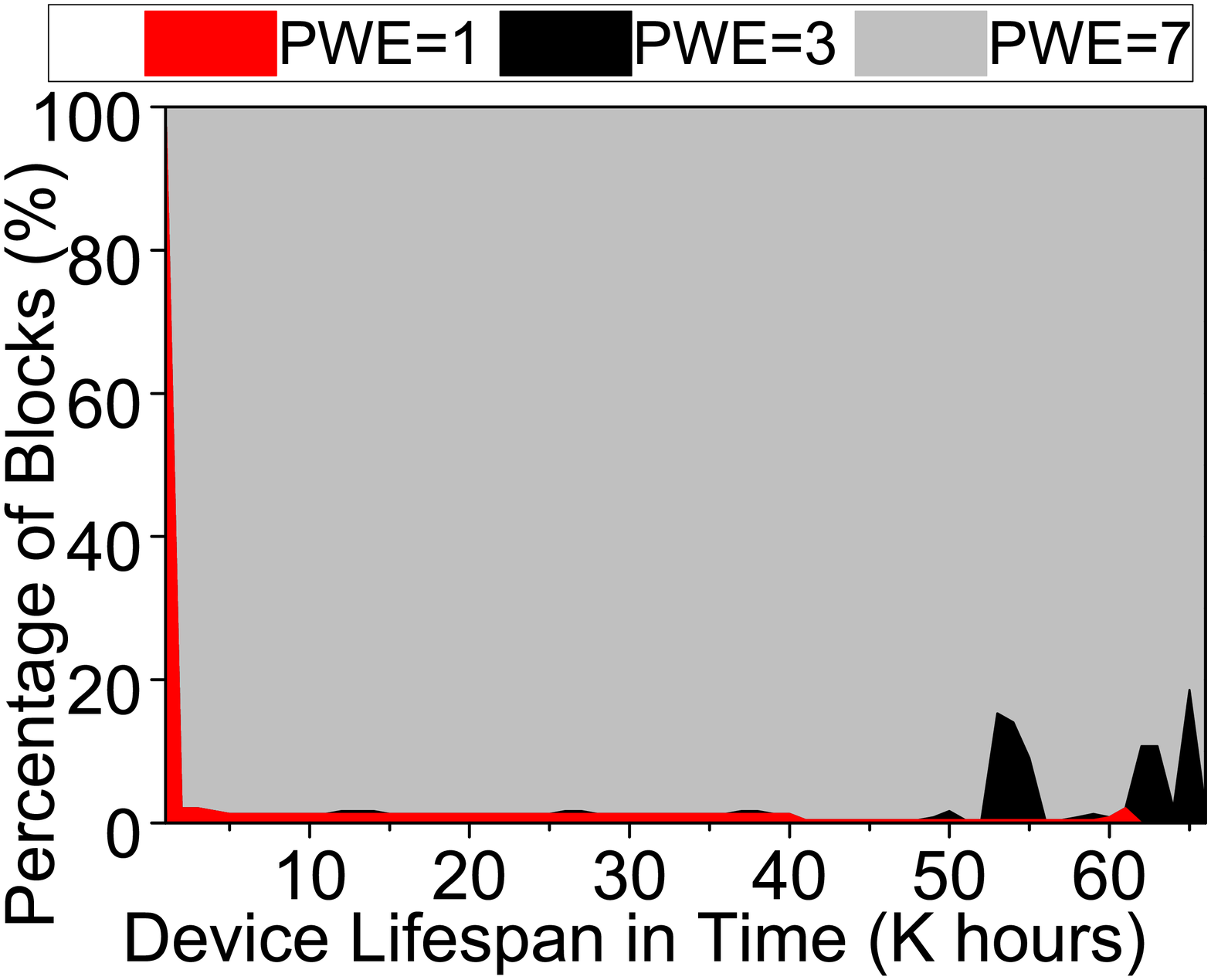}
		\caption{prn\textunderscore0}\label{fig:pwe-base-prn_0}
	\end{subfigure}
	\begin{subfigure}{.24\linewidth}
		\centering
		\includegraphics[width=0.99\linewidth]{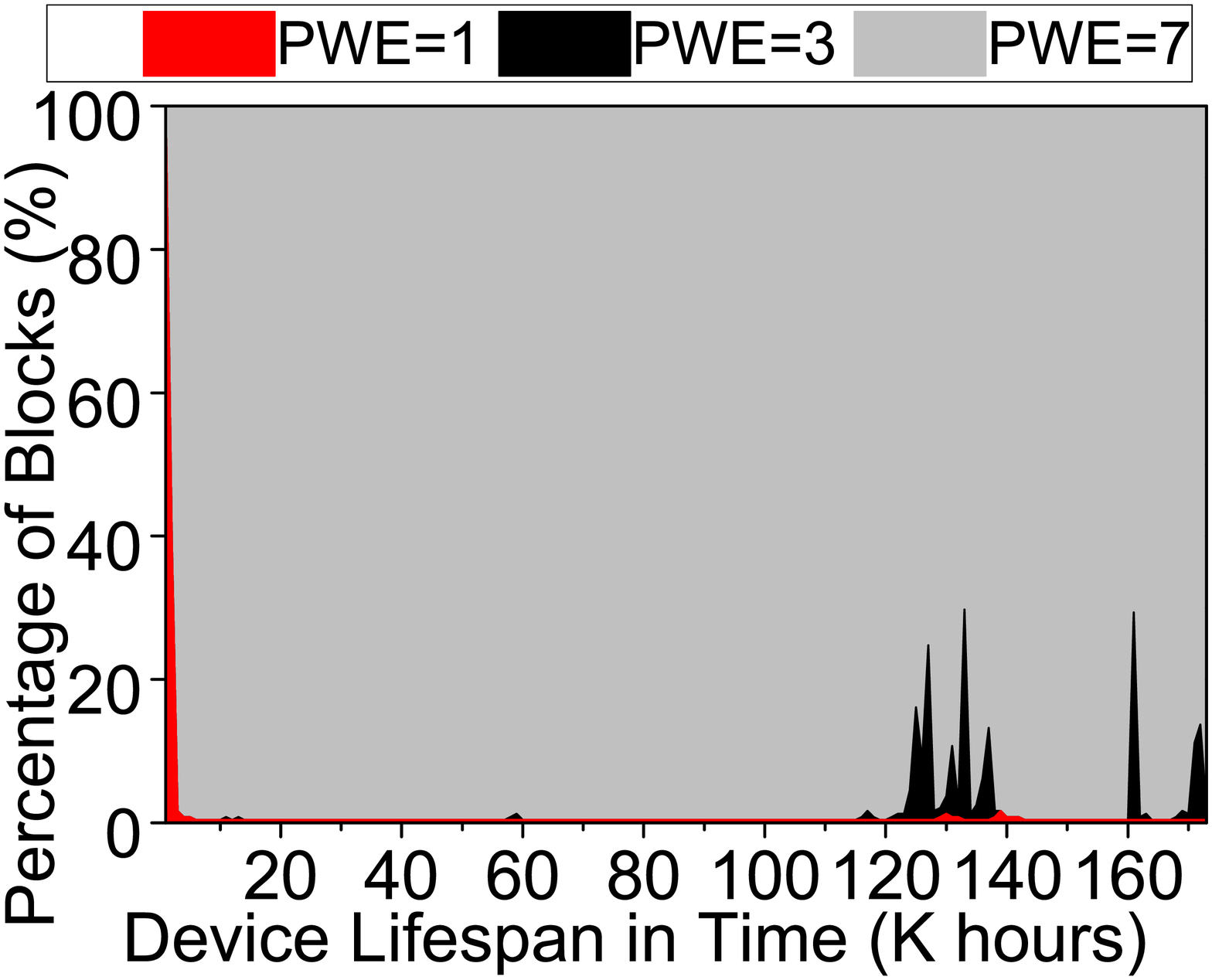}
		\caption{prn\textunderscore1}\label{fig:pwe-base-prn_1}
	\end{subfigure}
	\begin{subfigure}{.24\linewidth}
		\centering
		\includegraphics[width=0.99\linewidth]{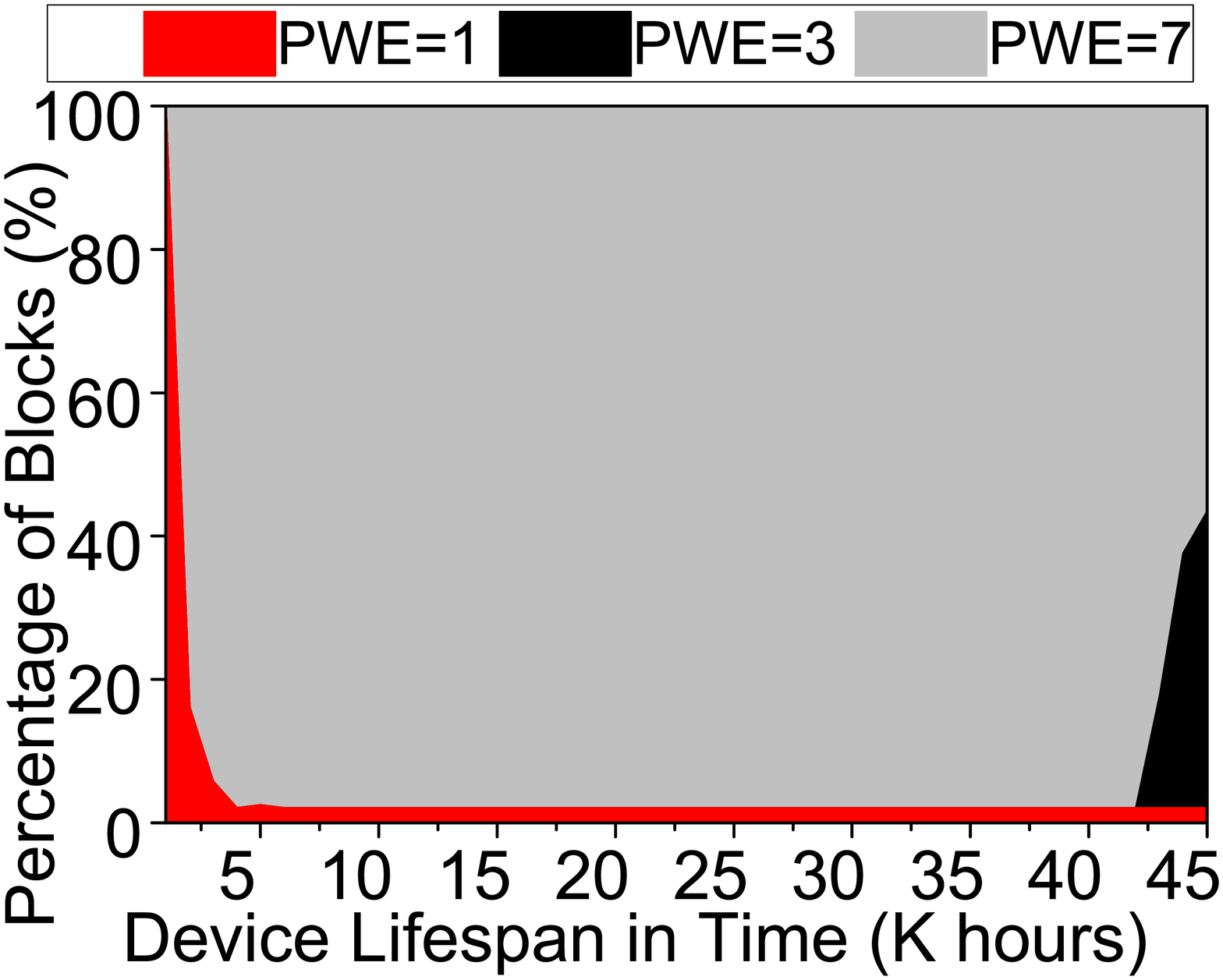}
		\caption{proj\textunderscore0}\label{fig:pwe-base-proj_0}
	\end{subfigure}
	\begin{subfigure}{.24\linewidth}
		\centering
		\includegraphics[width=0.99\linewidth]{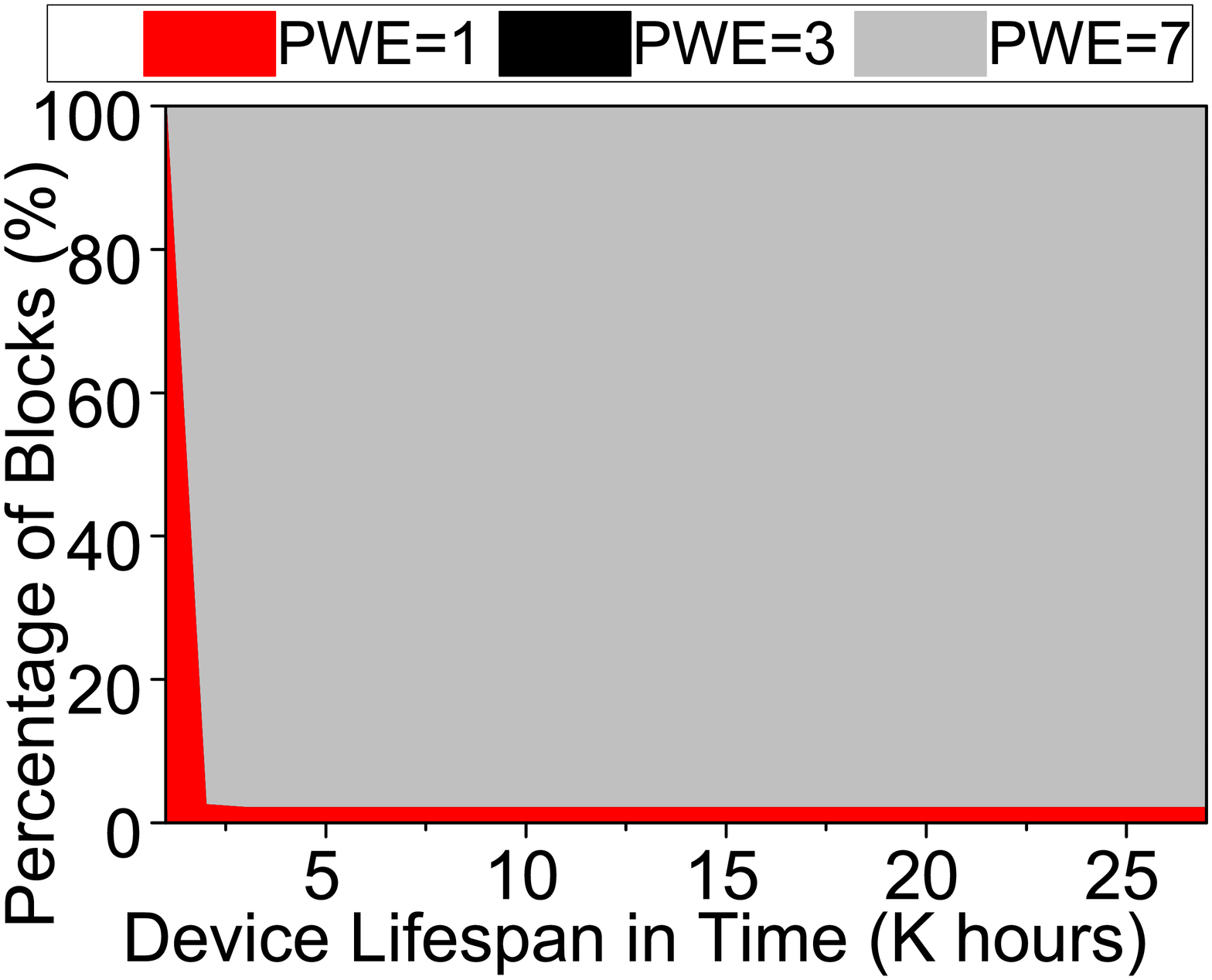}
		\caption{prxy\textunderscore0}\label{fig:pwe-base-prxy_0}
	\end{subfigure}
	\begin{subfigure}{.24\linewidth}
		\centering
		\includegraphics[width=0.99\linewidth]{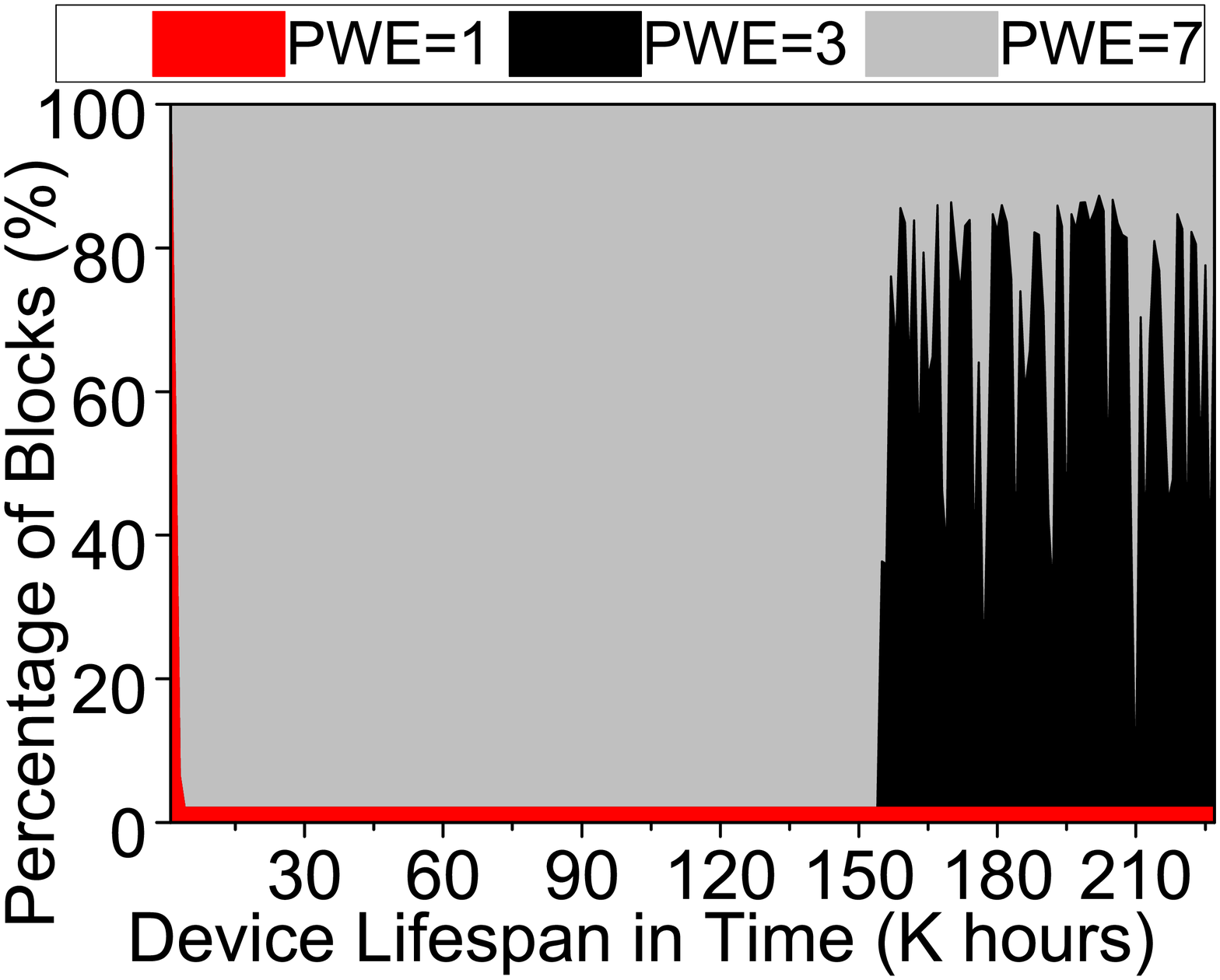}
		\caption{mds\textunderscore0}\label{fig:pwe-base-mds_0}
	\end{subfigure}
	\begin{subfigure}{.24\linewidth}
		\centering
		\includegraphics[width=0.99\linewidth]{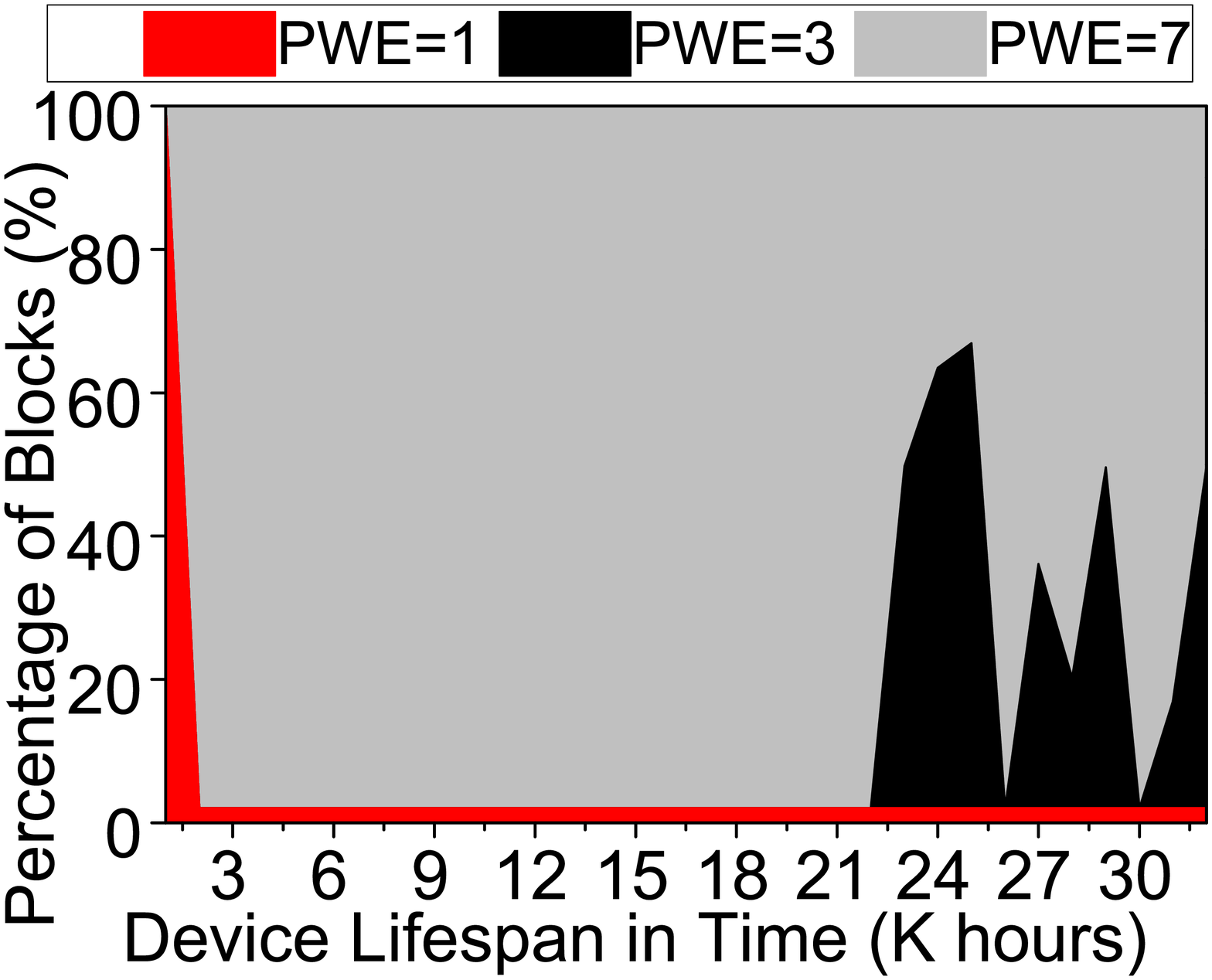}
		\caption{src1\textunderscore2}\label{fig:pwe-base-src1_2}
	\end{subfigure}
	\begin{subfigure}{.24\linewidth}
		\centering
		\includegraphics[width=0.99\linewidth]{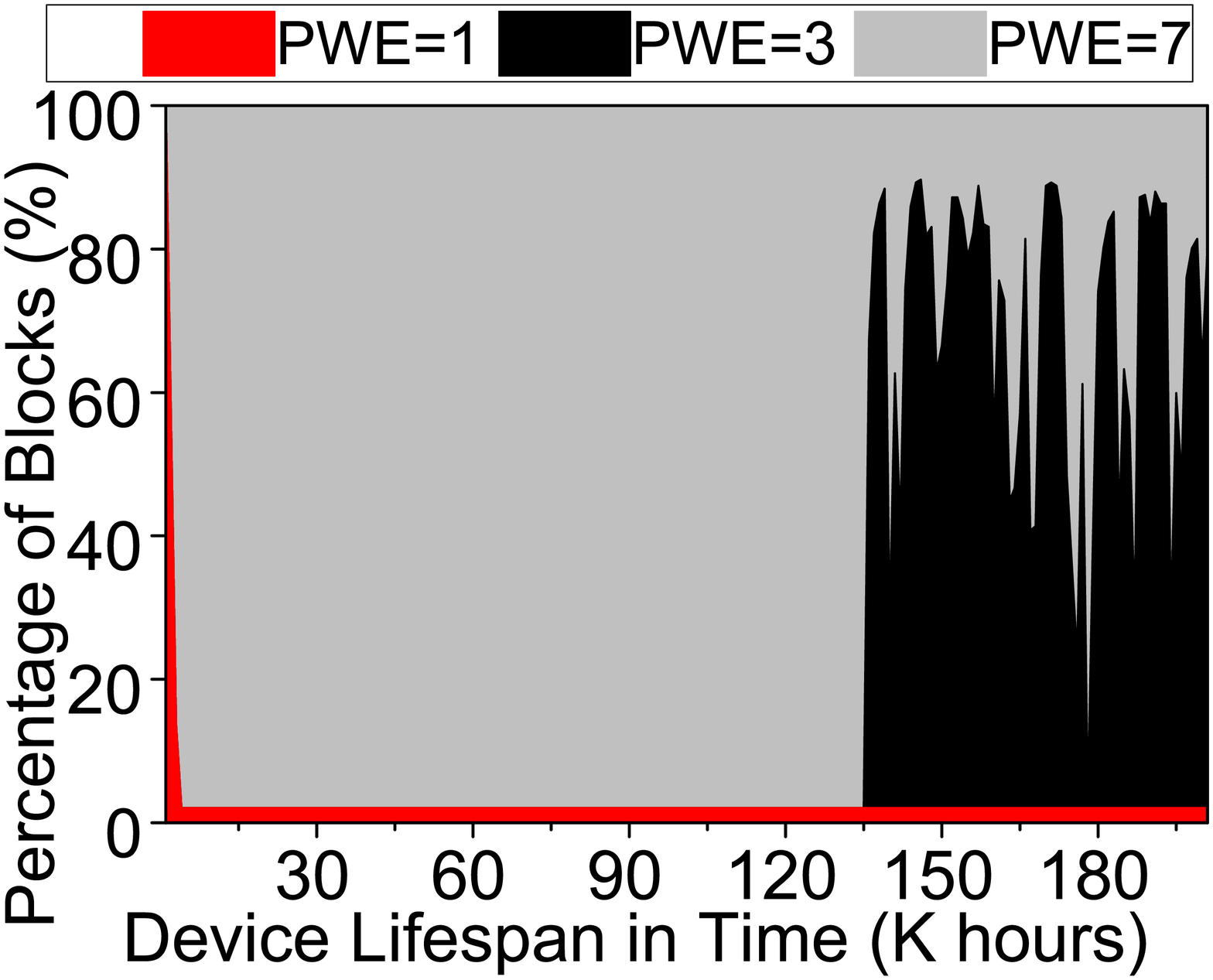}
		\caption{src2\textunderscore0}\label{fig:pwe-base-src2_0}
	\end{subfigure}
	\begin{subfigure}{.24\linewidth}
		\centering
		\includegraphics[width=0.99\linewidth]{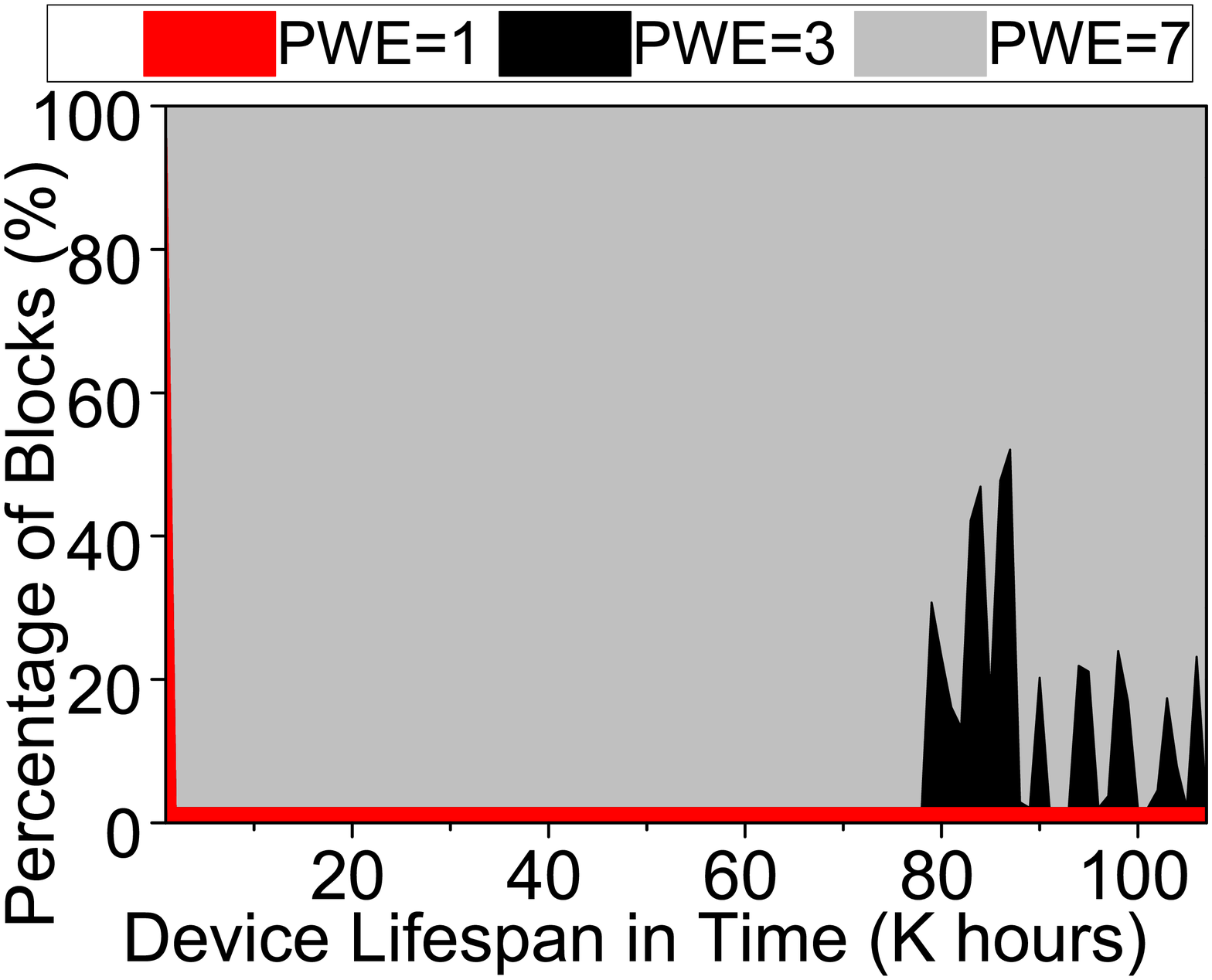}
		\caption{stg\textunderscore0}\label{fig:pwe-base-stg_0}
	\end{subfigure}
	\begin{subfigure}{.24\linewidth}
		\centering
		\includegraphics[width=0.99\linewidth]{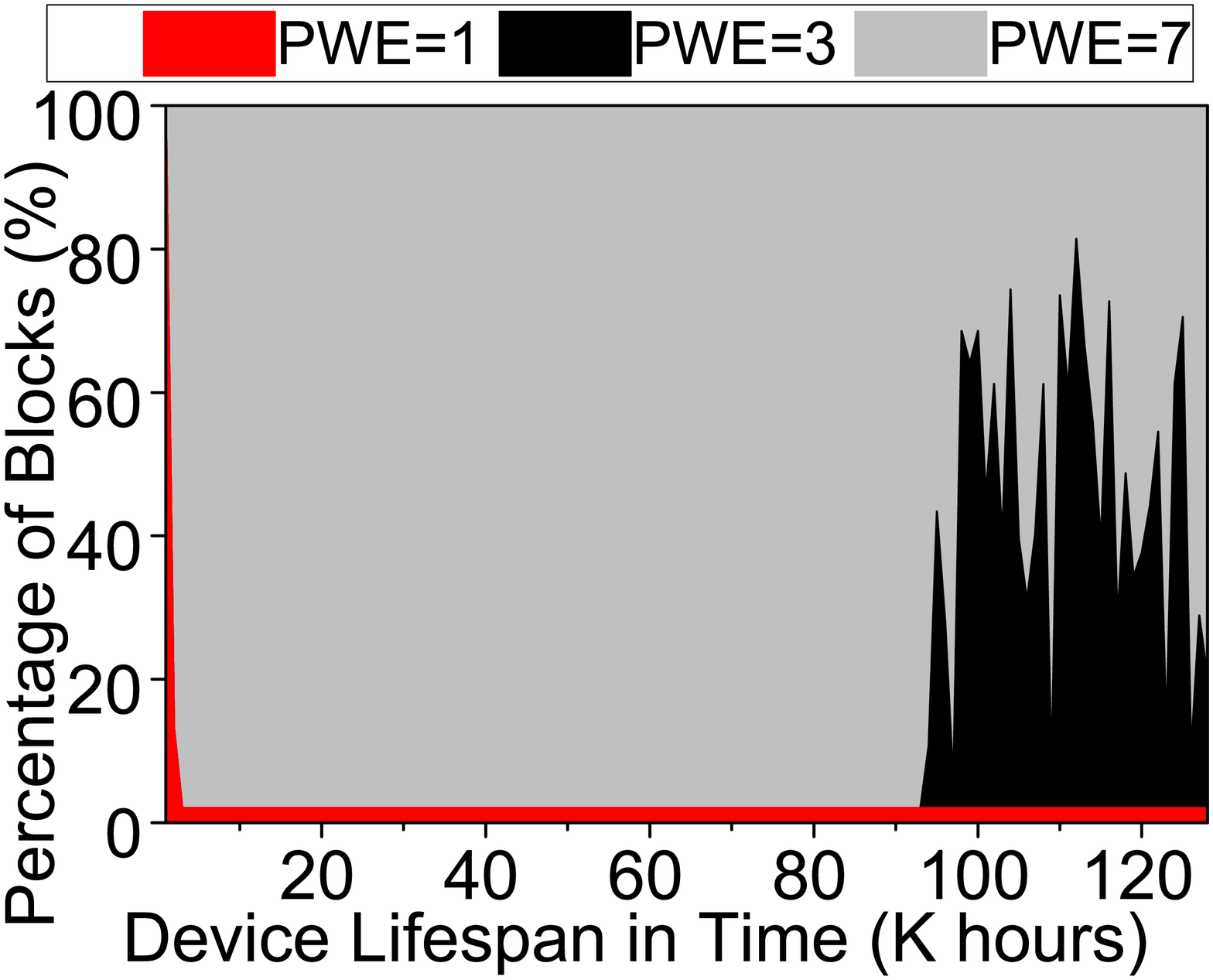}
		\caption{usr\textunderscore0}\label{fig:pwe-base-usr_0}
	\end{subfigure}
	\begin{subfigure}{.24\linewidth}
		\centering
		\includegraphics[width=0.99\linewidth]{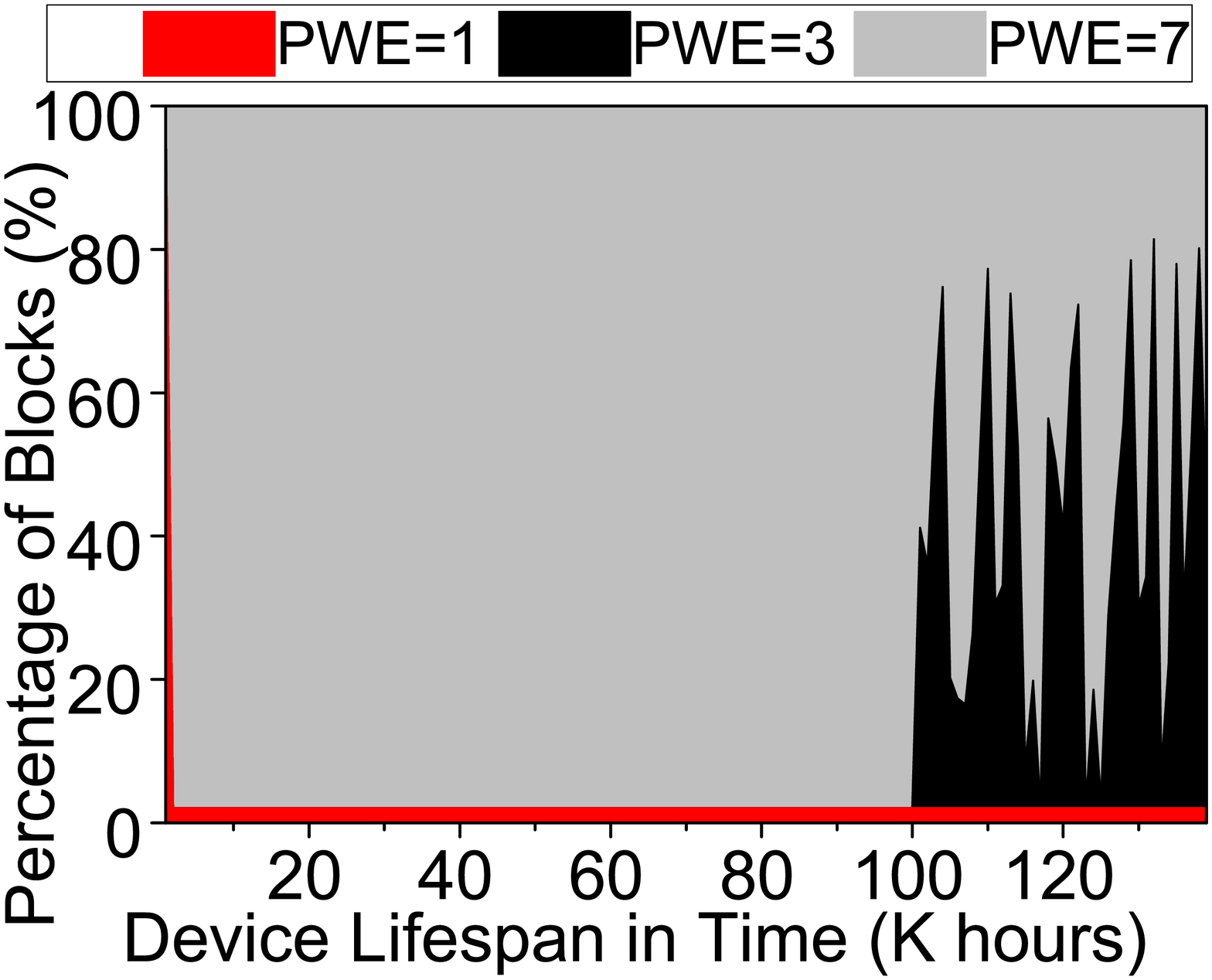}
		\caption{web\textunderscore0}\label{fig:pwe-base-web_0}
	\end{subfigure}
	\begin{subfigure}{.24\linewidth}
		\centering
		\includegraphics[width=0.99\linewidth]{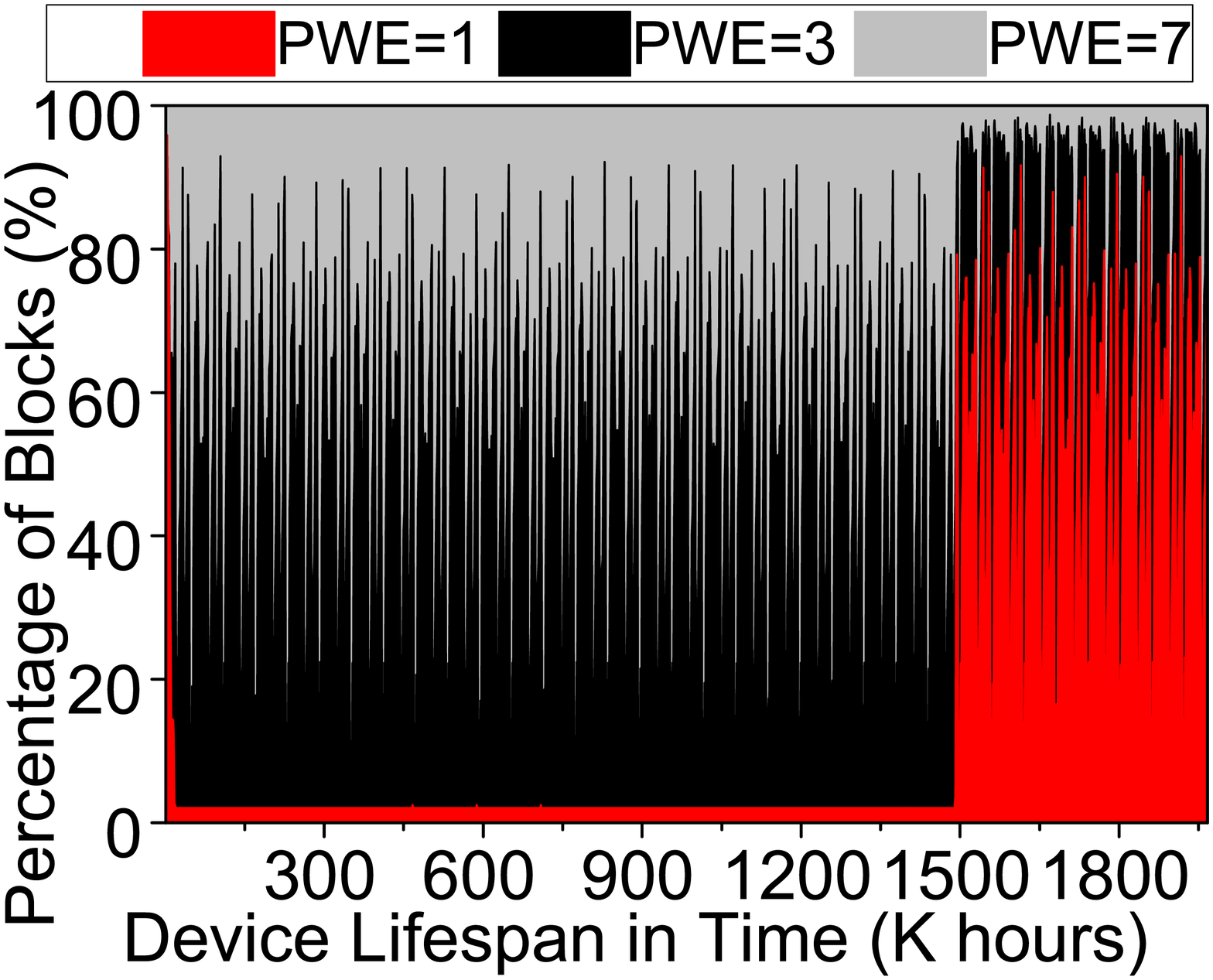}
		\caption{web\textunderscore1}\label{fig:pwe-base-web_1}
	\end{subfigure}
	\begin{subfigure}{.24\linewidth}
		\centering
		\includegraphics[width=0.99\linewidth]{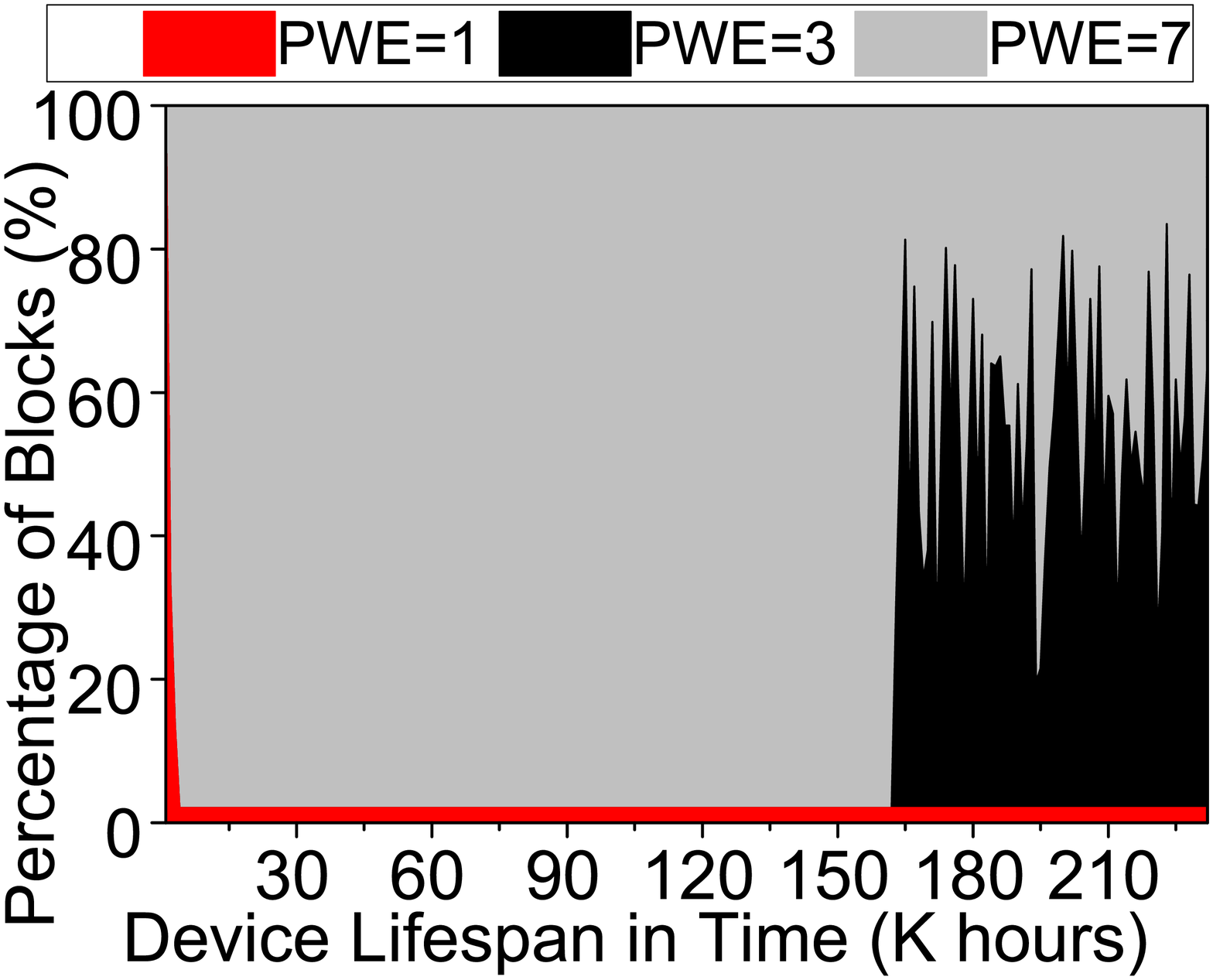}
		\caption{wdev\textunderscore0}\label{fig:pwe-base-wdev_0}
	\end{subfigure}
	\begin{subfigure}{.24\linewidth}
		\centering
		\includegraphics[width=0.99\linewidth]{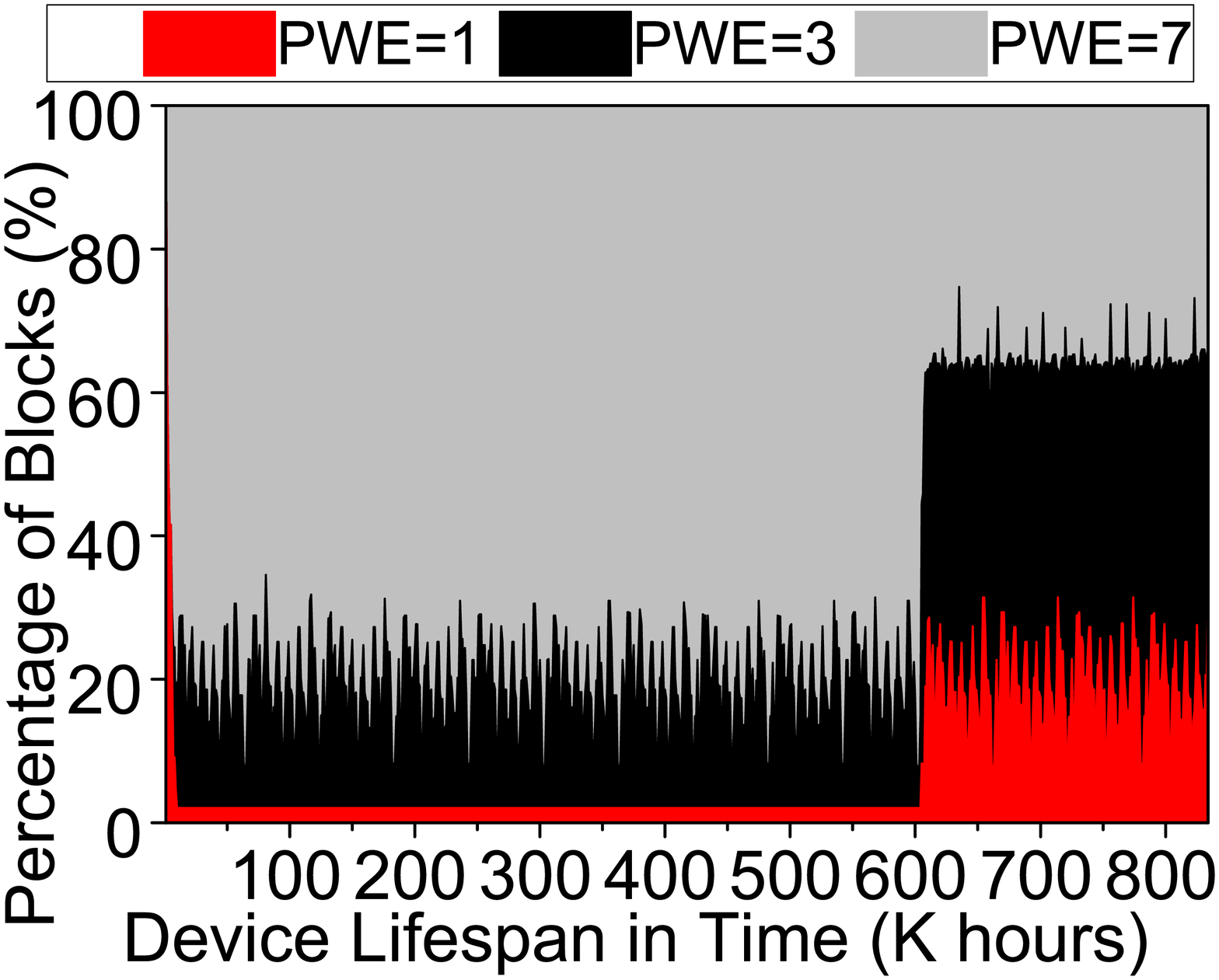}
		\caption{wdev\textunderscore2}\label{fig:pwe-base-wdev_2}
	\end{subfigure}
	\begin{subfigure}{.24\linewidth}
		\centering
		\includegraphics[width=0.99\linewidth]{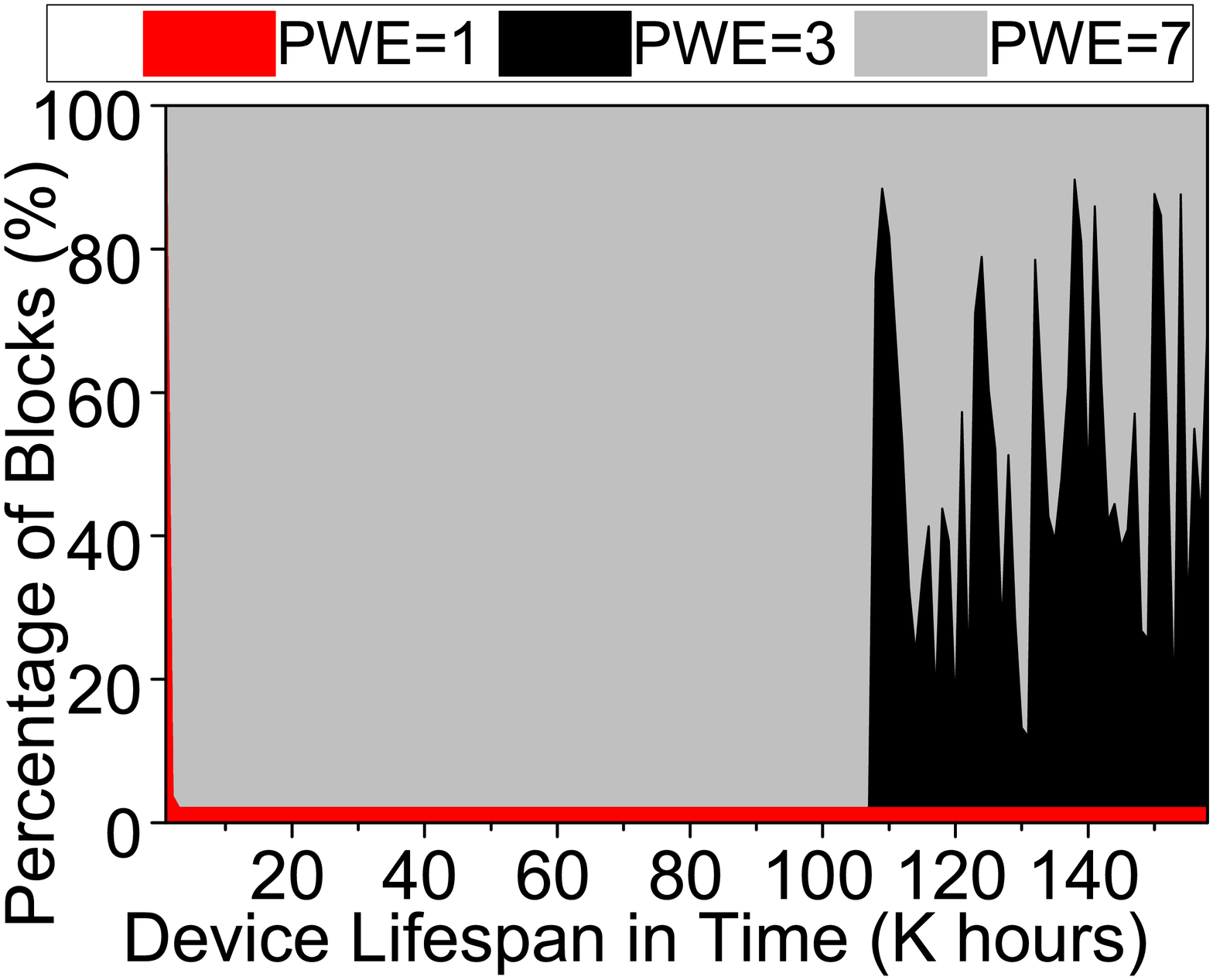}
		\caption{rsrch\textunderscore0}\label{fig:pwe-base-rsrch_0}
	\end{subfigure}
\caption{Percentages of blocks with 2, 4, and 8 states (whose PWEs are 1, 3, and 7, respectively) during the device lifespan.}
\label{fig:pwe-base}
\end{figure*}

Compared to the other workloads tested, \emph{web\textunderscore1} and \emph{wdev\textunderscore2} achieve lower lifetime improvements (5.1x and 4.9x, respectively). Since over 20\% of their data have retention times between 10 hours and 3 days (see Table \ref{tab:longevity}) and such block cannot be placed in blocks with 8-state mode (see Table \ref{tab:mode-assign-3-base}), they miss the opportunity for further increasing PWE and improving the storage lifetime. On the other hand, \emph{proj\textunderscore0} exhibits much higher lifetime enhancement than the average. This impressive result is due to two factors: (1) over 96\% of their retention times are below one hour, thus allowing almost all data to be stored in blocks with 8-state mode and maximizing PWE/lifetime. (2) Maintaining multiple active blocks efficiently separate highly-update data (with short retention times) from data with long retention times in the workload, which reduces its GC rate and cost. Note that the lower the GC rate/cost, the higher the lifetime gain. Furthermore, \emph{prxy\textunderscore0} with a similar distribution of retention times cannot lead to such a high improvement, as it cannot get the benefits of reducing the GC rate and cost. We discuss this in more detail in Section \ref{sec:eval-gc}.

\subsection{PWE Analysis}
\label{sec:eval-pwe}

The lifetime improvements brought by our scheme originate from the increase in PWE. Figure \ref{fig:pwe-base} provides the PWE analysis, which is the change in ratios of blocks with 2, 4, and 8 state-mode (whose PWEs are 1, 3, and 7, respectively) as the storage ages. In general, the larger the gray area (PWE=7), the more beneficial our scheme. Note however that, we do not directly compare across different workloads, since their lifespans in time (x-axis) are all different. We can observe a few common characteristics across the traces:
\begin{itemize}[leftmargin=*]
	\item \textbf{Thin red area along X-axis}: Throughout the storage lifespan, there are always a few blocks with 2-state mode due to two reasons: (1) a few blocks with 2-state mode (including one of the active blocks) are reserved to serve write data in need of a long retention guarantee. In addition, (2) such blocks have a tendency to maintain their mode (i.e., 2-state mode), as they are not likely to get invalid and erased.
	\item \textbf{Black/red cliffs around old ages}: When the storage gets older, the number of blocks whose PWE is ``3'' dramatically increases. This is because the data whose retention times range from 1 to 10 hours should be stored in blocks with 4-state mode from 30K P/Es, while they could be accommodated in 8-state mode blocks in early ages (see Table \ref{tab:mode-assign-3-base}). For the same reason, data whose retention times are between 10 hours and 3 days need to be placed in 2-state mode blocks instead of blocks with 4-state mode from 30K P/Es. This can be seen in \emph{web\textunderscore1} and \emph{wdev\textunderscore2}.
	\item \textbf{Laciniate black/red lines}: The ratios of block with different modes continue to change as time goes by. This indicates that a block frequently changes its mode, when it is erased and allocated as a new active block again. Thus, as the workload being executed moves from one phase to another, the storage can adapt to the change and adjust the ratios of blocks with different modes.
\end{itemize}


One can further see from Figures \ref{fig:pwe-base-web_1} and \ref{fig:pwe-base-wdev_2}, the reason why \emph{web\textunderscore1} and \emph{wdev\textunderscore2} achieve relatively lowered lifetime enhancements. Specifically, the percentages of blocks with black (PWE=3) and red (PWE=1) stays relatively high throughout the whole lifetime. This is because these workloads have a large portion (27\%) of data whose retention times are between 10 hours and 3 days and they occupy blocks with 4 and 2-state mode.

\begin{table*}[!t]
\caption{The data scrubbing ``rate'' (percentage of blocks on which the data scrubbing is invoked) and ``cost'' (the number of valid pages to be migrated per a block scrubbing).}
\label{tab:scrubbing-overhead}
\small
\begin{tabular}{|c|c|c|c|c|c|c|c|c|  }
\hline
\multicolumn{2}{|c|}{\textbf{Workload}} & \textbf{hm\textunderscore0} & \textbf{prn\textunderscore0} & \textbf{prn\textunderscore1} & \textbf{proj\textunderscore0} & \textbf{prxy\textunderscore0} & \textbf{mds\textunderscore0} & \textbf{src1\textunderscore2} \\
\hline\hline 
\multicolumn{2}{|c|}{\textbf{Scrub. Rate (\%)}} & 0.068 & 0.012 & 0.029 & 0.061 & 0.187 & 0.096 &	0.078 \\
\hline 
\multicolumn{2}{|c|}{\textbf{Scrub. Cost (cnt.)}} & 26.01 & 25.56 & 26.12 & 27.88 & 20.74 & 16.51 & 12.89	\\
\hline 
\textbf{src2\textunderscore0} & \textbf{stg\textunderscore0} & \textbf{usr\textunderscore0} & \textbf{web\textunderscore0} & \textbf{web\textunderscore1} & \textbf{wdev\textunderscore0} & \textbf{wdev\textunderscore2} & \textbf{rsrch\textunderscore0} & \textbf{AVG}	\\
\hline\hline
0.076 & 0.021 & 0.045 & 0.086 & 0.109 & 0.063 & 0.097 & 0.028 & 0.071 	\\
\hline
13.14 & 23.42 & 25.45 & 26.13 & 21.11 & 19.19 & 25.87 & 33.49 & 22.83 	\\
\hline
\end{tabular}
\end{table*}

\begin{figure}[!t]
\centering
\includegraphics[width=0.99\linewidth, bb= 0 0 864 432]{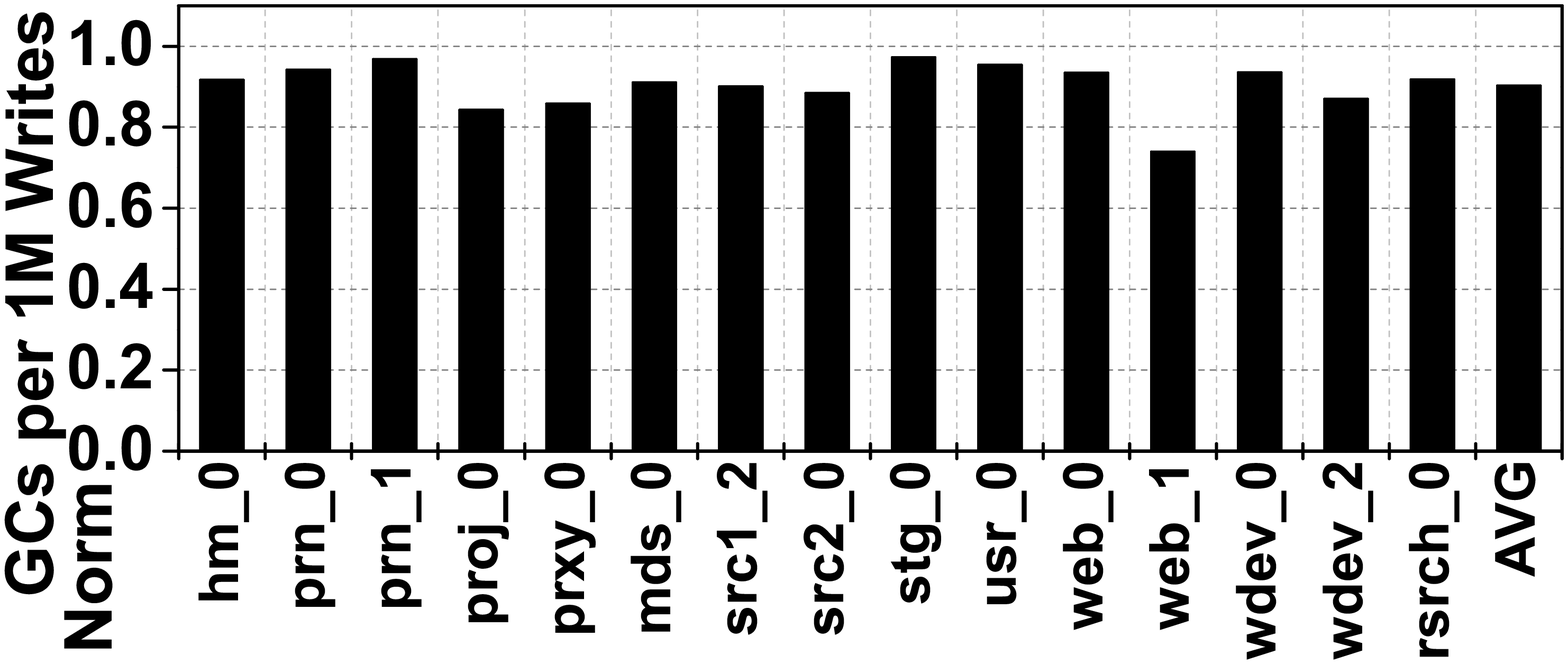}
\includegraphics[width=0.99\linewidth, bb= 0 0 864 360]{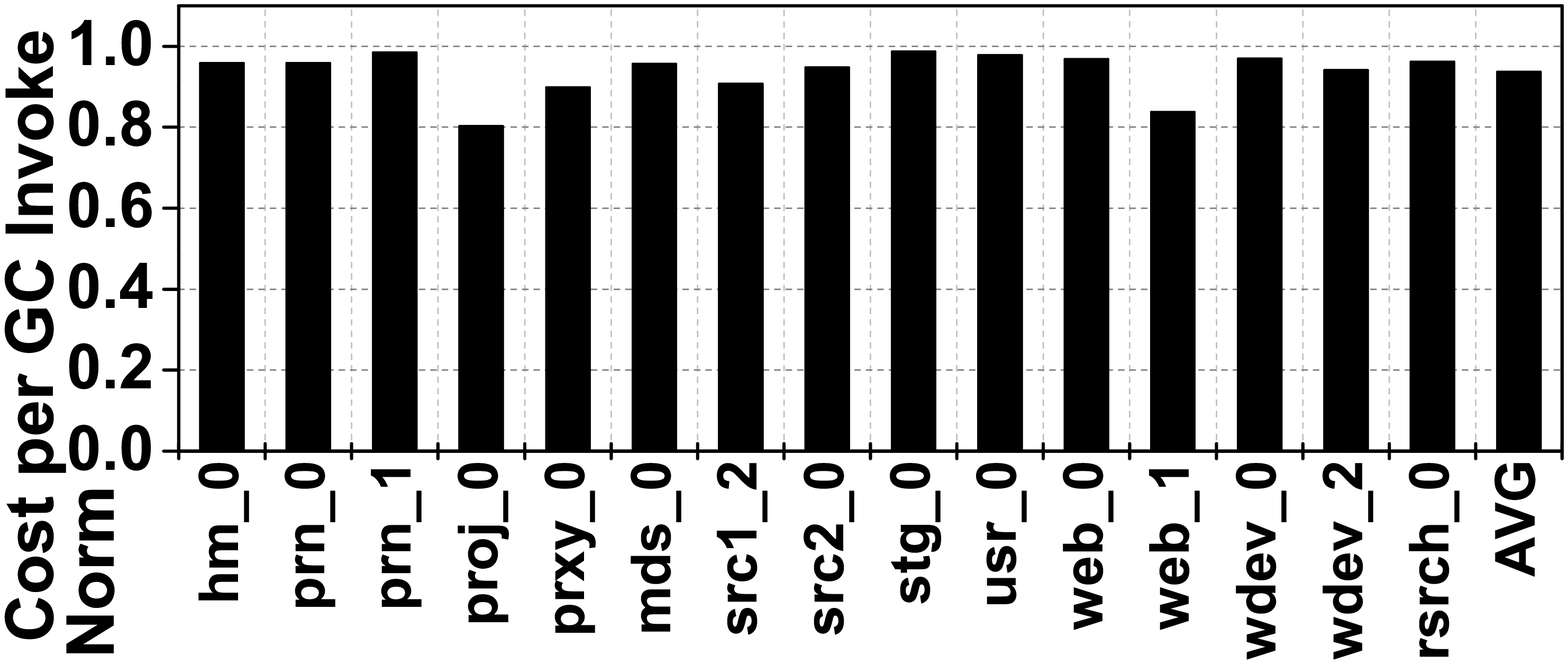}
\caption{The numbers of GC invocations in 1M writes and the costs of a GC invocation, normalized to the baseline SLC.}
\label{fig:gc-overhead}
\end{figure}

\subsection{GC Analysis}
\label{sec:eval-gc}

Figure \ref{fig:gc-overhead} provides the GC frequencies and the GC costs, both of which correctively analyze the GC overhead of D-SLC. Compared to the baseline SLC, our scheme decreases ``the number of GC invocations per one million writes'' and ``the cost per a GC invocation'' by 9.7\% and 6.3\%, on average. This small reduction in GC burdens helps our scheme indirectly save the storage lifetime and bandwidth, even though the effectiveness is not high.

This reduction in GC overheads comes from the isolation of hot data (which are highly updated) from cold data with long retention times. Note that our scheme provides multiple active blocks and groups data with similar retention guarantee together in a single block. Hence, when a GC is invoked, victim blocks (which are blocks with 8 states in most cases) have a tendency to include relatively fewer valid pages, since no data with long retention times is placed in them. As a result, the number of page migrations (i.e., the cost per a GC invocation) decrease, and in turn, the new pages are not wasted during the GC executions and the GC invocation frequency (i.e., the number of GC invocations per one million writes) is lowered.

The significant lifetime improvement in \emph{proj\textunderscore0} results from the largely-reduced GC overheads as well as its high PWEs. Surprisingly, \emph{proj\textunderscore0} drops ``the number of GC invocations per one million writes'' and ``the cost per a GC invocation'' by 16\% and 20\%, respectively. This indirect advance in lifetime helps \emph{proj\textunderscore0} with our scheme achieve a 8.7x of lifetime improvement, which is far beyond 7x when assuming all blocks whose PWE of ``7'' are used throughout the storage lifespan. One might note that \emph{web\textunderscore1} also experiences a significantly-reduced GC overheads. Unfortunately, this GC benefit does not lead to the high lifetime improvement (i.e., only 5.1x) in \emph{web\textunderscore1}. We want to emphasize that the lifetime enhancement in our scheme mainly comes from the increased PWEs and this additional GC overhead reduction is a secondary advantage.

\subsection{Scrubbing Overhead Analysis}
\label{sec:eval-scrub}

Table \ref{tab:scrubbing-overhead} presents the data scrubbing rate and cost, which collectively represent the data scrubbing overhead. The data scrubbing rate (``the percentage of blocks for which the data scrubbing is triggered as a fraction of the total number of allocated blocks'') is quite low (i.e., 0.071\%, on average). Furthermore, the data scrubbing cost (``the number of valid pages in a 128 page-block where the data scrubbing is triggered'') is also low (i.e., 22.83 / 128 pages, on average). This low data scrubbing rate is because most of target blocks are already erased when it comes to the deadline and there is no need to act for such blocks. Note that most (page) data in a block are invalidated before the deadline is reached, and such blocks where most pages are invalidated are the best candidates for the GC. Even though the target block is not erased and the data scrubbing is executed, most of its data are already invalidated, which results in low scrubbing costs.


Compared to other workloads, \emph{prxy\textunderscore0} shows a relatively high data scrubbing rate (0.187\%), even though its cost is still low. It is because the longevity of its data blocks vary. One can confirm the impact of this high scrubbing rate from Figure \ref{fig:lifetime-base}; the lifetime improvement of \emph{prxy\textunderscore0} is lowered a bit, compared to Oracle-D-SLC which is aware of the longevity of all data in advance. However, in general, the scrubbing overhead is too small to severely hurt the storage lifetime and bandwidth. 

One might wonder why workloads like \emph{web\textunderscore2} and \emph{wdev\textunderscore1}, where a large portion of data have long longevity, would experience low data scrubbing overheads.
For these two workloads, a large fraction of blocks have 1 hour to 3 days data longevity, and they are written in the 8-state mode block at first. However, their scrubbing overheads are also very small. This is because, once they are moved to 4-state or 2-state mode blocks by the scrubbing, the following updates on these blocks are written in the 4-state or 2-state active blocks, after which there is no more scrubbing activities on these data. Thus, the scrubbing overhead is not significant for these workloads after the state changes happen.

\begin{figure}
\centering
\includegraphics[width=0.99\linewidth, bb= 0 0 864 432]{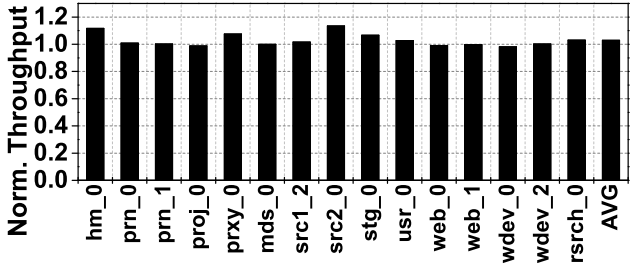}
\caption{The storage throughput of D-SLC, normalized to the baseline SLC.}
\label{fig:throughput}
\end{figure}

\subsection{Performance Analysis}

Figure \ref{fig:throughput} shows the storage throughput results, which are comparable to those of the baseline SLC. Some workloads experience improved throughput, while others lose a bit of their performance; overall, the storage throughput increases by 3\% on average. The important parameters that shape the storage throughput in our scheme are as follows:
\begin{itemize}[leftmargin=*]
	\item \textbf{Device read/write latencies}: If device latencies increase, storage throughput decreases and vice versa. However, our scheme provides read/write latencies close to the baseline SLC, as discussed in Section \ref{drift}. So, we assume that device latencies do not affect the throughput in our scheme.
	\item \textbf{Garbage collection overhead}: The higher the GC overhead, the lower the storage throughput. As evaluated in Section \ref{sec:eval-gc}, our scheme reduces the GC overheads a bit; the saved bandwidth in turn helps the storage throughput increase slightly.
	\item \textbf{Data scrubbing overhead}: This additional storage operation consumes storage bandwidth and has a negative impact on storage throughput. However, as discussed in Section \ref{sec:eval-scrub}, our scheme does not frequently invoke the data scrubbing, which minimizes the loss of the storage throughput.
\end{itemize}
Note that our scheme does \emph{not} have an impact on other critical parameters that might affect the storage performance. For example, the degree of storage parallelization (how many I/O requests the storage can process in parallel) and inter-arrival times (how frequently I/O requests are submitted to the storage) remain unchanged under our scheme and evaluation methodology.

%% file: eval-sense-sum.tex
\section{Sensitivity Analysis}
The efficiency of the proposed D-SLC design can be influenced by device parameters or configuration setup. To examine this, we performed a series of sensitivity studies.

\subsection{Different Voltage Drift Distances}
\label{sec:sense-drift}

\subsubsection{Non-Uniform Device Characteristics}

The voltage drift distance as a function of the retention time and the P/E cycles (discussed in Section \ref{drift}) can be a little longer or shorter, depending on the device characteristics. Specifically, it is affected by a wide variety of design factors (such as vendors, technology nodes, material-level characteristics), which makes a need to evaluate our scheme in different devices exhibiting varying drift patterns. Hence, in addition to the configuration evaluated in Section \ref{sec:eval-base}, we employ two more devices by changing the scaling constant (K) of Equation \ref{eq:drift-model}. The three evaluated systems in this experiment are as follows:

\begin{itemize}[leftmargin=*]
	\item \textbf{Weak}: In this device, the voltage state drifts longer under the same P/E cycles and retention times. K is set to $5 \times 10^{-4}$.  
	\item \textbf{Normal}: This is the configuration employed so far (Section \ref{sec:eval-base}). The scaling constant K is set to $4 \times 10^{-4}$.   
	\item \textbf{Strong}: The voltage state in this device drifts shorter under the same P/E cycles and retention times (K is set to $3 \times 10^{-4}$).   
\end{itemize}

These three devices have different mappings of state modes to their blocks for each pair of P/E cycles and retention times, which are listed in Table \ref{tab:mode-assign-3-sense-drift}. For example, \emph{Strong} device can store data whose retention times are between 1 and 10 hours in blocks with 8-state mode at any time (P/E cycle), whereas in \emph{Weak} device, such data should be placed only in blocks with 4-state mode after 10K P/Es.

\begin{table}[!h]
\caption{Block state mode assignment (2, 4, or 8-state mode) for different I/O retention times as a function of the block age in Weak / Normal / Strong devices.}
\label{tab:mode-assign-3-sense-drift}
\footnotesize
\begin{tabular}{|p{0.8in}||c|c|c|c|c|  }
\hline 
\textbf{Retention}		& \multicolumn{5}{|c|}{\textbf{Number of Erases to the Block} } \\ \cline{2-6}
\textbf{Time}	&	0$\sim$10K 	&	10$\sim$20K	&	20$\sim$30K	&	30$\sim$40K	&	40$\sim$50K	 \\ 
\hline\hline 
$\le$1 Hour			&	8/8/8	& 	8/8/8	&	8/8/8	&	8/8/8	&	8/8/8	\\
\hline 
1 $\sim$10 Hours		&	8/8/8	&   	4/8/8	&	4/8/8	&	4/4/8	&	4/4/8	\\
\hline 
10 Hours $\sim$ 3 Days	& 	4/4/8	&	4/4/4	&	2/4/4	&	2/2/4	&	2/2/4	\\
\hline 
$\ge$ 3 Days 		& 	2/2/2	&	2/2/2	&	2/2/2	&	2/2/2	&	2/2/2	\\
\hline
\end{tabular}
\end{table}

\begin{figure*}[t]
\centering
\includegraphics[width=0.98\linewidth, bb= 0 0 3312 720]{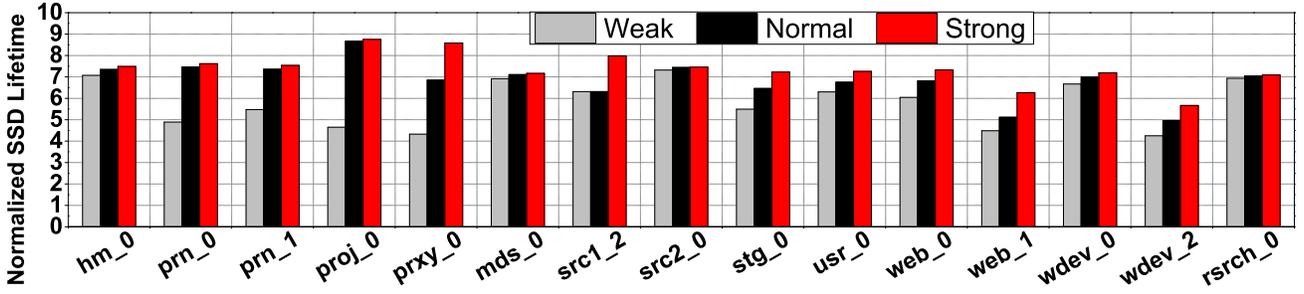}
\caption{Lifetime improvement brought by our scheme in Weak, Normal, and Strong devices, normalized to the baseline SLC.}
\label{fig:sense-drift-life}
\end{figure*}

\begin{figure*}
\centering
	\begin{subfigure}{.28\linewidth}
		\centering
		\includegraphics[width=0.99\linewidth]{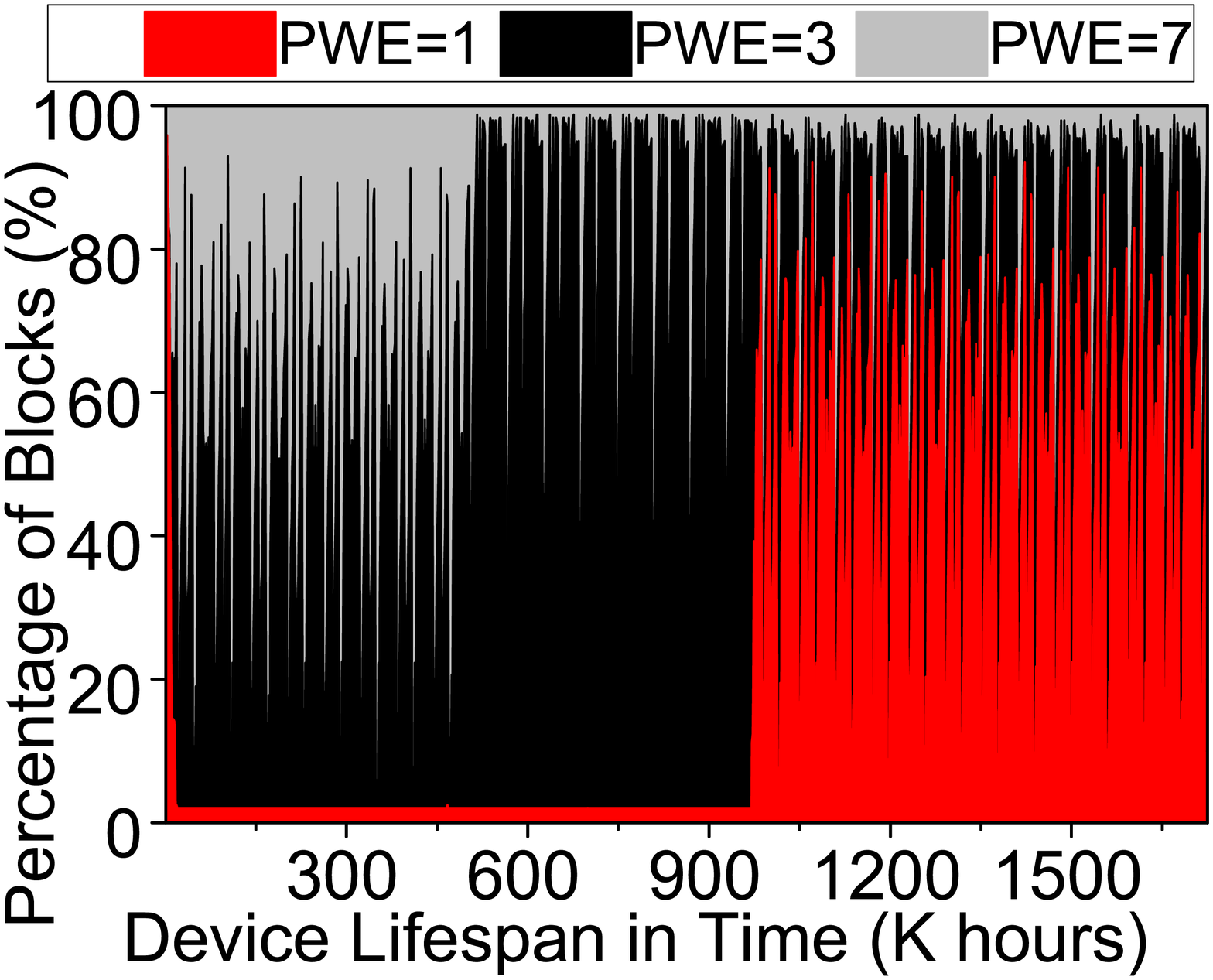}
		\caption{web\textunderscore1 in Weak}\label{fig:pwe-sense-drift-web_1-weak}
	\end{subfigure}
	\begin{subfigure}{.28\linewidth}
		\centering
		\includegraphics[width=0.99\linewidth]{pwe-web_1.eps}
		\caption{web\textunderscore1 in Normal}\label{fig:pwe-sense-drift-web_1-normal}
	\end{subfigure}
	\begin{subfigure}{.28\linewidth}
		\centering
		\includegraphics[width=0.99\linewidth]{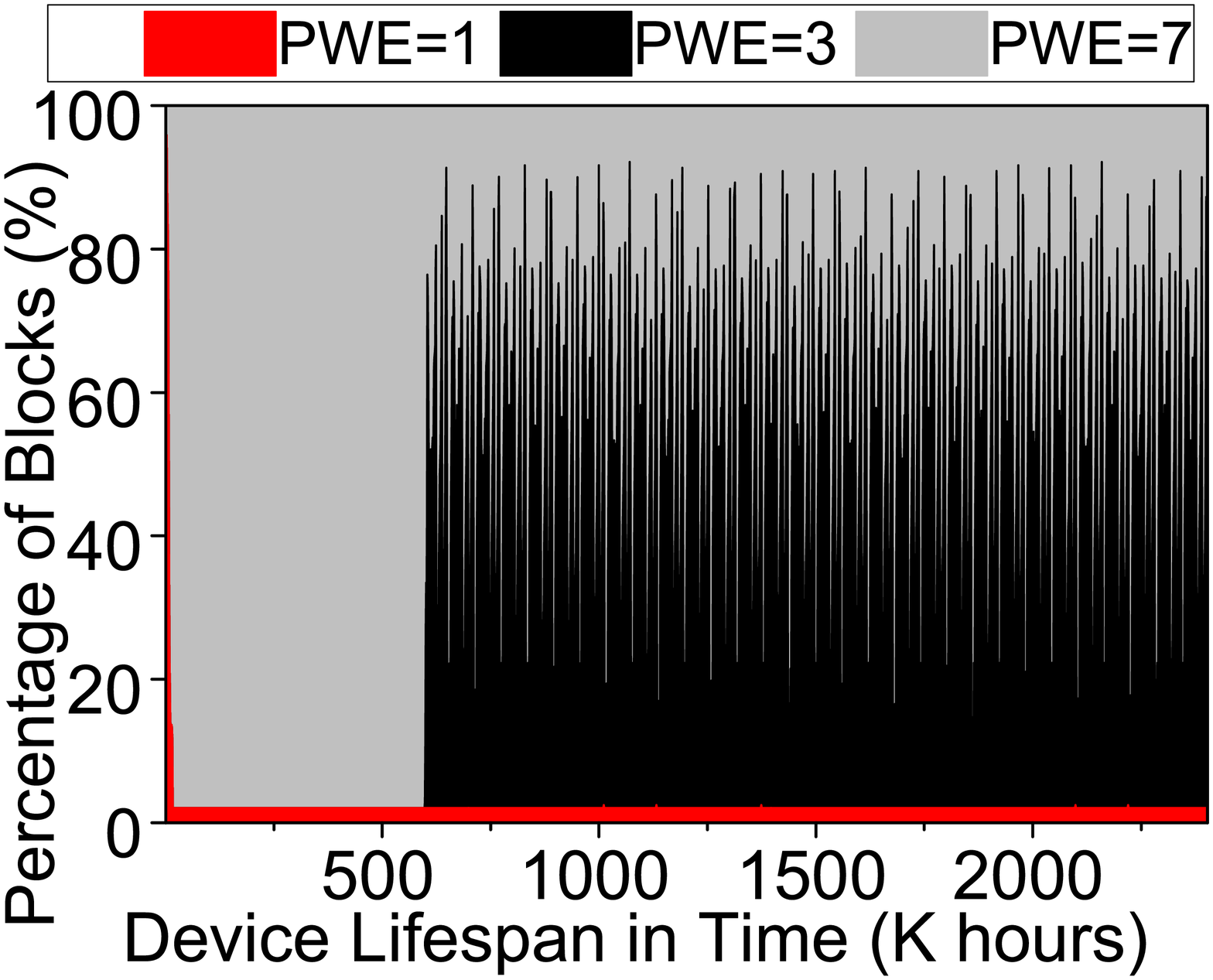}
		\caption{web\textunderscore1 in Strong}\label{fig:pwe-sense-drift-web_1-strong}
	\end{subfigure}
\caption{Percentages of blocks with 2, 4, and 8 states, when running \emph{web\textunderscore1} on Weak, Normal, and Strong devices.}
\label{fig:pwe-sense-drift-web_1}
\end{figure*}

\begin{figure*}
\centering
	\begin{subfigure}{.28\linewidth}
		\centering
		\includegraphics[width=0.99\linewidth]{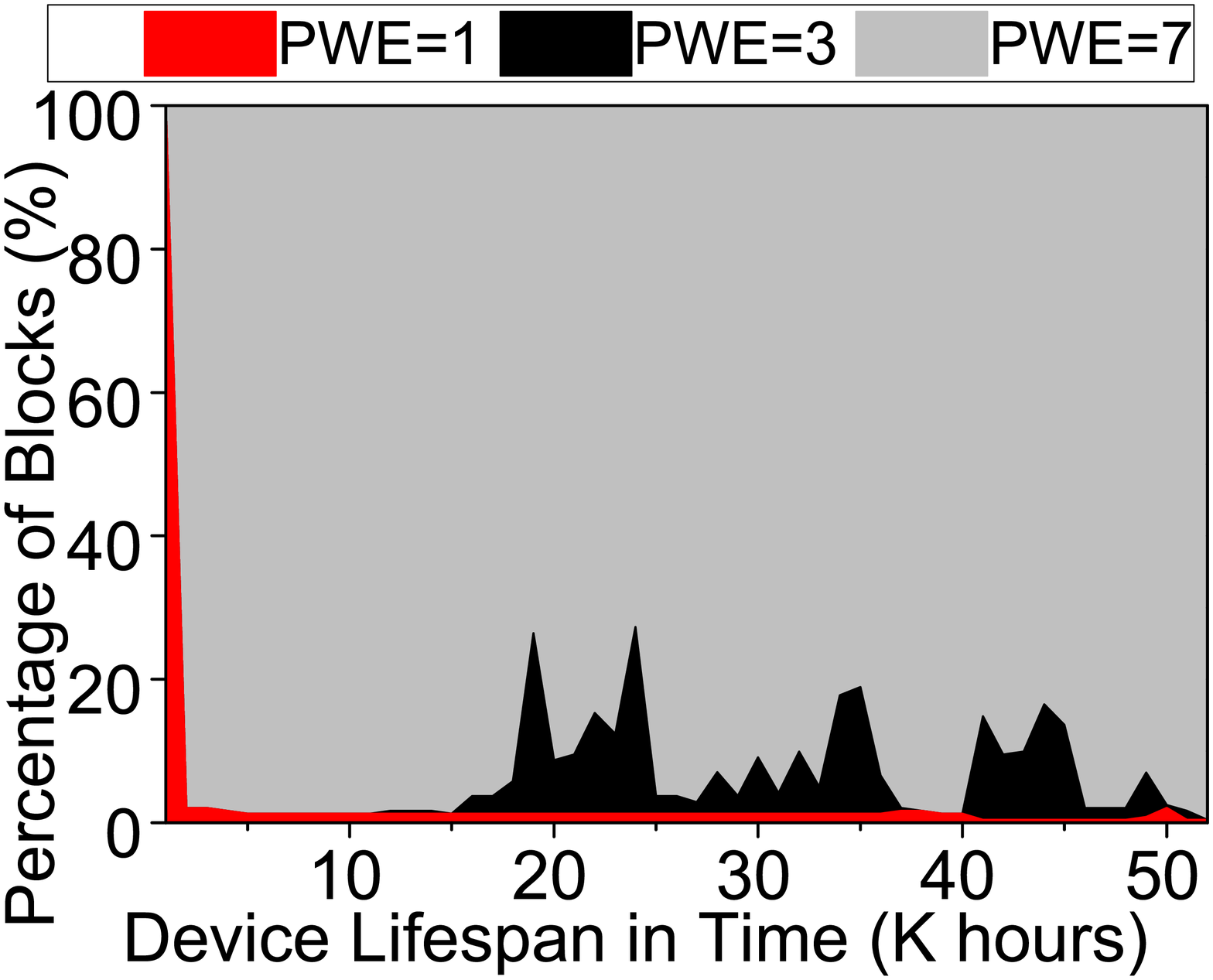}
		\caption{prn\textunderscore0 in Weak}\label{fig:pwe-sense-drift-prn_0-weak}
	\end{subfigure}
	\begin{subfigure}{.28\linewidth}
		\centering
		\includegraphics[width=0.99\linewidth]{pwe-prn_0.eps}
		\caption{prn\textunderscore0 in Normal}\label{fig:pwe-sense-drift-prn_0-normal}
	\end{subfigure}
	\begin{subfigure}{.28\linewidth}
		\centering
		\includegraphics[width=0.99\linewidth]{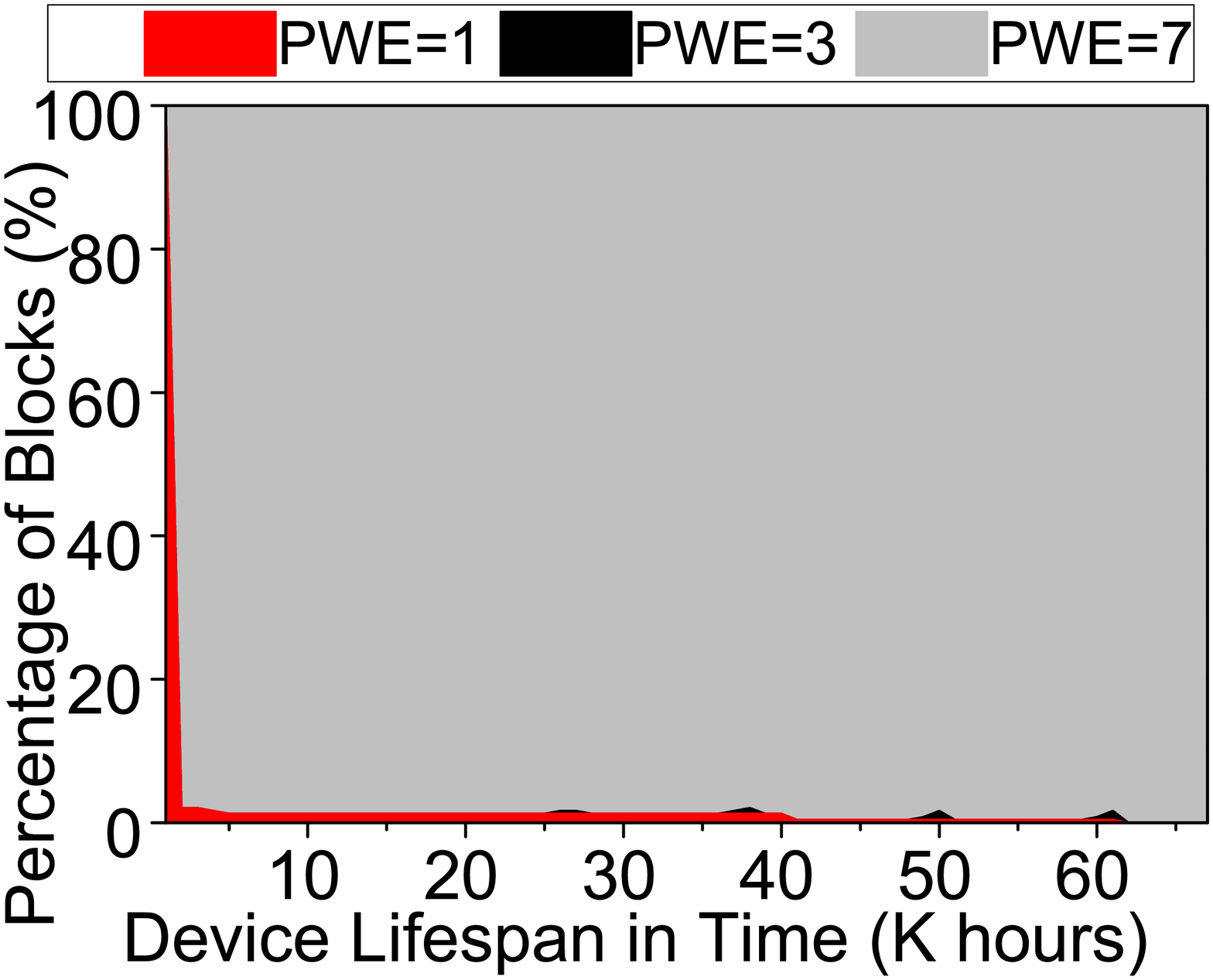}
		\caption{prn\textunderscore0 in Strong}\label{fig:pwe-sense-drift-prn_0-strong}
	\end{subfigure}
\caption{Percentages of blocks with 2, 4, and 8 states, when running \emph{prn\textunderscore0} on Weak, Normal, and Strong devices.}
\label{fig:pwe-sense-drift-prn_0}
\end{figure*}

\subsubsection{Effectiveness of Our Schemes under Different Devices.}

Figure \ref{fig:sense-drift-life} compares the lifetime improvements achieved by our scheme in three different devices. As can be seen, \emph{our scheme brings more lifetime improvements in stronger devices (where their voltage drifts are shorter) than in weaker devices (whose voltage drifts are longer)}. This is due to two main reasons:

\begin{itemize}[leftmargin=*]
	\item In all the devices, the inter-update times (retention times) of data remain unchanged. Note that this is an attribute of the workloads, regardless of device characteristics.
	\item In stronger devices, more and more data with the same retention times can be stored in blocks with more states. For example, according to Table \ref{tab:mode-assign-3-sense-drift}, the Strong device allows data with retention times between 10 hours and 3 days to be stored in blocks with 0 to 10K P/Es as 8-state modes, while such data should be accommodated in 4-state mode blocks until 10K P/Es in the Normal or Weak devices.
\end{itemize}


The PWE analysis shows how our scheme increases PWE values as the target device becomes drift-resistant (i.e., stronger). As a representative, Figures \ref{fig:pwe-sense-drift-web_1-weak}, \ref{fig:pwe-sense-drift-web_1-normal}, and \ref{fig:pwe-sense-drift-web_1-strong} show the percentage of blocks with 2, 4, and 8 states, when running \emph{web\textunderscore1} in Weak, Normal, and Strong devices, respectively. (Note that the PWE analysis for all other workloads can be referred in Appendix). One can see from these figures that the ratios of blocks with 4 states (black area) and with 2 states (red area) gradually decrease, as the drift resistance increases from Weak to Strong devices. Specifically, the Weak device needs blocks with 2 states to store data whose retention times range from 10 hours to 3 days from around 1,000K hours (i.e., 20K P/Es), whereas such data require 2-state mode blocks after around 1,500K hours (i.e., 30K P/Es) in the Normal device. On the other hand, no 2-state mode block is needed to store such data in Strong device; no red cliff is observed in Figure \ref{fig:pwe-sense-drift-web_1-strong}.

One might also observe that, in some workloads such as \emph{prn\textunderscore0} and \emph{proj\textunderscore0}, the lifetime improvements stay quite low in the Weak device. This phenomenon of these workloads can be explained by the PWE analysis. Figures \ref{fig:pwe-sense-drift-prn_0-weak}, \ref{fig:pwe-sense-drift-prn_0-normal}, and \ref{fig:pwe-sense-drift-prn_0-strong} show the percentage of blocks with three different states when executing \emph{prn\textunderscore0}, a representative of such workloads, in the Weak, Normal, and Strong devices, respectively. The ratio of blocks with 4 states is very low in the Normal device, and it is almost removed in the Strong device. In contrast, the percentage of 4-state mode blocks largely increases in the Weak device; the black area in Figure \ref{fig:pwe-sense-drift-prn_0-weak} appears remarkably. It is because the data whose retention times range 1 to 10 hours need 4-state mode blocks so early (i.e., after 10K P/Es), while such data need them after 30K P/Es in the Normal device and none of them throughout the lifespan in the Strong device.


\begin{table}
\caption{The assignment of state modes (2, 4, 5, 6, or 8-state mode) to blocks in our schemes supporting 2/3/4/5 modes.}
\label{tab:mode-assign-5-sense-mode}
\footnotesize
\begin{tabular}{|p{0.8in}||c|c|c|c|c|  }
\hline 
\textbf{Retention}		& \multicolumn{5}{|c|}{\textbf{Number of Erases to the Block} } \\ \cline{2-6}
\textbf{Time}	&	0$\sim$10K 	&	10$\sim$20K	&	20$\sim$30K	&	30$\sim$40K	&	40$\sim$50K	 \\ 
\hline\hline 
$\le$1 Hour			&	8/8/8/8 &	8/8/8/8 &	8/8/8/8 &	8/8/8/8 &	8/8/8/8 \\
\hline 
1 $\sim$10 Hours		&	8/8/8/8 &	8/8/8/8 &	8/8/8/8 &	\textbf{2/4/5/6} & \textbf{2/4/5/6} \\
\hline 
10 Hours$\sim$3 Days	&  \textbf{2/4/5/6} & \textbf{2/4/5/5} & \textbf{2/4/4/4} & 2/2/2/2 & 2/2/2/2 \\
\hline 
$\ge$ 3 Days 			& 	2/2/2/2 &	2/2/2/2 &	2/2/2/2 &	2/2/2/2 &	2/2/2/2 \\
\hline
\end{tabular}
\end{table}

\subsection{Different Numbers of State Modes}

\subsubsection{Varying the Granularity of Our Scheme}

So far, our scheme has employed ``three'' different modes, which are 2, 4, and 8-state modes. However, one might want to manage voltage drifts at finer granularities by employing additional modes such as 5 and 6-state modes. Thus, we also evaluate our scheme supporting different numbers of state modes. In particular, we compare the following four systems:

\begin{itemize}[leftmargin=*]
	\item \textbf{2-Mode}: It adds one additional mode (8-state) to the baseline SLC; so, this system has two modes (2 and 8-state modes).
	\item \textbf{3-Mode}: This system is what we considered throughout the paper, which has three state modes (2, 4, and 8-state modes).
	\item \textbf{4-Mode}: One more state mode is added to the Mode-3, which has four state modes (2, 4, 5, and 8-state modes).
	\item \textbf{5-Mode}: Managing voltage drift more precisely, it supports five different modes (2, 4, 5, 6, and 8-state modes).
\end{itemize}

Increasing the number of supporting modes changes the mapping of their state modes to blocks. Table \ref{tab:mode-assign-5-sense-mode} provides the different mappings of state modes to their blocks for each pair of P/E cycles and retention times in the four evaluated systems.

\begin{figure}
\centering
\begin{subfigure}{.49\linewidth}
	\centering
	\includegraphics[width=0.99\linewidth, bb= 0 0 648 504]{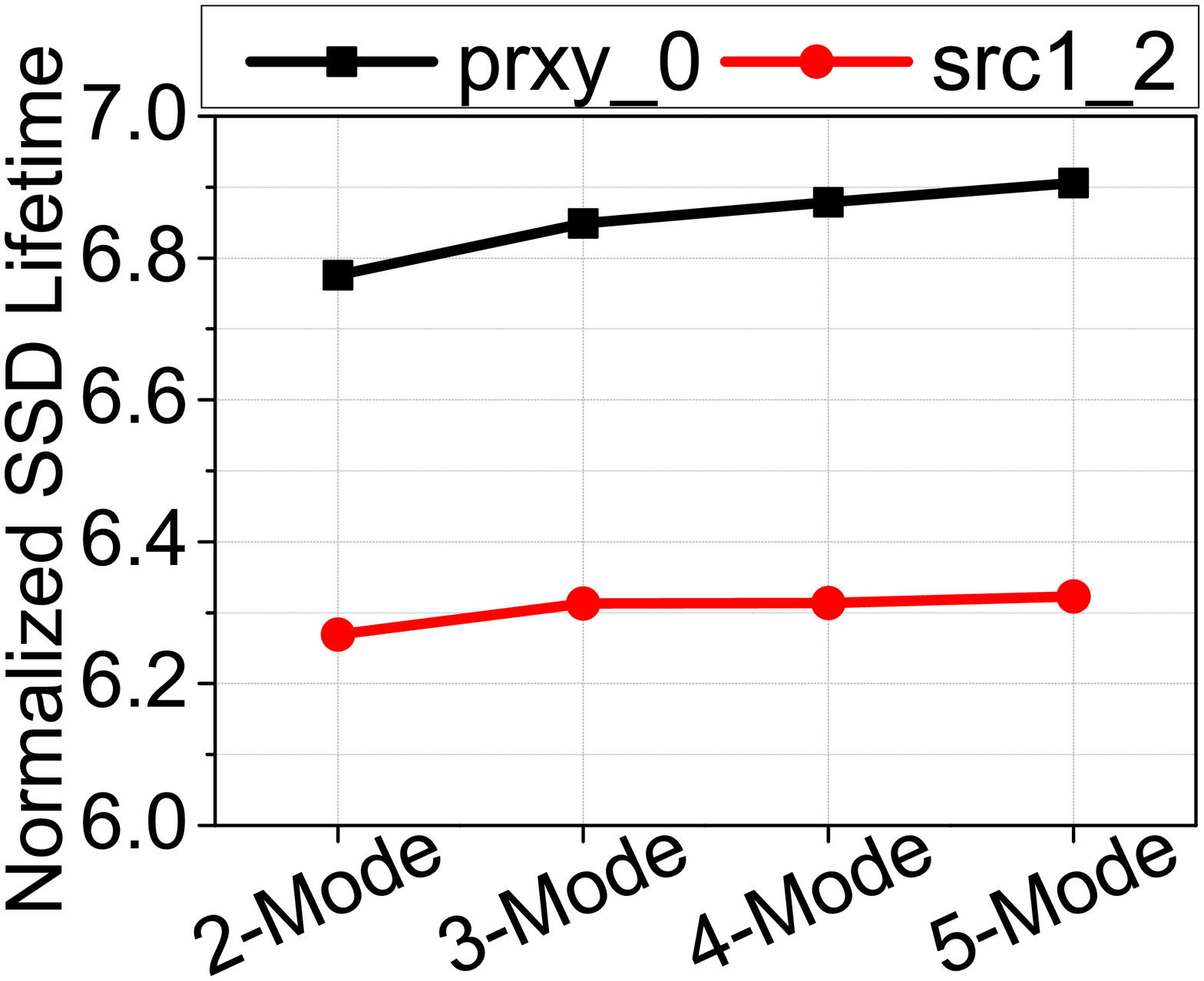}
	\caption{Low-beneficial workloads}\label{fig:sense-mode-life-same}
\end{subfigure}
\begin{subfigure}{.49\linewidth}
	\centering
	\includegraphics[width=0.99\linewidth, bb= 0 0 648 504]{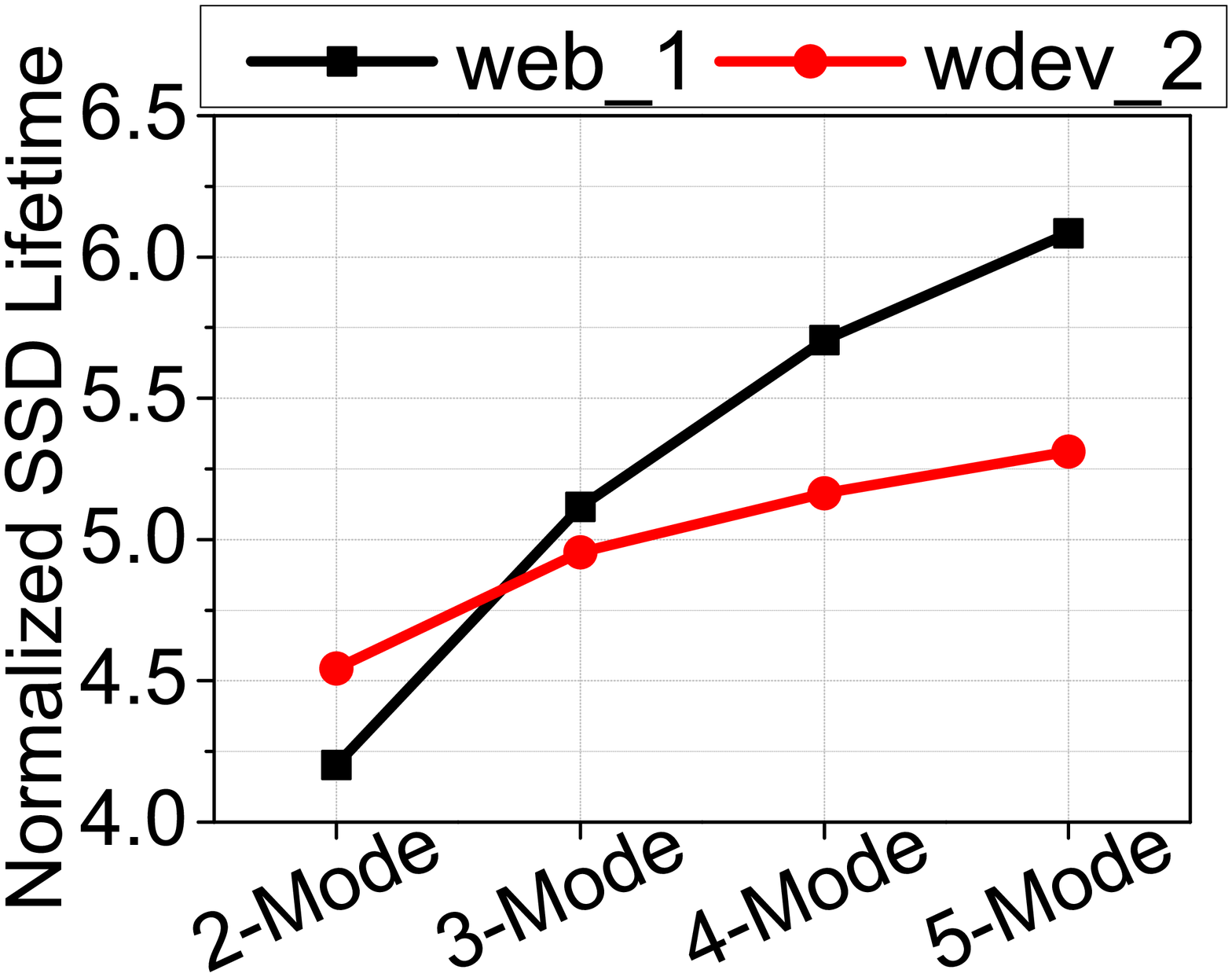}
	\caption{High-beneficial workloads}\label{fig:sense-mode-life-increase}
\end{subfigure}
\caption{Lifetime improvement brought by device under varying numbers of states, normalized to the baseline SLC.}
\label{fig:sense-mode-life}
\end{figure}

\begin{figure*}
\centering
	\begin{subfigure}{.24\linewidth}
		\centering
		\includegraphics[width=0.99\linewidth]{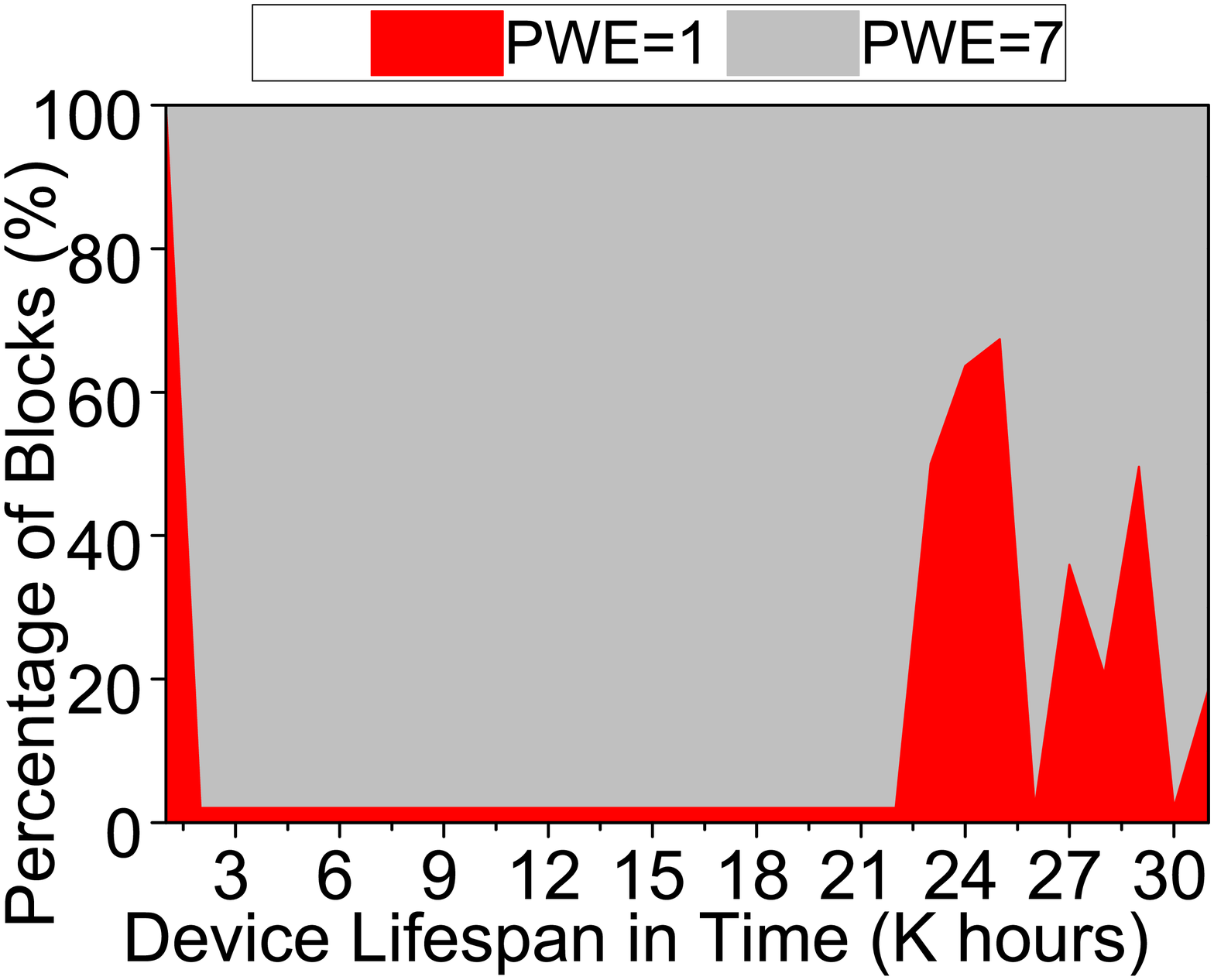}
		\caption{src1\textunderscore2 in 2-Mode}\label{fig:pwe-sense-mode-src1_2-m2}
	\end{subfigure}
	\begin{subfigure}{.24\linewidth}
		\centering
		\includegraphics[width=0.99\linewidth]{pwe-src1_2.eps}
		\caption{src1\textunderscore2 in 3-Mode}\label{fig:pwe-sense-mode-src1_2-m3}
	\end{subfigure}
	\begin{subfigure}{.24\linewidth}
		\centering
		\includegraphics[width=0.99\linewidth]{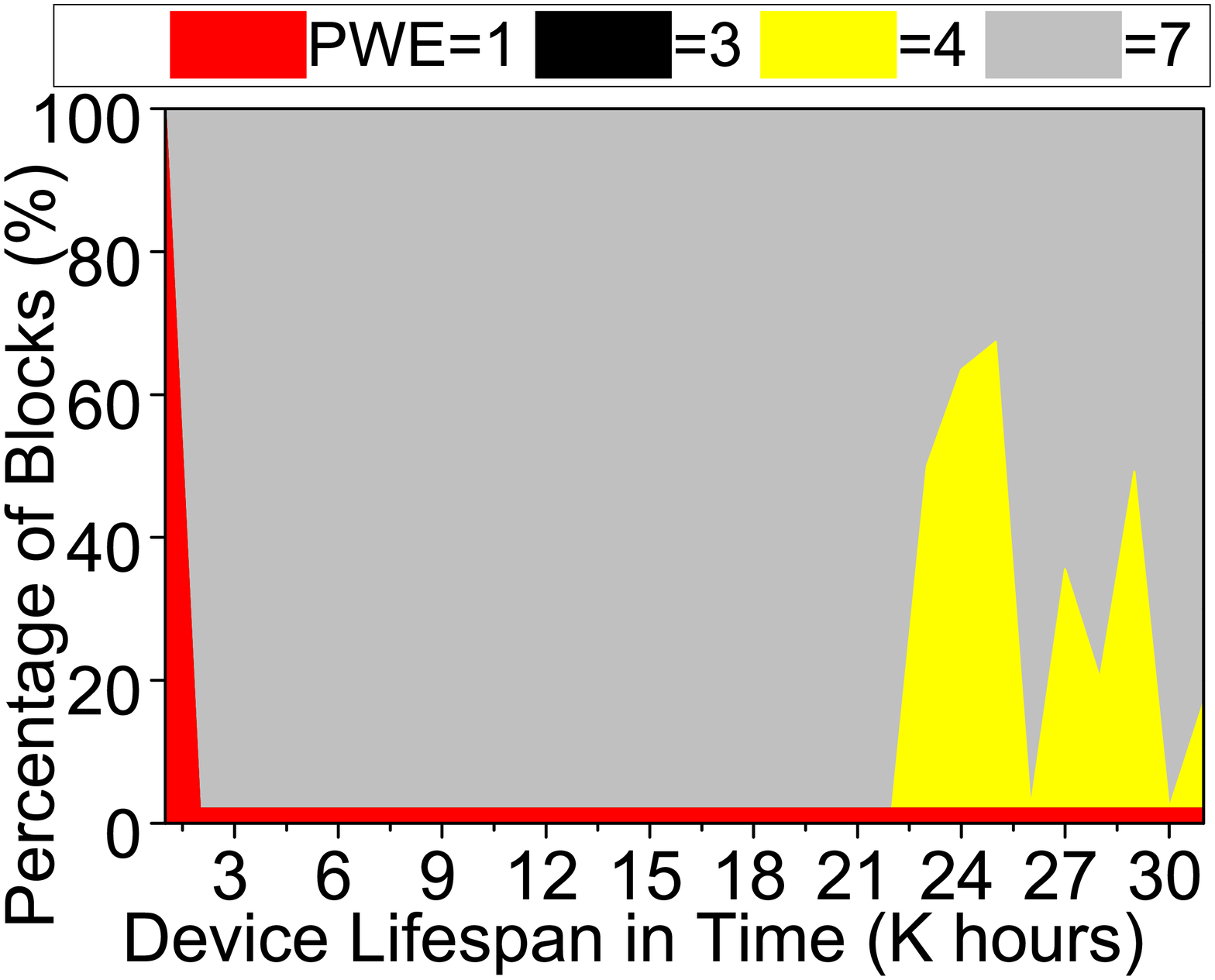}
		\caption{src1\textunderscore2 in 4-Model}\label{fig:pwe-sense-mode-src1_2-m4}
	\end{subfigure}
	\begin{subfigure}{.24\linewidth}
		\centering
		\includegraphics[width=0.99\linewidth]{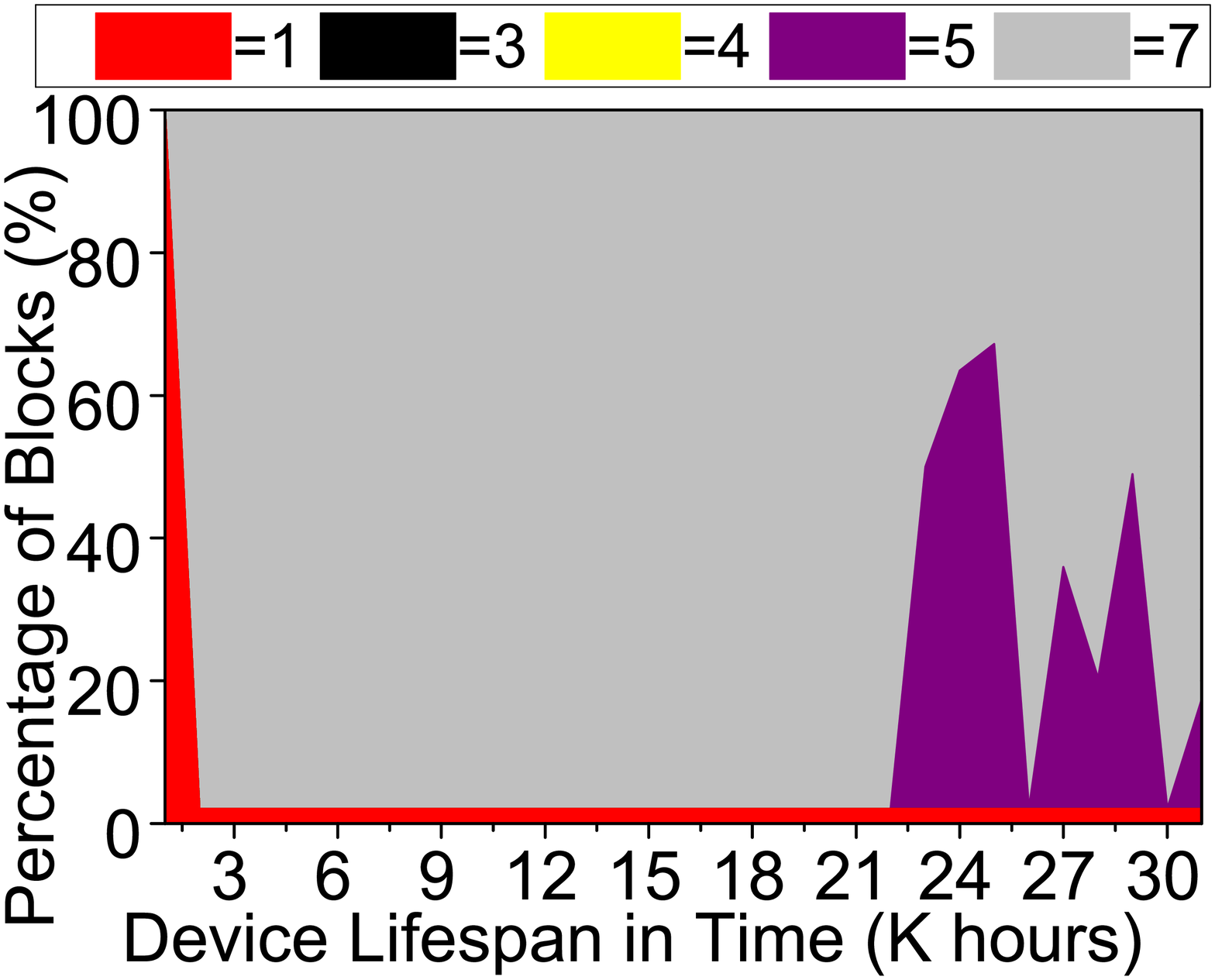}
		\caption{src1\textunderscore2 in 5-Model}\label{fig:pwe-sense-mode-src1_2-m5}
	\end{subfigure}
\caption{Percentages of blocks with different states, when running \emph{src1\textunderscore2} in 2-Mode, 3-Mode, 4-Mode, and 5-Mode devices.}
\label{fig:pwe-sense-mode-src1_2}
\end{figure*}

\begin{figure*}
\centering
	\begin{subfigure}{.24\linewidth}
		\centering
		\includegraphics[width=0.99\linewidth]{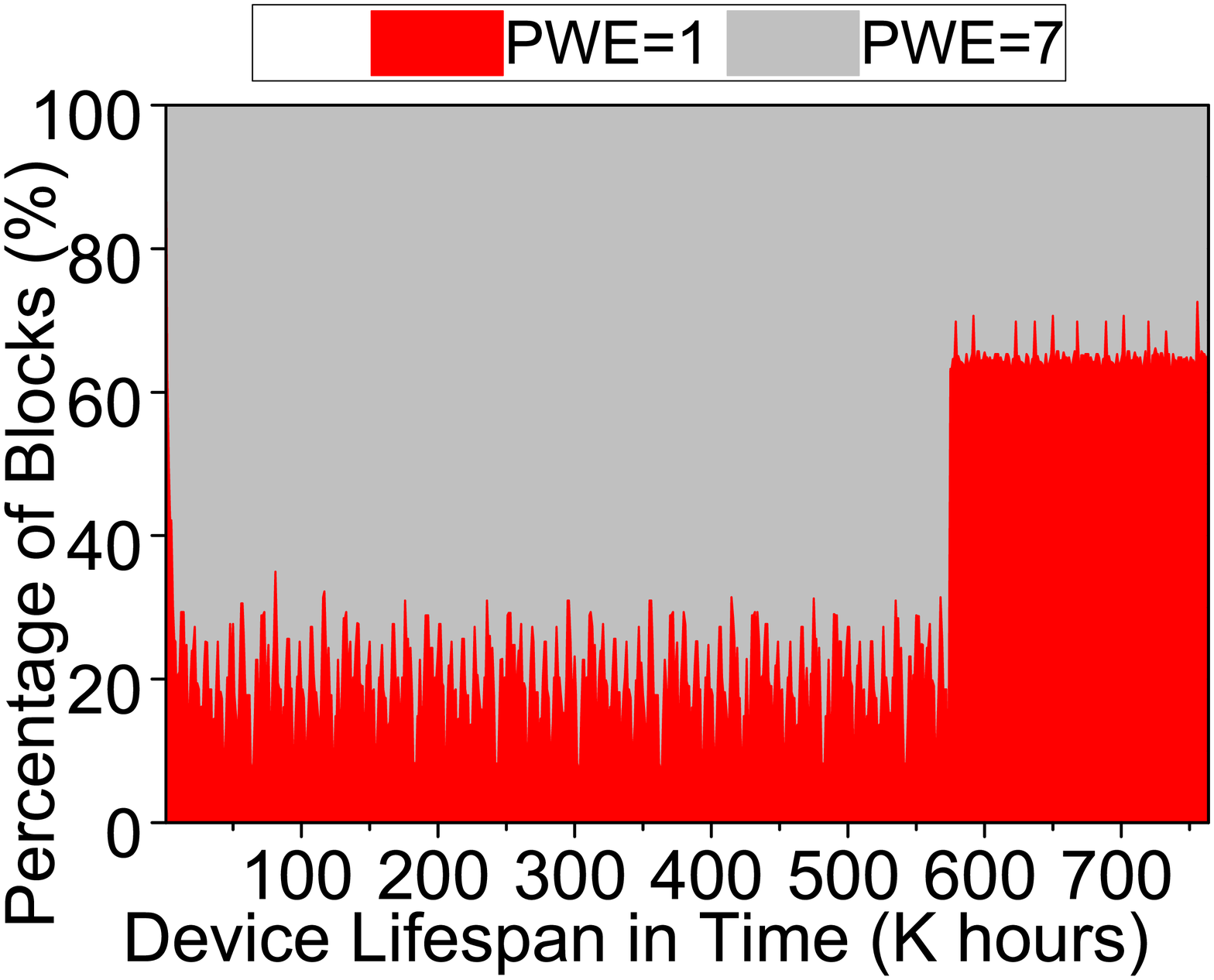}
		\caption{wdev\textunderscore2 in 2-Mode}\label{fig:pwe-sense-mode-wdev_2-m2}
	\end{subfigure}
	\begin{subfigure}{.24\linewidth}
		\centering
		\includegraphics[width=0.99\linewidth]{pwe-wdev_2.eps}
		\caption{wdev\textunderscore2 in 3-Mode}\label{fig:pwe-sense-mode-wdev_2-m3}
	\end{subfigure}
	\begin{subfigure}{.24\linewidth}
		\centering
		\includegraphics[width=0.99\linewidth]{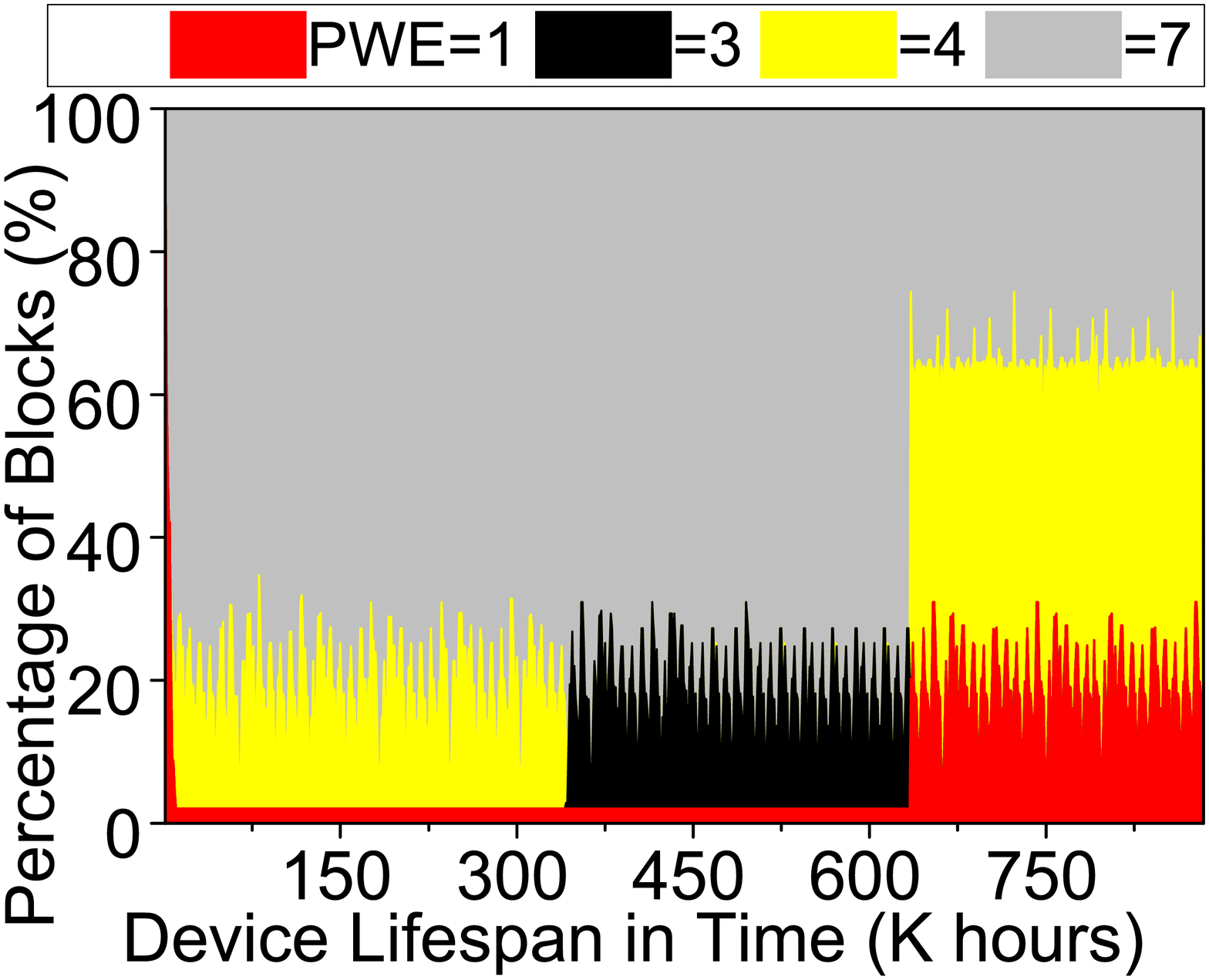}
		\caption{wdev\textunderscore2 in 4-Mode}\label{fig:pwe-sense-mode-wdev_2-m4}
	\end{subfigure}
	\begin{subfigure}{.24\linewidth}
		\centering
		\includegraphics[width=0.99\linewidth]{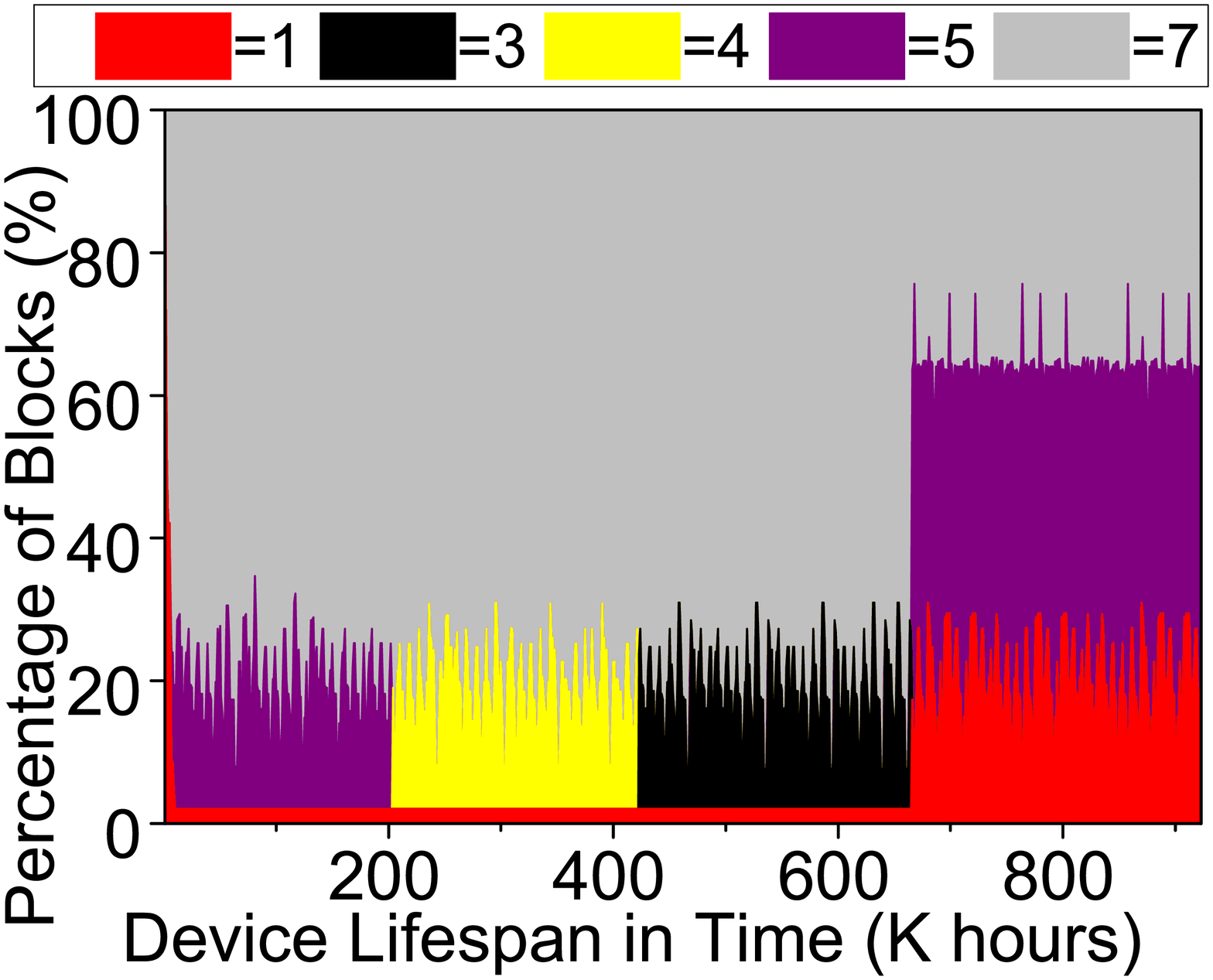}
		\caption{wdev\textunderscore2 in 5-Mode}\label{fig:pwe-sense-mode-wdev_2-m5}
	\end{subfigure}
\caption{Percentages of blocks with different states, when running \emph{wdev\textunderscore2} in 2-Mode, 3-Mode, 4-Mode, and 5-Mode devices.}
\label{fig:pwe-sense-mode-wdev_2}
\end{figure*}

\subsubsection{Effectiveness of Our Schemes under Varying Granularities}

In general, the more voltage states, the longer the device lifetime. However, some workloads significantly benefit from increasing the number of modes, while the lifetime gain is negligible in others. Figure \ref{fig:sense-mode-life} shows the lifetime improvement achieved by the 2-Mode, 3-Mode, 4-Mode, and 5-Mode devices in four representative workloads, which are categorized into two groups. (the results for all others are shown in Appendix).

\begin{itemize}[leftmargin=*]
	\item \textbf{Low-beneficial workloads}: As shown in Figure \ref{fig:sense-mode-life-same}, workloads in this category (e.g., prxy\textunderscore0, src1\textunderscore2) benefit less (or negligible) from the increasing number of modes. It is because majority retention times in these workloads are below 10 hours, while the difference among the 2, 3, 4, and 5-Mode devices is the assignment of different voltage modes (2, 4, 5, and 6-state mode, respectively) to blocks whose P/Es are between 30K and 50K (see the last two mapping of the second row in Table \ref{tab:mode-assign-5-sense-mode}).
	\item \textbf{High-beneficial workloads}: For workloads in this group (e.g., web\textunderscore1, wdev\textunderscore2), increasing the number of supporting modes leads to a significant device lifetime improvement (see Figure \ref{fig:sense-mode-life-increase}). These workloads include a large amount of data whose retention times are between 10 hours and 3 days, which can be placed in blocks with finer-granularity state modes such as 6 and 5-state mode in the 4-Mode/5-Mode devices. In contrast, in the 2-Mode/3-Mode devices, such data are accommodated in blocks with 2 and 4-state mode. (compare the third row of Table \ref{tab:mode-assign-5-sense-mode}).
\end{itemize}

The PWE analysis shown in Figure \ref{fig:pwe-sense-mode-src1_2} illustrates why a low-beneficial workload (src1\textunderscore2) cannot fully draw the full potential of increasing the number of modes, whereas Figure \ref{fig:pwe-sense-mode-wdev_2} illustrates how a high-beneficial workload (wdev\textunderscore2) experiences a significantly-increased lifespan by supporting more modes. In \emph{src1\textunderscore2}, the cliff at the latter of its lifespan represents that the blocks ``where data whose retention times range from 1 to 10 hours are stored'' change their modes, when their P/E cycles go beyond 30K. Such data can use blocks with 2 (red), 4 (black), 5 (yellow), and 6 (purple) states in the 2-Mode, 3-Mode, 4-Mode, and 5-Mode devices, respectively. Consequently, these small differences do not results in a significant lifetime gain. In contrast, \emph{wdev\textunderscore2} includes a lot of data whose retention times are between 10 hours and 3 days, and such data can be stored in blocks with more states (or higher PWEs) in early P/E cycles (i.e., from 0 to 30K) in the 4-Mode and 5-Mode devices. As a result, one can observe from the 5-Mode device (Figure \ref{fig:pwe-sense-mode-wdev_2-m5}) that the 20\% of total blocks gradually change their modes (i.e., red, black, yellow, and purple areas) throughout the device lifespan, which results in much higher PWEs, compared to the continuous low and unchanged PWE (i.e., ``1'') for the same blocks in the 2-Mode device (see the red area in Figure \ref{fig:pwe-sense-mode-wdev_2-m2}).

%% file: related-works.tex
\section{Related Work}
Retention time relaxation has been considered as an attractive optimization option for flash memories. Prior work exploit this capability for different purposes and design trade-offs. We categorize the related works in two groups:
\begin{enumerate}[leftmargin=*]
\item \textbf{Using retention time relaxation for enhancing write performance~\cite{relax-1,relax-2}:} The flash write latency based on the ISPP \cite{ispp-1} is mainly determined by two components, namely, (i) the number of ISPP loops and (ii) staircase-up amplitude. In practice, storing data for long retention times needs a number of ISPP loops (and in turn a long write latency) by forming the threshold voltage in the exact ranges. Instead, the works in this group attempt to reduce the number of ISPP (and the write latency) by placing the target threshold less-accurately. Note that our D-SLC targets lifetime enhancements and tries to keep the write latency similar or very close to the baseline SLC (by keeping the number of ISPP pulses and pulse' durations similar to the baseline).
 \item \textbf{Using retention time relaxation for enhancing flash lifetime~\cite{relax-3}:} Similar to D-SLC, WARM~\cite{relax-3} optimizes flash lifetime by taking advantage of retention relaxation. However, there are substantial differences between the two approaches. WARM begins with \emph{a retention-relaxed flash memory which employs refresh mechanism} to avoid data loss. Motivated by the high overhead of refresh for hot data (i.e., those with longevity less than the refresh period), they propose an algorithm for hot data detection and design separate pools of hot and cold blocks for the efficient refresh management. In contrast, D-SLC is a generic design and the baseline should not necessarily be a retention-relaxed flash nor a flash with the refresh support. Furthermore, instead of employing an algorithm to estimate the data longevity, D-SLC includes a heuristic mechanism (based on data scrubbing), which is able to put data with similar data longevity history in the same block. More importantly, D-SLC writes multiple bits into a cell during one erase cycle, while WARM allows just a single-bit write in each erase cycle (as the baseline SLC does). Note that D-SLC improves the lifetime by increasing PWE, whereas WARM (still keeping the PWE one) achieves it by removing unnecessary refreshes. Therefore, D-SLC and WARM can be combined for further lifetime improvement.
\end{enumerate}

%% file: conclusion.tex
\section{Conclusions}
Despite the advances in non-volatile memory technologies, flash-based SCMs are still widely used by commercial computing systems, ranging from laptop to desktop to enterprise systems, to hide the performance-cost gap between DRAM and HDD.
However, the limited endurance seems to be the main issue for flash-based SCMs, and is the target of our design and optimization in this paper.
Specifically, we make three main contributions in this paper:
First, by quantifying data longevity in an SCM, we show that a majority of the data stored in a solid-state SCM do not require long retention times provided by flash memory.
Second, by relaxing the guaranteed retention time, we propose a novel mechanism, named Dense-SLC (D-SLC), which enables us perform multiple writes into a cell during each erase cycle for lifetime extension.
Third, we discuss the required changes in the FTL in order to exploit these characteristics for extending the lifetime of solid-state part of an SCM.
Using an extensive simulation-based analysis of a flash-based SCM, we demonstrate that our proposed D-SLC is able to significantly improve device lifetime (between 5.1$\times$ and 8.6$\times$) with no performance overhead and also very small changes in the FTL software.